\documentclass[11pt]{article}
\usepackage[utf8]{inputenc}
\usepackage{graphicx}
\usepackage{rviewport}
\usepackage{hyperref}
\usepackage{xcolor}
\usepackage{cite}
\usepackage{xspace}
\usepackage{url}
\usepackage{amsmath}
\usepackage{amssymb}
\usepackage{mathpazo}
\usepackage{tocloft}
\usepackage{wrapfig}
\usepackage{tikz}
\usepackage{caption}
\usepackage{paralist}

\usetikzlibrary{decorations.pathmorphing}
\definecolor{darkBlue}{rgb}{0, 0, 0.8}
\hypersetup{
   bookmarks=true,         % show bookmarks bar?
   colorlinks = true,
   citecolor  = darkBlue,
   urlcolor   = darkBlue,
   linkcolor  = darkBlue
}
\def\flashtalkDone#1#2{%\cleardoublepage
  \subsubsection[#1: #2]{\textbf{\textsc{#1}}:  {\sl #2}}}

\def\gcm{g/cm$^2$\xspace}
\def\deg{$^\circ$\xspace}
\def\Xmax{\ensuremath{X_\text{max}}\xspace}
\def\q{$\bullet$~}
\newcommand{\xmax}[0]{\Xmax}

\newcommand{\fn}[1]{\uppercase{\expandafter\firstletter#1\relax}.\,}
\def\firstletter#1#2\relax{#1}

\def\a#1#2#3{\def\test{#3}{\scriptsize #1 #2\ignorespacesafterend}\ifx\test\empty\else~{\scriptsize(#3)}\fi\nolinebreak,}
\def\aEnd#1#2#3{\def\test{#3}{\scriptsize #1 #2\ignorespacesafterend}\ifx\test\empty\else~{\scriptsize(#3)}\fi\nolinebreak.}

\tikzstyle{mybox} = [draw=black, fill=white, very thick,
    rectangle, rounded corners, inner sep=10pt, inner ysep=10pt]
\tikzstyle{fancytitle} =[fill=black, text=white, font=\bfseries]
\newcommand{\infobox}[3]{%
  \begin{wrapfigure}{r}{8cm}
\begin{tikzpicture}
\node [mybox] (box){%
    \begin{minipage}{0.45\textwidth}
    $\,$ #2
    \end{minipage}
};
\node[fancytitle, right=10pt] at (box.north west) {#1};
\end{tikzpicture}
#3
\end{wrapfigure}
}
\graphicspath{{flashtalks/plots/}}

\title{Ideas and Requirements for the\\ Global Cosmic-Ray Observatory}

\date{\today}
\begin{document}
%\maketitle

\begin{center}
\textbf{\Large Ideas and Requirements for the\\[0.5\baselineskip] Global Cosmic-Ray Observatory (GCOS)}\\[\baselineskip]

\today
\end{center}
%%%%%%%%%%%%%%%%%%%%%%%%%%%%
{\parindent0mm

%\par\noindent\rule{\textwidth}{0.4pt}
%\textsl{Participants of the 2021 workshop:}
%\input{authors-nijmegen}

\par\noindent\rule{\textwidth}{0.4pt}
\textsl{\footnotesize Participants of the 2022 workshop:}
\a{Kevin}{Almeida Cheminant}{Institute of Nuclear Physics Polish Academy of Sciences}
\a{Jaime}{Alvarez-Muniz}{IGFAE \& Univ. Santiago de Compostela}
\a{Rafael}{Alves Batista}{Universidad Autónoma de Madrid}
\a{Luis}{Anchordoqui}{Lehman College, City University of New York}
\a{Jose}{Bellido}{The University of Adelaide}
\a{Mario}{Bertaina}{Univ. \& INFN Torino, Italy}
\a{Sonali}{Bhatnagar}{Dayalbagh educational institute}
\a{Pierre}{Billoir}{LPNHE Sorbonne Universite}
\a{Teresa}{Bister}{RWTH Aachen University}
\a{Martina}{Bohacova}{FZU}
\a{Washington}{Carvalho Jr}{Radboud University}
\a{Lorenzo}{Cazon}{IGFAE - USC}
\a{Alan}{Coleman}{Bartol Research Institute, University of Delaware}
\a{Fabio}{Convenga}{KIT}
\a{Kai}{Daumiller}{KIT}
\a{Bruce}{Dawson}{University of Adelaide}
\a{Luca}{Deval}{KIT}
\a{Armando}{di Matteo}{INFN Torino}
\a{Ralph}{Engel}{KIT}
\a{Johannes}{Eser}{The University of Chicago}
\a{Ke}{Fang}{University of Wisconsin-Madison}
\a{Glennys R.}{Farrar}{New York University}
\a{Anatoli}{Fedynitch}{Institute of Physics, Academia Sinica}
\a{Thomas}{Fitoussi}{KIT}
\a{Tomas}{Fodran}{Radboud University}
\a{Toshihiro}{Fujii}{Osaka Metropolitan University}
\a{Keitaro}{Fujita}{ICRR, University of Tokyo}
\a{Maria Vittoria}{Garzelli}{Universitaet Hamburg}
\a{Noemie}{Globus}{University of California Santa Cruz and Flatiron Institute}
\a{Quanbu}{Gou}{Institute of High Energy Physics, Chinese Academy of Sciences}
\a{Balakrishnan}{Hariharan}{Tata Institute of Fundamental Research}
\a{Andreas}{Haungs}{KIT}
\a{Ryo}{Higuchi}{RIKEN}
\a{Bohdan}{Hnatyk}{Taras Shevchenko National University of Kyiv, Astronomical Observatory}
\a{Tim}{Huege}{KIT}
\a{Jörg}{Hörandel}{Radboud University}
\a{Matthias}{Kadler}{JMU Würzburg}
\a{Karl-Heinz}{Kampert}{University Wuppertal}
\a{Donghwa}{Kang}{KIT}
\a{Abha}{Khakurdikar}{Radboud University}
\a{Eiji}{Kido}{RIKEN Cluster for Pioneering Research}
\a{Ramesh}{Koirala}{Nanjing University}
\a{Chuizheng}{Kong}{Nanjing University}
\a{John}{Krizmanic}{NASA/GSFC}
\a{Shivam}{Kulshrestha}{Dayalbagh Educational Institute, India}
\a{Viktoria}{Kungel}{Colorado School of Mines}
\a{Agnieszka}{Leszczyńska}{University of Delaware}
\a{Ruoyu}{Liu}{Nanjing University}
\a{Quentin}{Luce}{KIT}
\a{Allan}{Machado Payeras}{KIT, UNICAMP}
\a{Volodymyr}{Marchenko}{Astronomical Observatory of the Jagiellonian University}
\a{Analisa}{Mariazzi}{IFLP CONICET Universidad Nacional de La Plata}
\a{Ioana}{Mari\c{s}}{ULB}
\a{Eric}{Mayotte}{Colorado School of Mines}
\a{Kohta}{Murase}{Penn State University}
\a{Marco}{Muzio}{Penn State University}
\a{Ana Laura}{Müller}{ELI Beamlines, Institute of Physics, Czech Academy of Sciences}
\a{David}{Nitz}{Michigan Tech / Radboud}
\a{Toshiyuki}{Nonaka}{Institute for Cosmic Ray Research}
\a{Shoichi}{Ogio}{ICRR, The University of Tokyo}
\a{Foteini}{Oikonomou}{Norwegian University of Science and Technology}
\a{Angela V}{Olinto}{University of Chicago}
\a{Hitoshi}{Oshima}{ICRR, the University of Tokyo}
\a{Ek Narayan}{Paudel}{University of Delaware}
\a{Thomas}{Paul}{Lehman College, City University of New York}
\a{Vincent}{Pelgrims}{Institute of Astrophysics - FORTH}
\a{Lorenzo}{Perrone}{Università del Salento and INFN Lecce}
\a{Bjarni}{Pont}{Radboud University Nijmegen}
\a{Alessio}{Porcelli}{Universidad de Antofagasta}
\a{Julian}{Rautenberg}{University of Wuppertal}
\a{Felix}{Riehn}{LIP}
\a{Markus}{Risse}{University of Siegen}
\a{Markus}{Roth}{KIT}
\a{Francesco}{Salamida}{University of L'Aquila and INFN LNGS}
\a{Andrea}{Santangelo}{IAAT University of Tuebingen}
\a{Eva}{Santos}{FZU}
\a{Fred}{Sarazin}{Colorado School of Mines}
\a{Viviana}{Scherini}{Università del Salento and INFN Lecce}
\a{Harald}{Schieler}{KIT}
\a{David}{Schmidt}{KIT}
\a{Harm}{Schoorlemmer}{NIKHEF/RU}
\a{Christoph}{Schäfer}{KIT}
\a{Olga}{Sergijenko}{Main Astronomical Observatory of NASU}
\a{Dennis}{Soldin}{KIT}
\a{Mauricio}{Suarez-Duran}{Université Libre de Bruxelles}
\a{Kaoru}{Takahashi}{ICRR}
\a{Masahiro}{Takeda}{ICRR, University of Tokyo}
\a{Yuichiro}{Tameda}{Osaka Electro-Communication University}
\a{Olena}{Tkachenko}{KIT}
\a{Takayuki}{Tomida}{Shinshu University}
\a{Petr}{Travnicek}{FZU}
\a{Michael}{Unger}{KIT}
\a{Arjen}{van Vliet}{Khalifa University}
\a{Darko}{Veberi\v{c}}{KIT}
\a{Tonia}{Venters}{NASA GSFC}
\a{Valerio}{Verzi}{INFN Roma Tor Vergata}
\a{Jakub}{Vicha}{FZU}
\a{Alan  A.}{Watson}{University of Leeds}
\a{Alexey}{Yushkov}{Institute of Physics AS CR, Prague}
\aEnd{Pengfei}{Zhang}{Xidian University}

\par\noindent\rule{\textwidth}{0.4pt}
\textsl{\footnotesize Participants of the 2023 workshop:}
\a{Juan Antonio}{Aguilar Sánchez}{ULB}
\a{Markus}{Ahlers}{Niels Bohr Institute}
\a{Ingo}{Allekotte}{CNEA Argentina}
\a{Kevin}{Almeida Cheminant}{Institute of Nuclear Physics Polish Academy of Sciences}
\a{Rafael}{Alves Batista}{Universidad Autónoma de Madrid}
\a{Gioacchino Alex}{Anastasi}{Università di Catania \& INFN Catania}
\a{Hari Haran}{Balakrishnan}{Tata Institute of Fundamental Research}
\a{Sonali}{Bhatnagar}{Dayalbagh Educational Institute}
\a{Kathrin}{Bismark}{KIT}
\a{Martina}{Bohacova}{FZU}
\a{Carla}{Bonifazi}{ICAS - ICIFI - UNSAM}
\a{Washington}{Carvalho Jr}{IMAPP Radboud University Nijmegen}
\a{Antonella}{Castellina}{INFN}
\a{Paramita}{Dasgupta}{ULB}
\a{Bruce}{Dawson}{University of Adelaide}
\a{Luca}{Deval}{KIT}
\a{Armando}{di Matteo}{INFN Torino}
\a{Rita}{de Cassia Dos Anjos}{UFPR}
\a{Yunos}{El Kaderi}{Cergy-Paris University}
\a{Francesco}{Fenu}{KIT}
\a{Thomas}{Fitoussi}{KIT}
\a{Benjamin}{Flaggs}{Bartol Research Institute, University of Delaware}
\a{Toshihiro}{Fujii}{OMU}
\a{Keitaro}{Fujita}{ICRR, the University of Tokyo}
\a{Hazal}{Goksu}{Max-Planck-Institut für Kernphysik (Max Planck Institute for Nuclear Physics)}
\a{Steffen}{Hahn}{KIT}
\a{Ryo}{Higuchi}{RIKEN}
\a{Bohdan}{Hnatyk}{Taras Shevchenko National University of Kyiv, Astronomical Observatory}
\a{Jörg}{Hörandel}{Radboud University}
\a{Tim}{Huege}{KIT}
\a{Daisuke}{Ikeda}{Kanagawa University}
\a{Gina}{Isar}{Institute of Space Science - ISS, Bucharest-Magurele, Romania}
\a{Karl-Heinz}{Kampert}{University Wuppertal}
\a{Matthias}{Kleifges}{KIT}
\a{Ioana}{Mari\c{s}}{ULB}
\a{John}{Matthews}{University of Utah}
\a{Eric}{Mayotte}{Colorado school of Mines}
\a{Peter}{Mazur}{Fermilab}
\a{Athina}{Meli}{North Carolina A\&T State University}
\a{François}{Montanet}{LPSC Grenoble UGA/CNRS/IN2P3}
\a{Marco}{Muzio}{Pennsylvania State University}
\a{Lukas}{Nellen}{Instuto de Ciencias Nucleares, UNAM}
\a{Shoichi}{Ogio}{ICRR, The University of Tokyo}
\a{Foteini}{Oikonomou}{Norwegian University of Science and Technology}
\a{Hitoshi}{Oshima}{ICRR, the University of Tokyo}
\a{Rami}{Oueslati}{STAR Institute IFPA Group ULiège}
\a{Jannis}{Pawlowsky}{Bergische Universität Wuppertal}
\a{Vincent}{Pelgrims}{IA-FORTH / Phys. Dept. Univ. of Crete}
\a{Bjarni}{Pont}{Radboud University Nijmegen}
\a{Alessio}{Porcelli}{Universidad de Antofagasta}
\a{Julian}{Rautenberg}{University of Wuppertal}
\a{Markus}{Roth}{KIT}
\a{Alexandra}{Saftoiu}{"Horia Hulubei" National Institute for Physics and Nuclear Engineering, Romania"}
\a{Takashi}{Sako}{ICRR, the University of Tokyo}
\a{Shunsuke}{Sakurai}{Osaka Metropolitan University}
\a{Francesco}{Salamida}{University of L'Aquila and INFN LNGS}
\a{Fred}{Sarazin}{Colorado School of Mines}
\a{Frank}{Schroeder}{University of Delaware + KIT}
\a{Olga}{Sergijenko}{MAO NASU \& AGH UST \& AO KNU}
\a{Heungsu}{Shin}{Osaka Metropolitan University}
\a{Masahiro}{Takeda}{Institute for Cosmic Ray Research, University of Tokyo}
\a{Yuichiro}{Tameda}{Osaka Electro-Communication University}
\a{Michael}{Unger}{KIT}
\a{Arjen}{van Vliet}{Khalifa University}
\aEnd{Orazio}{Zapparrata}{ULB}

\par\noindent\rule{\textwidth}{0.4pt}
\textsl{\footnotesize Further GCOS supporters:}
\a{Fraser}{Bradfield}{Osaka Metropolitan University}
\a{Yuko}{Ikkatai}{Kanazawa University}
\a{Robin}{James}{College of William and Mary}
\a{C.}{Koyama}{ICRR, University of Tokyo}
\a{John N.}{Matthews}{University of Utah}
\a{Hiroaki}{Menjo}{Nagoya University}
\a{Marcus}{Niechciol}{University of Siegen}
\a{Yutaka}{Ohira}{The University of Tokyo}
\a{H.S.}{Shin}{Osaka Metropolitan University}
\aEnd{Federico}{Urban}{CEICO, FZU Prague}

\par\noindent\rule{\textwidth}{0.4pt}

\begin{center}
\footnotesize edited by
{J\"org R. H\"orandel} (Radboud University),
{Ioana C. Mari\c{s}} (ULB),
and
{Michael Unger} (KIT). \\[\baselineskip]
%on behalf of the GCOS Supporters, attending the workshops ...
\end{center}

}
%%%%%%%%%%%%%%%%%%%%%%%%%%%%%%%%%
\begin{abstract}
  After a successful kick-off meeting in
  2021~\cite{firstGCOSworkshop}, two
  workshops in 2022 and 2023 on the future
  Global Cosmic-Ray Observatory (GCOS) focused mainly on a straw man
  design of the detector and science possibilities for astro- and
  particle physics~\cite{secondGCOSworkshop, thirdGCOSworkshop}. About 100 participants gathered for in-person and
  hybrid panel discussions. In this report, we summarize these
  discussions, present a preliminary straw-man design for GCOS and
  collect short write-ups of the flash talks given during the focus
  sessions.
\end{abstract}

\newpage

{
  \small
\tableofcontents
}

\newpage

\section{Introduction}
Nature is providing particles at enormous energies, exceeding $10^{20}$~eV -- orders of magnitude beyond the capabilities of human-made facilities  like the Large Hadron Collider at CERN.
%%%%%%%%%%%%%%%%%%%%%%%%%%%%%%%%%%%%
\infobox{Science Targets of GCOS}{
\begin{compactitem}
   \item discovery of UHE accelerators
   \item charged-particle astronomy
   \item UHE neutrinos and photons
   \item BSM physics
   \item cosmic magnetism
   \item multi-messenger studies
\end{compactitem}
}{\vspace{-0.8cm}}
%%%%%%%%%%%%%%%%%%%%%%%%%%%%%%%%%%%%
\noindent
At the highest energies the precise particle types are not yet known, they might be ionised atomic nuclei or even neutrinos or photons. Even for heavy nuclei (like e.g.\ iron nuclei) their Lorentz factors $\gamma=E_\text{tot}/m c^2$ exceed values of $\gamma>10^9$.
The existence of such particles imposes immediate, yet to be answered questions~\cite{AlvesBatista:2019tlv,Sarazin:2019fjz}:
\q What are the physics processes involved to produce these particles?
\q Are they decay or annihilation products of Dark Matter?
\cite{Alcantara:2019sco,Aloisio:2015lva}
If they are accelerated in violent astrophysical environments: \q How is Nature being able to accelerate particles to such energies?
\q What are the sources of the particles? Do we understand the physics of the sources?
\q Is the origin of those particles connected to the recently observed mergers of compact objects -- the gravitational wave sources?\cite{Lipari:2017qpu,Branchesi:2016vef,Gergely:2007ny,Gergely:2008dw,Gergely:2010xr,Tapai:2013jza}
The highly-relativistic particles also provide the unique possibility to study (particle) physics at it extremes:
\q Is Lorentz invariance (still) valid under such conditions?
\cite{Klinkhamer:2008ss,Klinkhamer:2017puj,Aloisio:2000cm,Cowsik:2012qm,Martinez-Huerta:2016azo,Lang:2020geh}
\q How do these particles interact?
\q Are their interactions described by the Standard Model of particle physics?
When the energetic particles interact with the atmosphere of the Earth, hadronic interactions can be studied in the extreme kinematic forward region (with pseudorapidities $\eta>15$) \cite{PierreAuger:2012egl}.

The Global Cosmic-ray Observatory (GCOS) is a planned large-scale
facility designed to study ultra-high-energy cosmic particles,
including cosmic rays, photons, and neutrinos. Its main objective is
to precisely characterize the properties of the most energetic
particles in the universe and to pinpoint their mysterious
origins. Featuring an aperture that is twenty times larger than current
observatories, GCOS aims to begin operations after 2030, coinciding
with the gradual phase-out of existing detectors~\cite{Coleman:2022abf}.

\begin{figure}[h]
  \centering
  \includegraphics[width=0.625\linewidth]{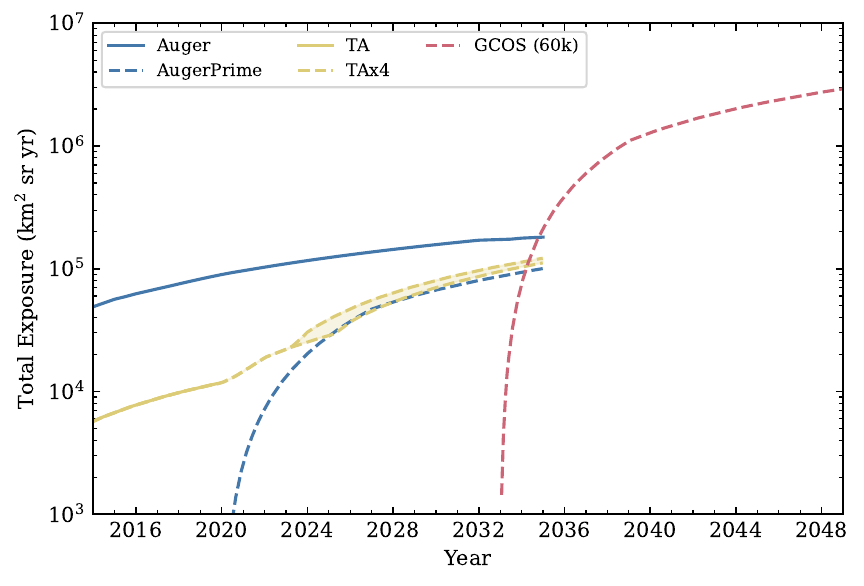}
  \caption{Expected exposures of GCOS (dashed red line) and existing
    air shower arrays as function of time. A band is shown to indicate
    the exposure for various deployment schedules for TA$\times$4. The
    solid blue line denotes the total Auger exposure and the exposure
    collected with the upgraded AugerPrime detectors is indicated by
    the blue dashed line. Adapted from~\cite{Coleman:2022abf}.}

\end{figure}

\section{A Straw-Man Design for GCOS}
\subsection{General Considerations}
To probe the fundamental nature and origin of ultrahigh-energy cosmic
rays a major leap in the available exposure for detecting air showers is needed.
We therefore target a detector
%%%%%%%%%%%%%%%%%%%%%%%%%%%%%%%%%%%%
\infobox{GCOS Requirements}{
\begin{compactitem}
   \item total area: 60\,000 km$^2$
   \item number of sites $\geq$ 2
   \item  trigger threshold: $10^{19}$~eV
   \item  high-quality threshold: $3\times 10^{19}$~eV
   \item  $\sigma_E$: 10\%, $\sigma_{\ln A}$: 1, $\sigma_\theta$: 1$^\circ$
   \item high duty cycle, low maintenance
\end{compactitem}
}{\vspace{-0.7cm}}
\noindent
 design which would allow us to collect the equivalent exposure of the Pierre Auger
 Observatory of 20 equivalent years within just one year of data taking. This corresponds to an area of 60\,000~km$^2$. To achieve a full sky
 coverage, the GCOS will require at least two observational sites
 strategically placed around the globe, ideally at intermediate
 geographical latitudes of $\pm35^\circ$, see
 Sec.~\ref{sec:dimatteo2}. This setup will ensure comprehensive
 monitoring of cosmic rays from all directions, providing a complete
 picture of the cosmic-ray sky.  The workhorse of the observatory
 will be its particle detector (PD) consisting of an array of stations
 that sample the density of electromagnetic particles (electrons,
 positrons, photons) and muons on the ground.  The energy and mass
 scale of this array will be set by a fluorescence detector (FD)
 and/or a radio detector (RD).

%%%%%%%%%%%%%%%%%%%%%%%%%%%%%%%%%%%%
We aim at a threshold for 100\% trigger efficiency at 10~EeV above
which the end of the UHECR spectrum can be studied, including the
instep feature, the flux suppression, and potentially a flux recovery
at extremely high energies. Around the threshold energy, most events will be measured with
just the minimum number of three particle detector stations.  The
energy threshold for high-quality data is at 30~EeV and above, where
at least a 5-fold coincidence of particle detectors will be triggered
by an air shower.  Above this energy, we aim at an energy resolution
of better than 10\% per event, a muon number resolution for mass
measurements of better than 10\%, and an $X_{\rm max}$ resolution of
better than 30 g/cm².  The angular resolution for the direction of
cosmic rays is expected to be better than $1^\circ$.

\subsection{Particle Detector}
%%%%%%%%%%%%%%%%%%%%%%%%%%%%%%%%%%%%
\noindent
To cover the required area of GCOS with a minimal number of particle
detector stations, a trade-off between the spacing between detectors and the covered area has to be taken into account. With the footprints of the vertical air showers on the ground not exceeding 20 km$^2$ a spacing of 2.2~km provides an upper limit to achieve an optimal trigger efficiency (Sec. \ref{sec:maris1}). To
\infobox{Particle Detector Design}{
\begin{compactitem}
   \item layered water Cherenkov detectors
   \item emag.\ and muonic EAS component
   \item  detector spacing: 2.2~km
   \item  number of stations: 18\,000
   \item  $\sigma_{S}$: 10\%,  $\sigma_{N_\mu}$: 10\%, $\sigma_{\Xmax}$: 30\,\gcm
\end{compactitem}
}{\vspace{-0.5cm}}
\noindent
cover the planned surface of 60\,000~km$^{2}$ about 18\,000 detectors are required.
It is becoming more and more clear that the nature of the ultra-high-energy cosmic rays can be obtained from advanced techniques based on machine learning. These techniques, applied to surface detector data, currently provide resolutions on the $X_{\text{max}}$ reconstruction almost as good as the direct observations with telescopes. The more information one can provide to train the networks the better is expected to be the performance and simultaneous determination of the $X_{\text{max}}$ and the muon number will increase the resolution on the mass determination. Therefore the main requirement for a surface particle detector is to have sensitivity for the separation of the electromagnetic and muonic components of the air showers.  Moreover, as a large number of detectors are needed to cover the area, the detectors need to be robust and cost-effective. The two techniques used for measuring the particles reaching the ground are based either on scintillators, water-Cherenkov detectors, or a combination of both.

To have good detection efficiency and enough effective area for inclined air-showers, the detectors should be built to cover a large solid-angle phase space.  While different configurations of scintillators can be imagined, the water-Cherenkov detectors are a rather simple solution for this.

A simple water-Cherenkov detector, such as the one used at the Pierre Auger Observatory cannot distinguish directly between the electromagnetic and muonic components, but it has been proven that machine learning techniques might provide very good resolution in extracting the muonic component from the total signal~\cite{PierreAuger:2021nsq} for certain distances to the air-shower axis and energies of the primary particles. The reconstruction of the number of muons in air showers can be further improved by separating horizontally the optical volume of the detectors in two: the upper layer would be more sensitive to the electromagnetic component, while the bottom layer would contain more light produced by muons. By having these different signals, one can extract via a set of linear equations the individual air-shower components~\cite{Letessier-Selvon:2014sga} (Sec.~\ref{sec:maris2}). A combination of scintillators (with photon conversion) and water-Cherenkov detectors can be also investigated.

The array configuration is not obvious, as it might depend on the
deployment sites, but classical triangular or rectangular grids are a
very good solutions. Variable grid spacing, optimizing the cost, and
the trigger efficiencies might be also an option. The size of the
detectors has not been yet optimized, nevertheless, the dimensions
need to be realistic for a large deployment, with a maximal
diameter of about 3~m and a height of less than 2~m.
In the next years, we will learn also from gamma-ray experiments that employ different sizes of water-Cherenkov detectors, including segmented optical volumes, like the planned large detectors of SWGO~\cite{Chiavassa:2024rew}. From a probably future experiment, PEPS~\cite{Maris:2023anl} that proposes similar sizes of relatively small detectors and besides the gamma-ray studies provides a GCOS prototype array, and from the combination of scintillators and water-based detectors at the Pierre Auger Observatory we will gain further insights.
%%%%%%%%%%%%%%%%%%%%%%%%%%%%%%%%%%%%
\subsection{Fluorescence Detector}
The longitudinal development of air showers can be observed directly
using fluorescence telescopes. The integral of the profile provides a
model-independent measurement of the calorimetric energy of a shower
and the depth \Xmax at which the measured profile reaches its maximum
is proportional to the logarithm of the mass of the primary
particle. The main purpose of the FD in GCOS will be the calibration
of the absolute energy scale of the PD with less than 10\% uncertainty
and to provide a calibration of its \Xmax scale. The required coverage
is $\geq$50\% above the quality energy threshold of the PD, i.e.\ at
$>30$~EeV. The resolution should be similar to the one achieved with
current FDs: $<10$\% energy resolution, 15~\gcm \Xmax resolution and
0.5\deg angular resolution. Current FDs operate with a duty cycle of
$\lesssim 15\%$ during clear and moonless nights, but using SiPM
cameras, the GCOS FD will be able to safely operate with higher duty
cycles of up to 35\%.

Two layouts of the FD are considered. Layout A of one 30\,000~km$^2$
site of GCOS is illustrated in the figure above. The telescopes are located
at two sites (``Mastercard'' layout in Sec.~\ref{sec:dawson}).  The
light- and dark-blue areas denote the FD coverage of the array at $10^{20}$~eV with
low-elevation telescopes (Sec.~\ref{sec:unger}) and single-pixel
telescopes (Sec.~\ref{sec:fujii} and~\ref{sec:tameda}),
respectively. Laser facilities, that are needed for a continous monitoring of
the aerosol content of the atmosphere, are indicated as red points.
%%%%%%%%%%%%%%%%%%%%%%%%%%%%%%%%%%%%
\infobox{Fluorescence Detector Tasks}{
\begin{compactitem}
   \item calibration of PD energy scale
   \item calibration of PD mass scale
   \item $\sigma_E < 10\%$, $\sigma_{\ln A} \sim 0.4$
   \item duty cycle 20\%
\end{compactitem}
}{
  \centering
  \includegraphics[clip, rviewport=0 0.96 1 1,width=0.8\linewidth]{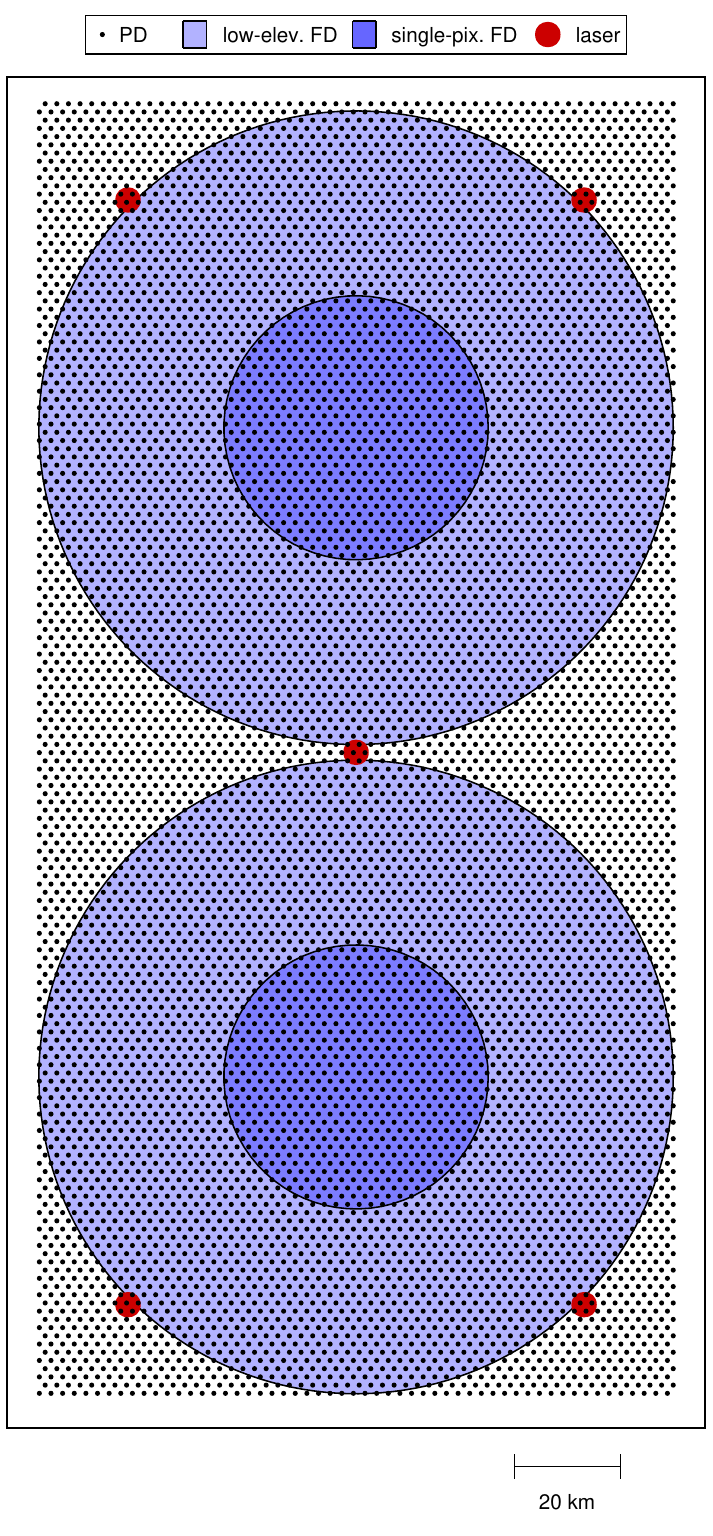}\\
  \phantom{bb}\includegraphics[angle=90,clip, rviewport=0 0.0 1 0.95,width=0.95\linewidth]{gcos-fd.pdf}\\
 \footnotesize{Layout A of the FD at one 30\,000 km$^2$ GCOS site.}
          {\vspace{-0.4cm}}
 }
As can be seen, this layout minimizes the overlap
between the areas viewed by FD stations at the expense of stereoscopic
observations. The low-elevation telescopes detect air showers with
full coverage of the neccessary slant depth range at distances between
20 and 60~km ($A \approx 2\ \times 10\,000~\text{km}^2$). Nearby showers
at $R<20$~km ($A \approx 2\ \times 1\,200~\text{km}^2$) are observed by
both telescope types simultaneously, but only the single-pixel
telescopes provide enough elevation coverage to observe the shower
maximum of nearby shallow showers. The geometry of air showers is
determined from the combined data of low-elevation and single-pixel
telescopes and the timing at ground from at least one station of the
PD with an energy threshold of $\gtrsim 10^{18.5}$~eV.

As an alternative setup we consider layout B in which only
single-pixel telescopes are deployed. In this case $\geq 15$ sites
need to be operated and again there is minimal overlap between the
sites. The air shower geometry is taken from the PD, raising the enery
threshold of the FD to the PD trigger threshold of $10^{19}$~eV.

%\newpage
\subsection{Radio Detector}\label{RDsec}
%%%%%%%%%%%%%%%%%%%%%%%%%%%%%%%%%%%%
\infobox{Radio Detector Potential}{
\begin{compactitem}
   \item $\sigma_{\rm e/m}<10$\%
   \item independent energy scale, $\sigma_{\rm E}<10\%$
   \item hybrid $\mu$ \&  e/m  $\rightarrow$ mass
   \item interferometry $\rightarrow$ mass
\end{compactitem}
}{\vspace{-0.8cm}}
%%%%%%%%%%%%%%%%%%%%%%%%%%%%%%%%%%%%
The particle detector array can be complemented by a radio detector array.
Radio detection provides a very clean and
accurate measurement of the e/m shower component. E.g.\ for the Auger Radio Detector an accuracy of 6\% has been obtained from full end-to-end simulations \cite{Huege:2023pfb}.
It is important to point out that this uncertainty is independent of
the cosmic-ray particle type.
The fact that $E_{\rm em}$ is reconstructed from radio data with no significant
dependence on the incoming cosmic ray particle type makes it a very suitable energy estimator for use in the discrimination of air showers induced by
different primary particles~\cite{Aab:2015vta,Aab:2016eeq,Huege:2023pfb}.
\begin{wrapfigure}{r}{5cm}
 \includegraphics[width=5cm]{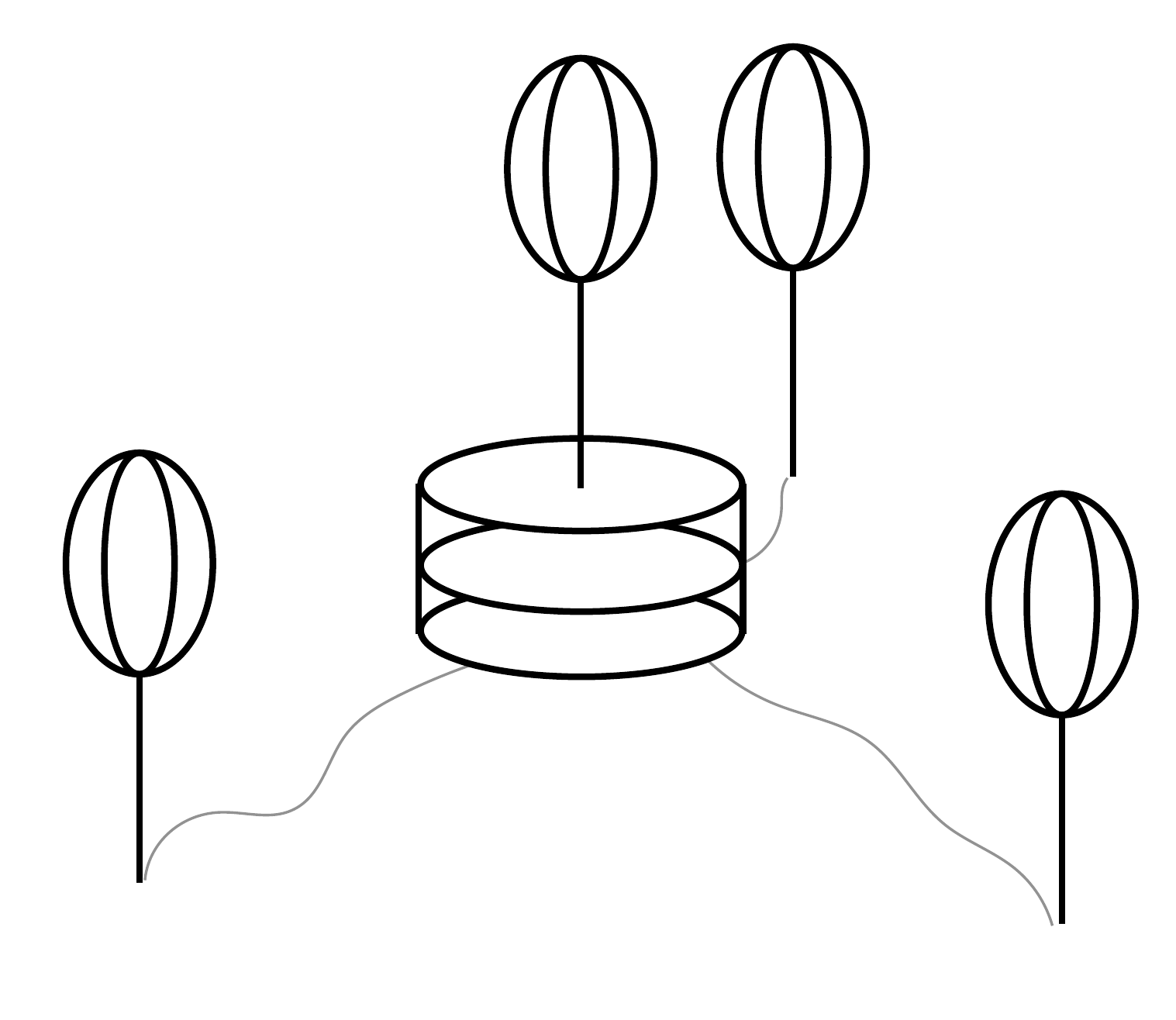}
 {Conceptual sketch of a layered water-Cherenkov detector with satellite radio antennas.}
\end{wrapfigure}
\noindent
The total cosmic-ray energy can be reconstructed with an accuracy of better than 10\%.
Highly inclined air showers\cite{Aab:2018ytv} with zenith angles
$\Theta>60$\deg traverse a big amount of atmosphere until they are detected.
The thickness of the atmosphere in horizontal direction ($\Theta=90$\deg) amounts
to about 35
times the column density of the vertical atmosphere. Thus, the e/m shower
component is mostly absorbed and only muons are detected with the particle detectors.
The atmosphere is transparent for radio emission in our band (tens to hundreds MHz)
and radio measurements are an ideal tool for a calorimetric measurement of the
electromagnetic component in horizontal air showers.
The combination of electromagnetic and muonic information is used to derive the mass/particle type of the incoming cosmic ray. Another method
with enormous potential to determine the mass/particle type of the incoming cosmic ray is interferometry \cite{Schoorlemmer:2020low}.
This requires sub-ns time resolution, which can be achieved, e.g.\ by using time synchronization signals from the Galileo satellite GPS system.\\

\noindent
\textbf{Technical considerations} for a Radio Detector:
The relatively steep lateral distribution of the radio emission requires a detector spacing around 1\,500~m or smaller. This makes radio detection in particular interesting for inclined air showers with zenith angles $\Theta>60$\deg.
Due to cost constraints, the spacing of the particle detector array (see above) will  probably be of the order of 2~km or above. A smaller grid size for the radio detector array could be achieved by introducing satellite radio stations to each particle detector station. They could be rather simple, only being comprised of a radio antenna and a low-noise pre-amplifier, connected to the particle detector station by cables in the ground. Thus, the satellite radio stations could be rather cheap and easy to deploy. The particle detector station could contain the main electronics (filter amplifier and digitization) and communication systems. The lower bound of the frequency range is somewhere around 30~MHz, at lower frequencies the Galactic emission is too strong, outshining all shower signals. Detecting a broad frequency range up to several hundreds MHz would allow to extract precise shower information from a few stations only \cite{Barwick:2016mxm}. Another topic of optimization is the number of polarization directions to be measured, two (north-south and east-west) or in addition also a third (vertical) component.
% notes from workshop:
%\begin{compactitem}
%   \item 10\% resolution for electromagnetic energy
%   \item establish absolute energy scale with less than 10\% uncertainty
%   \item for HAS: hybrid reconstruction wcd: $\mu$, RD: e/m (energy) $\rightarrow$ mass
%   \item sub-ns time resolution (new Galileo GPS) $\rightarrow$ interferometry $\rightarrow$ \Xmax
%\end{compactitem}
%\begin{compactitem}
 %  \item maximum spacing around 1500 m
  % \item multiple antennas for one SD station
   %\item placement in the middle of the triangle
   %\item  two/three polarisation system
%\end{compactitem}

\subsection{Electronics}
A significant challenge in the development of GCOS lies in the design and production of compact and robust electronic systems that require minimal maintenance. The concept of autonomous detection stations inherently involves exposing the electronics to extreme environmental conditions. Consequently, the equipment must endure harsh environments, including daily and seasonal temperature fluctuations and strong electric fields during thunderstorms.
The optimal solution is likely to integrate all functionality onto a single board or even a custom-designed chip.

Another challenge pertains to the communication system. The autonomous stations need to transmit their data to a central data acquisition system. To establish an efficient trigger, it is essential to gather information from neighbouring stations. Relaying all data through a central system necessitates substantial bandwidth. A more efficient approach appears to involve a next-neighbour communication system, where stations exchange information locally with their immediate neighbours. This information can then be used to form a trigger, which is communicated to the central data acquisition system via a long-range communication channel. This channel is also utilised to transmit the event data.

\subsection{Further Considerations}

\subsubsection*{High precision infill array}
In addition to the standard PD array as outlined above, one might consider an infill array with a higher density of detectors. This increased station density results in a lower energy threshold, allowing for the precise detection of air showers. Such an array would be instrumental in studying high-energy hadronic interactions in the atmosphere, thereby addressing key questions in particle physics. Furthermore, a dense array would also prove valuable for calibration purposes.

\subsubsection*{Existing sites}
The detection sites for GCOS will be strategically distributed across the globe, with a minimum of two locations—one in the Northern Hemisphere and one in the Southern Hemisphere. The ideal placement for these sites lies at latitudes between approximately 35° and 40° north and south, respectively.

\begin{wrapfigure}{r}{6cm}
 \includegraphics[width=6cm]{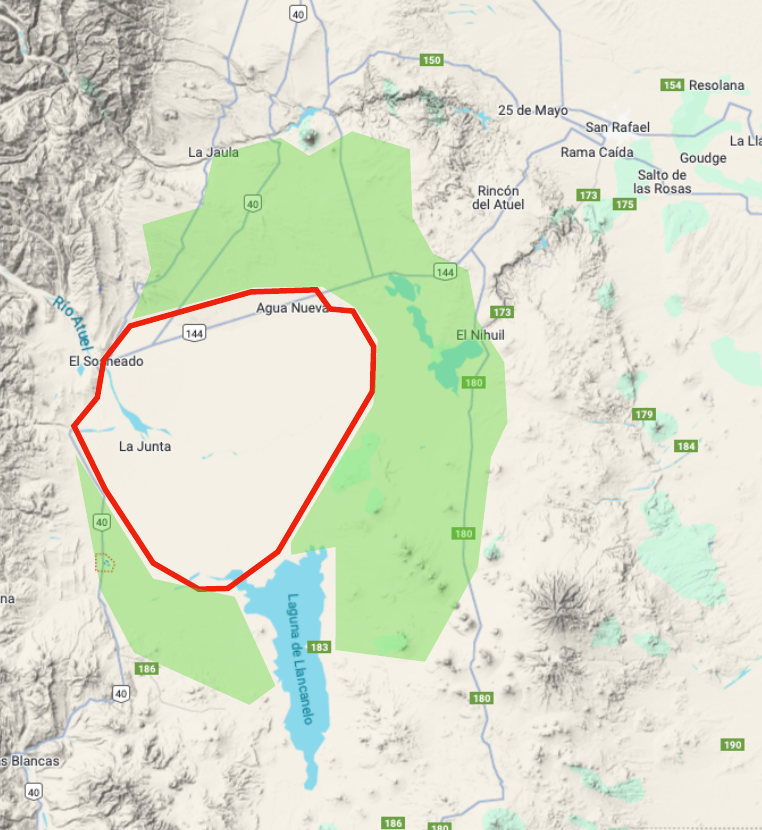}
 \caption{Illustration of a potential extension of the Pierre Auger Observatory.}
\end{wrapfigure}
Existing large-scale facilities for observing ultra-high-energy cosmic rays, such as the Pierre Auger Observatory in Argentina (at 35° S) and the Telescope Array in Utah (at 39° N), are positioned in optimal locations. These sites could function as infill arrays, benefiting from increased station density and, consequently, lower energy detection thresholds. Given the established infrastructure at both locations, they present excellent foundations for the expansion of larger GCOS arrays.

\subsubsection*{Beyond the Pierre Auger Observatory}
The Pierre Auger Observatory in Argentina could serve as potential core for a larger GCOS array.
If one requires a "flat" area at an altitude around 1\,500~m.a.s.l.\ one could extend the existing array maybe by a factor of two or three. Thus, an area of the order of almost 10\,000~km$^2$ could be reached, i.e.\ $1/6$ of the above mentioned total GCOS PD area.

\cleardoublepage
\section{Contributions}
\subsection{Science case}

\flashtalkDone{Ana Laura M\"uller}{The Importance of a Full-Sky Observatory for UHECR Sources}
Starburst galaxies (SBGs) and active galactic nuclei (AGNs) have been the most discussed sources for ultra-high-energy cosmic ray (UHECR) acceleration in recent years. Among these, SBGs are slightly more favoured ($4.2\sigma$) than AGNs ($3.3\sigma$) based on correlation studies of arrival directions, although no study has achieved statistical significance for any specific source type. There is also a notable indication of correlation with the region of Centaurus A, which is part of the hotspot observed by the Pierre Auger Collaboration \cite{PierreAuger:2018qvk,PierreAuger:2022axr}. Nevertheless, interpreting the Centaurus A region remains particularly challenging due to the high density of sources of different natures within several degrees. In the northern hemisphere, the hotspot found by the Telescope Array Collaboration is located near M82, another starburst galaxy \cite{TelescopeArray:2014tsd}. Despite these observational hints, the theoretical link between UHECRs and their sources remains uncertain.

Superwinds in SBGs were initially considered the most plausible regions for accelerating iron nuclei to energies as high as $10^{20}$~eV \cite{Anchordoqui:1999cu,Anchordoqui:2018vji,Anchordoqui:2020otc}. However, recent studies suggest that these environments may not be capable of reaching such extreme energies \cite{Romero:2018mnb,Muller:2020vdm,Aguilar-Ruiz:2020bzq,Peretti:2021yhc}. This has shifted attention to stellar sources within SBGs, given their high star formation rates. Nevertheless, a critical question remains: Why do UHECR correlations seem to be specific to SBGs, rather than appearing across all galaxies? Could this be related to the frequency of specific processes, or are there additional, less understood factors influencing the observations? Bell \& Matthews recently argued that SBGs might fail to meet the minimum energy requirements needed to produce UHECRs. They proposed that the hotspot observed near M82 could instead be caused by UHECRs originating from Centaurus A in the past and reflected by M82 \cite{Bell:2021pkk}.

A full-sky cosmic ray observatory like GCOS would be invaluable in addressing all these uncertainties mentioned above. Observing both hotspots with the same instrument would eliminate systematic biases and enable more robust comparisons. Such an observatory would also provide comprehensive coverage for correlation studies, facilitating investigations of UHECR associations with full-sky catalogues of SBGs and AGNs or any other potential source classes. Additionally, current studies often rely heavily on electromagnetic radiation as a proxy for evaluating the significance of possible sources. However, it is important to note that UHECRs are not required to directly account for the observed electromagnetic emissions of either SBGs or AGNs at any wavelength. This raises the possibility that some of the observed correlations might be coincidental rather than causative.

An observatory capable of providing single-event mass composition information would be essential in determining whether the composition of UHECRs is consistent across hotspots or specific source catalogues. It would also enable searches for multiplets, which remain challenging to perform due to the limitations of existing data. For example, differences in the composition near Centaurus A could shed light on the complex interplay of multiple sources in that region and help disentangle their respective contributions. Furthermore, identifying even a few single events with lighter compositions at the highest energies, contrary to the expected predominance of intermediate or heavy nuclei, could indicate a more complex scenario where the UHECR spectrum's end is the result of a variety of source types rather than being dominated by a single class.

\flashtalkDone{Arjen van Vliet}{Galactic and extragalactic magnetic fields with GCOS}
GCOS will most likely be able to identify at least some nearby UHECR sources. When that happens, the measurements of GCOS can also be used to determine the environment the UHECRs traveled through from their sources to Earth. As UHECRs are charged particles, they get deflected by the magnetic fields that they traverse along the way. Magnetic fields are prevalent across the entire Universe, and both large-scale Galactic and extragalactic magnetic fields can significantly affect the trajectories of UHECRs. The deflection of a UHECR from the direction towards its original source in the sky, together with the energy and charge of the UHECR, gives a direct measure of the strength and direction of the magnetic fields it encountered along the way (see e.g.~\cite{Bray:2018ipq, vanVliet:2021deg, Neronov:2021xua, Shaw:2022lqd}). This deflection, energy, and charge will all be detectable by GCOS.

Magnetic fields play vital roles in astrophysics, from star formation, galaxy formation, and cluster dynamics to cosmic-ray acceleration and propagation. However, the structure and strength of both Galactic and extragalactic magnetic fields remain long-standing unresolved problems (see e.g.~\cite{Widrow:2002ud, 2017ARA&A..55..111H, AlvesBatista:2021sln} for reviews). Cosmic magnetic fields can be probed in several ways, by synchrotron emission, Faraday rotation, Zeeman splitting, polarized thermal dust emission, polarisation of optical starlight and gamma-ray cascades. However, these methods mostly probe magnetic fields in regions with an abundance of emission from stars, dust, or electrons. With GCOS and the detection of UHECR sources, a much-needed additional way of investigating cosmic magnetic fields will open up. 

How strongly UHECRs will get deflected when propagating through the Universe depends on their energy, their charge, the source distance and the strength and coherence length of the magnetic fields they encounter on the way. In general, small-scale turbulent magnetic fields will cause a spread around the original source position. On the other hand, large-scale structured fields will cause a shift in the original source position in one specific direction. The combination of these effects can be investigated with UHECR simulation frameworks like CRPropa~3~\cite{AlvesBatista:2016vpy, AlvesBatista:2022vem}.

Currently, the evidence for anisotropies at intermediate angular scales for UHECRs with $E > 40$~EeV is at the $4 \sigma$ level~\cite{PierreAuger:2022axr}. The most significant correlation is found for a catalog of starburst galaxies. This significance is mainly driven by an excess around the Centaurus region, with several nearby starburst galaxies located in the Centaurus region. The optimal Fisher search radius for this catalog of starburst galaxies is 15 degrees, with the closest starburst galaxies at distances of 3~to~4~Mpc. Similar optimal search radii and nearest source distances are obtained if active galactic nuclei (AGN) are considered as potential sources. This suggests that UHECRs can be deflected significantly from their original source positions within relatively short distances, indicating the need for relatively strong Galactic or extragalactic magnetic fields. A first study in this direction indicates the need for extragalactic magnetic field strengths of $B > 0.2$~nG (for coherence lengths of 1~Mpc)~\cite{vanVliet:2021deg}. GCOS will provide a clearer identification of possible UHECR source candidates including a better handle on the charge of cosmic rays. This has the potential to significantly advance our knowledge of the Galactic and extragalactic magnetic fields.

\flashtalkDone{Armando di Matteo}{How isotropic can the UHECR sky {\itshape really} be?}
The propagation length of ultra-high-energy cosmic rays (UHECRs) is limited to a few hundred megaparsecs or less at the highest energies, and the Universe is not homogeneous on such scales, hence the distribution of sources of UHECRs is expected to leave imprints in that of their arrival directions.  Magnetic fields can rotate and distort the picture, but the amplitude of the largest-scale anisotropies, namely the dipole and quadrupole, are expected to mostly survive: coherent deflections can only alter their directions, not their amplitudes, and the effect of turbulent deflections of size~$\Delta\theta_\mathrm{turb}$ is~$\mathcal{O} \left( \ell^2 \Delta\theta_\mathrm{turb}^2/2 \right)$ on the $2^\ell$-upole amplitude, hence minor except for~$\Delta\theta_\mathrm{turb}$ of several tens of degrees (see Ref.~\cite{Eichmann:2020adn} for more detailed estimates).

In Ref.~\cite{diMatteo:2017dtg} we used these facts to obtain a lower bound on the expected dipole and quadrupole amplitude in the UHECR arrival direction distribution.  We assumed that there are no UHECR sources within 5~Mpc of us, all galaxies in the XSCz catalog \cite{2MASS:2006qir} at distances between 5~and 250~Mpc are equal sources of UHECRs (and used Lynden--Bell weights \cite{Koers:2009pd} to account for expected unobserved dim galaxies near bright distant galaxies), and that at distances longer than 250~Mpc there is a homogeneous isotropic distribution of UHECR sources with the same average density as at shorter distances.  (This does not mathematically guarantee a lower bound because it is possible in principle for anisotropies from nearby sources to cancel out ones from distant sources, but that would require an unlikely coincidence.)  We considered several different UHECR mass compositions and two different regular Galactic magnetic field (GMF) models \cite{Pshirkov:2011um,Jansson:2012pc} in order to find out which would result in the least anisotropies, and we used an upper bound \cite{Pshirkov:2013wka} for the turbulent GMF.

\begin{figure}[b]
    \centering
    \includegraphics[width=0.49\textwidth]{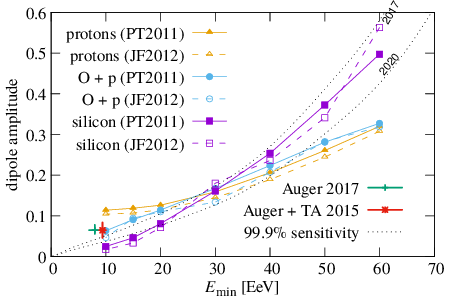}
    \hfil
    \includegraphics[width=0.49\textwidth]{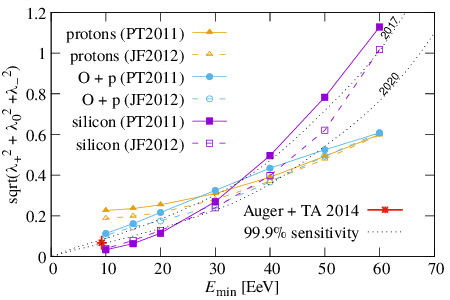}
    \caption{Dipole (left) and quadrupole (right) amplitude predictions \cite{diMatteo:2017dtg}}
    \label{fig:dipole}
\end{figure}

Our predictions are shown in Fig.~\ref{fig:dipole}.
At an energy threshold of 30~EeV, we expect a dipole amplitude~$\left|\mathbf{d}\right|$ of at least~13\% and a quadrupole amplitude~$4\pi C_2/C_0$ of at least~19\textperthousand.  For comparison, in the last report of the Auger--TA joint working group on arrival directions \cite{TelescopeArray:2021ygq}, in the $E_\mathrm{Auger} \in [32~\mathrm{EeV}, +\infty), E_\mathrm{TA} \in [40.8~\mathrm{EeV}, +\infty)$~bin we found~$\left|\mathbf{d}\right| = (11.6 \pm 3.8_\mathrm{stat} \pm 1.1_\mathrm{syst})\%$ and~$4\pi C_2/C_0 = (15.5 \pm 8.9_\mathrm{stat} \pm 2.4_\mathrm{syst})$\textperthousand, where the systematic uncertainties are due to the relative calibration of the Auger and TA energy scales.  Hence, assuming that GCOS can achieve negligible systematic uncertainties (e.g.\ by using the same type of detectors in both hemisperes) and reduce statistical uncertainties by $\sqrt{10}$~times from the last Auger--TA joint dataset, we can expect it to detect a dipole moment with $\sim 11\sigma$~significance and a quadrupole moment with $\sim 7\sigma$~significance.

Similar studies about medium-scale anisotropy searches, such as correlations with nearby galaxies \cite{TelescopeArray:2021gxg}, are underway in the Auger--TA joint working group and their results will be presented at upcoming conferences.

\flashtalkDone{Luis Anchordoqui}{GCOS, SHDM and UV physics}
\paragraph{What will GCOS data tell us about  dark matter?}

The nature of dark matter is a longstanding problem in modern day
physics~\cite{Feng:2010gw}. Due to the model independent
theoretical upper limit coming from
perturbative unitarity~\cite{Griest:1989wd}, a relatively low mass window below 100~TeV is open for weakly
interacting massive particles (WIMPs)~\cite{Steigman:1984ac}. As a consequence, the main experimental
efforts for dark matter
searches have been focussed on this low mass window. However, the lack of
experimental evidence of WIMPs after decades of
endeavour~\cite{PerezdelosHeros:2019lps} provides a motivation to
consider alternative models of dark matter. Grand Unified Theories (GUTs) predict the existence of $X$-particles
with masses around the GUT scale of $\sim 10^{16}~{\rm GeV}$. If the
lifetime of the $X$-particles is comparable to or larger than the age
of the universe, these particles are natural candidates for
super-heavy dark matter (SHDM), and their decays could contribute to UHECR fluxes today~\cite{Berezinsky:1997hy,Kuzmin:1997jua,Kuzmin:1999zk}. To estimate the flux of detectable particles one
needs to evaluate both the particle physics and astrophysical
factors. The particle physics factor is built in the fragmentation
function of the SM particles produced by the $X$-decay. The final
state stable particle spectra are obtained by solving the DGLAP equations
numerically~\cite{Gribov:1972rt,Gribov:1972ri,Dokshitzer:1977sg,Altarelli:1977zs}. There
is a general agreement among the various computational schemes
proposed to describe the secondary spectra of SM particles produced
via
$X$-decay~\cite{Birkel:1998nx,Berezinsky:2000up,Sarkar:2001se,Barbot:2002gt,Aloisio:2015lva}. The
astrophysical factor is determined by the distribution of dark matter
particles in the Galaxy, and it has associated $\sim 10\%$ uncertainty~\cite{Guepin:2021ljb}.
Null results of searches in Auger data lead to limits on the
$X$-lifetime~\cite{Alcantara:2019sco,Anchordoqui:2021crl}, and advances in
constraint-based GUT modeling~\cite{PierreAuger:2022jyk, PierreAuger:2022ubv, PierreAuger:2023vql}. Moreover,
since the mass scale of SHDM is ${\cal O} ({\rm GUT})$, GCOS will be a {\it unique
background free dark matter indirect detection experiment}: a clear
detection of an extreme energy photon would be momentous discovery.

\paragraph{What will GCOS data tell us about UV-IR mixing?}

There is growing evidence that a vast class of quantum field theories
(QFTs), which are totally consistent as low-energy (IR) effective
theories, do not have consistent UV completions with gravity
included. Such QFTs that cannot be embedded into a UV complete quantum
gravity theory are said to reside in the Swampland, in contrast to the
effective field theories that are low energy limits of string theory
and inhabit the Landscape~\cite{Vafa:2005ui}. This sorting of QFTs by
their consistency with gravity has become an unexpectedly powerful
theoretical tool, offering potential solutions to the problems of
fine-tuning. In particular, the smallness of the
cosmological constant in Planck units ($\Lambda \sim 10^{-122} M_{\rm
  Pl}^4$) might indicate that we are near
an infinite field distance limit~\cite{Lust:2019zwm,Montero:2022prj}. It is precisely in this asymptotic limit
where quantum gravity (QG) effects become important at scales much
below $M_{\rm Pl}$ and moreover QG can have an effect on the IR. Strictly speaking, physics would become strongly coupled to
gravity at $M_{\rm UV} \sim \lambda^{-1/3} \Lambda^{1/12} M_{\rm Pl}^{2/3}$,
with $\lambda \sim 10^{-3}$. This seemingly simple fact is behind a
very rich
phenomenology~\cite{Vafa:2024fpx,Antoniadis:2024sfa}. For
example, it is tempting to speculate whether $M_{\rm UV}$ could be
the carrier of the observed cutoff in the UHECR
spectrum~\cite{Montero:2022prj,Anchordoqui:2022ejw}, and to explore the interplay between the low-energy gravity scale and the GZK
horizon of UHECRs~\cite{Greisen:1966jv,Zatsepin:1966jv}. Along this line, a recent study showed that the allowed source-to-source variance of the maximum energy of
cosmic rays must be small to describe Auger
data~\cite{Ehlert:2022jmy}. This intriguing finding is at
odds with the typically high variances of intrinsic properties for the
most commonly assumed astrophysical
sources~\cite{Anchordoqui:2018qom}, and does not exclude the idea of a universal
cutoff energy regulated by $M_{\rm UV}$~\cite{Muzio:2023hdj,Muzio:2023fng}. To distinguish the
conjectured $M_{\rm UV}$ cutoff from the universal
GZK cutoff we
should study the individual spectra of nearby
sources~\cite{Anchordoqui:2022ejw,Noble:2023mfw}. Therefore, a {\it must}
requirement of GCOS design considerations should be {\it to achieve
 a gargantuan exposure for a high-statistics data sample with
 sensitivity to baryonic composition}.

\flashtalkDone{No\'emie Globus}{Treasure Maps for Extreme Energy Cosmic Rays}
Both Pierre Auger Observatory and Telescope Array reported a handful of "extreme energy cosmic rays" (EECR)
 $\gtrsim150$~EeV. In \cite{GFB2023} we proposed  a new methodology, using our current knowledge of the
Galactic Magnetic Field (GMF) models, the attenuation factor due to the GZK effect, and the source distribution
in the local universe to build "treasure maps" of the most promising directions to detect multiplets of EECR
events in the future. There are  important benefits of using EECR to constraint the origin of the UHECR
sources:
\begin{enumerate}
\item[{\small $\bullet$}]  Individual EECRs sources become more prominent, relative to the background, as the
GZK horizon diminishes.  (We define the GZK horizon as the distance from which 95\% of the cosmic rays of a
given energy and composition have been lost.)   The EECR horizon is limited to  our local supercluster,
$\sim$40 Mpc, for EECR starting as proton and iron nuclei at 150~EeV. For intermediate mass EECR such as CNO,
the horizon is not larger than $\sim2$~Mpc, \textit{i.e.} limited to the size of the local group. The
observational capability for the mass composition thus carries critical information about the EECR source
distance.

\item[{\small $\bullet$}] EECR have the highest rigidities (around 30~EV if they are CNO, and almost 10~EV if
they are iron). At these  rigidities, the EECR suffer less deflection and time delay in the GMF.  Multiplets of
EECR events are more likely to be detected in “magnetic windows” of the sky where the temporal dispersion due
to the Galactic and extragalactic magnetic fields  is small enough to detect at least one doublet of events
within the typical observation time of a UHECR observatory. The “magnetic window” has to back-project to source
candidates, which is not guaranteed when the sources are scarce due to the limited GZK horizon.

\item[{\small $\bullet$}] The detection of EECR multiplets would help us to understand the nature of their
sources.   We showed that an analysis of the EECR multiplet arrival times would allow us to distinguish between
transient and continuous sources \cite{GFB2023}. Essentially, time arrival of events with a similar rigidity
can be drawn independently from the asymmetrical distribution of time delays and compared with the data. (We
refer to the paper \cite{GFB2023} for more details and the methodology.) This requires a reasonable knowledge
of the mass of the cosmic-ray events which can be achieved by GCOS in the future (see section \ref{sec:flaggs}).
\end{enumerate}

The GMF can magnify certain directions because the large scale field can both increase and decrease the
cosmic-ray flux through magnetic focusing/lensing
(see section \ref{sec:futherScience})
. Due to the GMF turbulent component, EECR
arriving from the same direction in the sky have slightly different travel times leading to a temporal
dispersion $\tau_{d,\text{GMF}}$. We use different GMF models to  provide an estimate of the magnification
factor $M$ and the temporal dispersion $\tau_{d,\text{GMF}}$ in the Galaxy as a function of the direction of
the source in the sky and the rigidity. (The total temporal dispersion is denoted as $\tau_{d}$ and combine the
temporal dispersions due to the Galactic and extragalactic magnetic fields.)

\begin{figure}[h]
\centering
\includegraphics[width=0.45\textwidth]{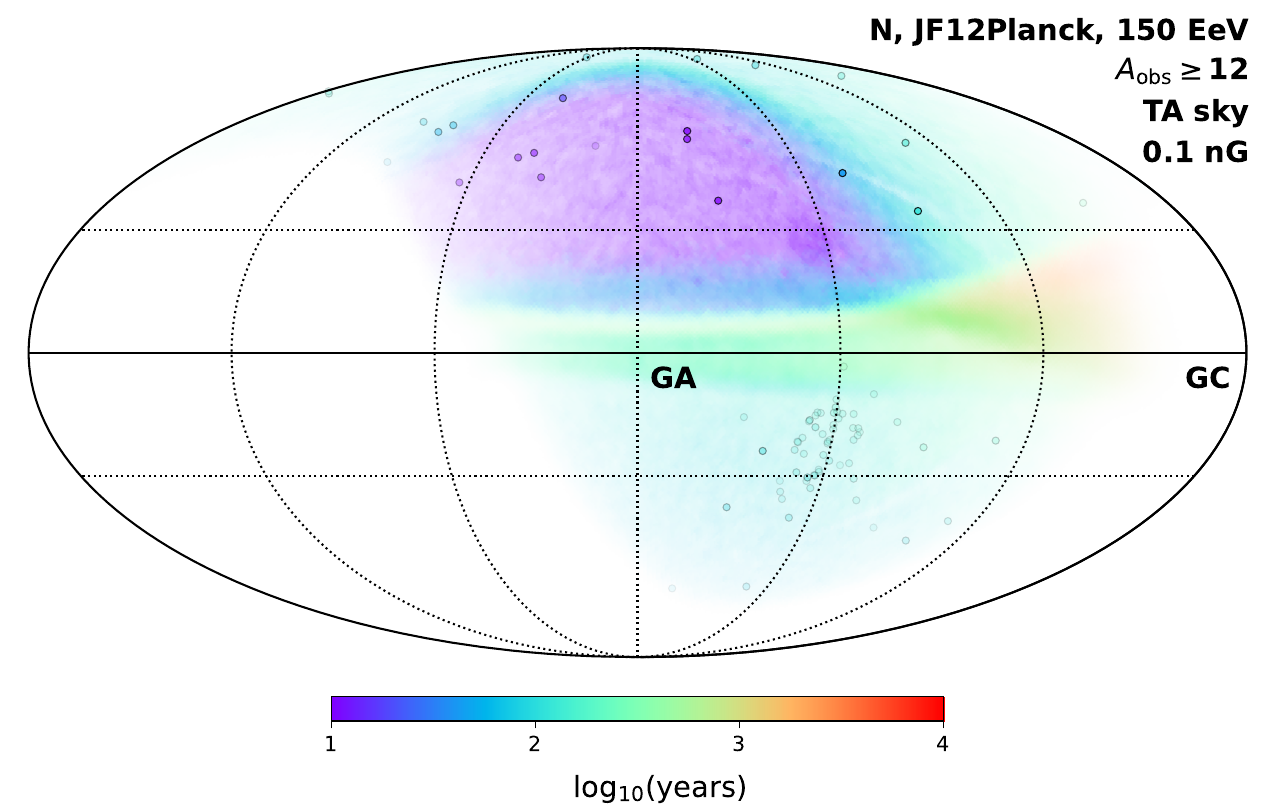}
\includegraphics[width=0.45\textwidth]{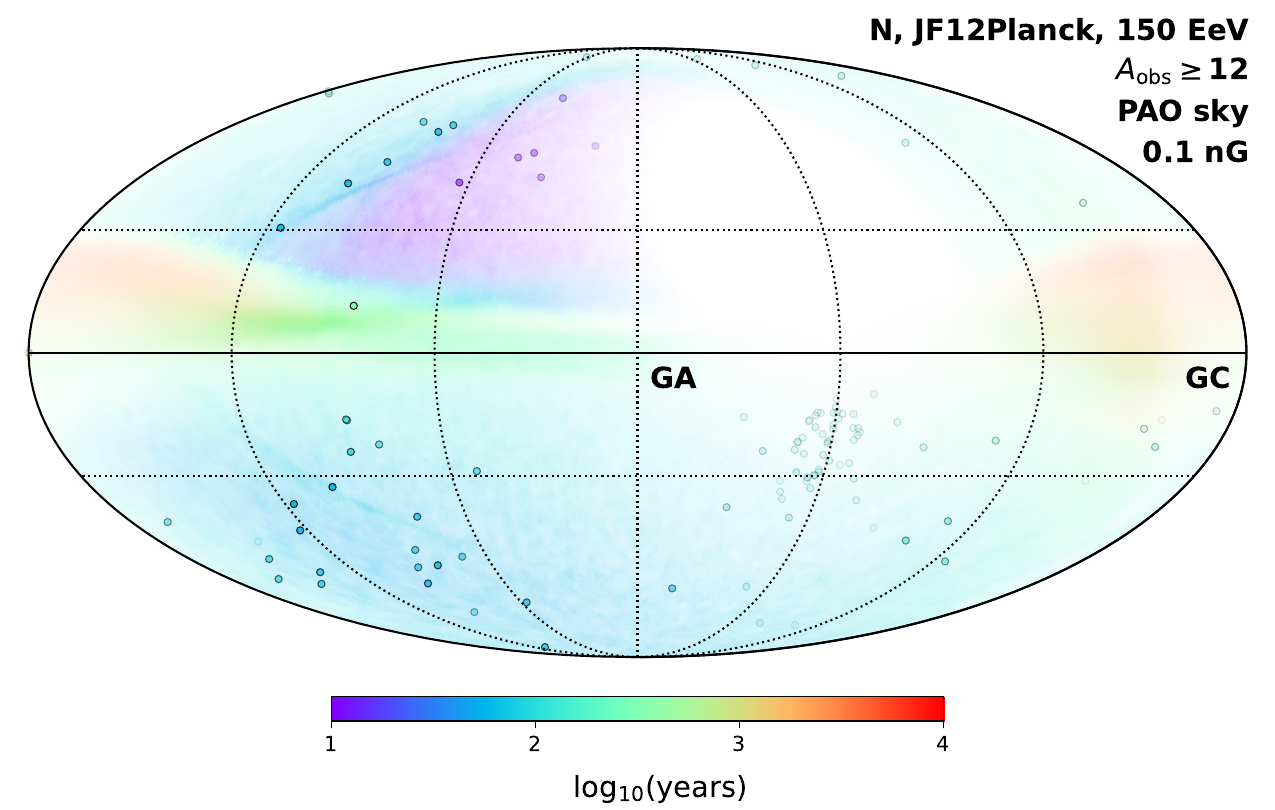}

\includegraphics[width=0.45\textwidth]{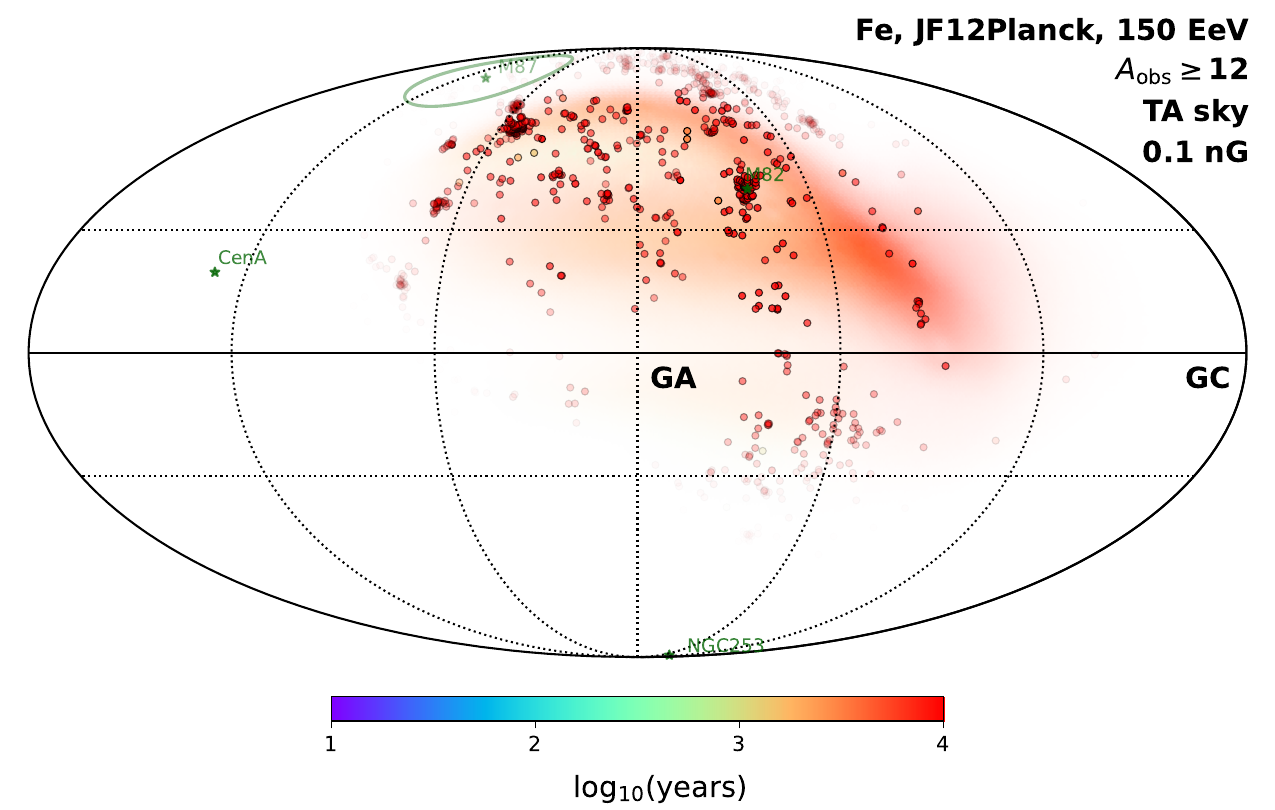}
\includegraphics[width=0.45\textwidth]{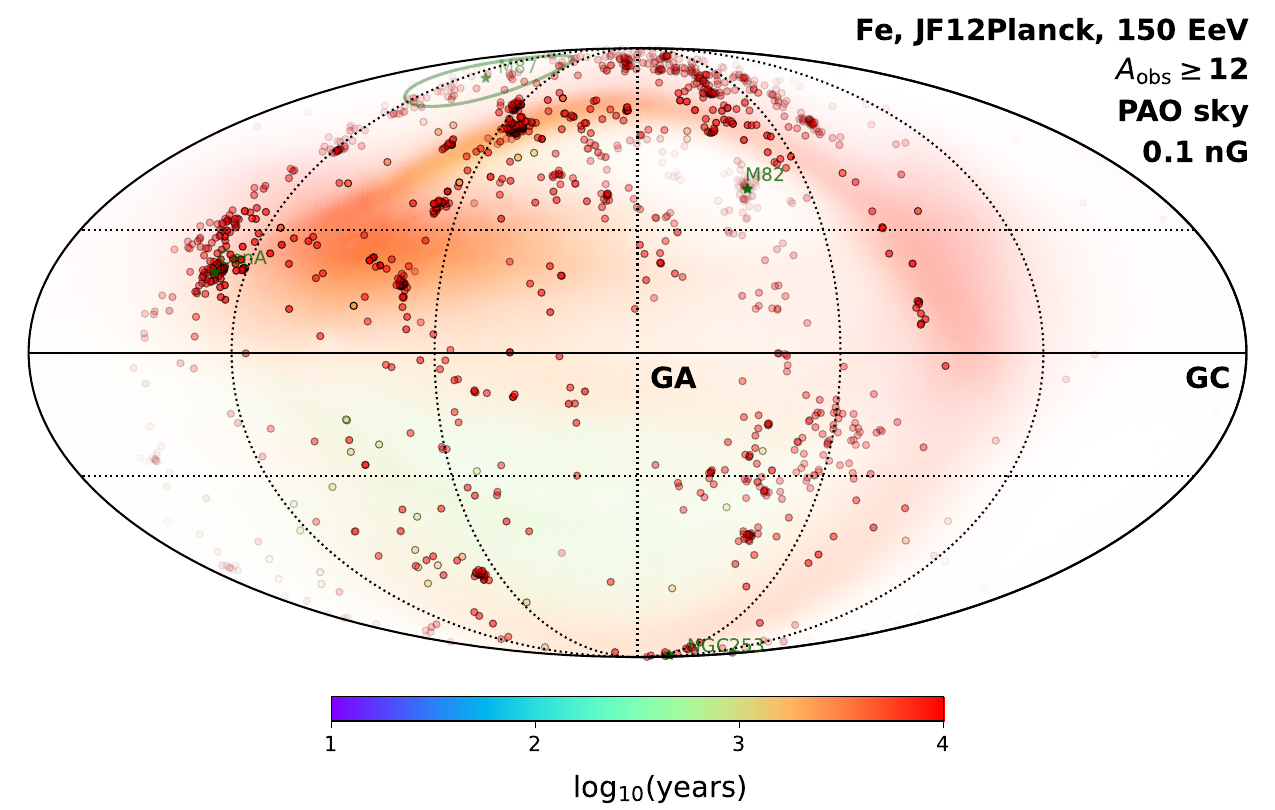}

\caption[]{EECR at 150 EeV  ``treasure maps'' for TA (left) and PAO (right), for nitrogen (top), iron (bottom)
for the JF12Planck  GMF model. The host galaxy candidates are shown as circles, their color corresponds to the
total temporal dispersion in the Galactic and extragalactic magnetic fields, $\tau_{d}$, and their transparency
channel is set to $\min(a_\text{GZK}, M)$ where $a_\text{GZK}$ is the attenuation factor due to the GZK effect.
Sources farther than $d_{95\%}$ are not shown for clarity. The color of the gradients is assigned to $\tau_{d,
\text{GMF}}$. The gradient's opacity is controlled by the magnification factor $\min(M, 1)$, which include the
detector exposure function.  The source marker opacity is set to $\min(a_\text{GZK}, M)$, i.e.~within the
detector's exposure the markers fade mostly due to $a_\text{GZK}$ and due to $M$ outside of it. We consider
here the case $\a_\text{obs}\geq 12$ at detection, to demonstrate the impact of on the observed volume in case of a
composition-sensitive detector that can provide a sub-sample of nucleus-like events. Figures are taken from
\cite{GFB2023}.}
\label{fig:TM_150}
\end{figure}

Examples of "treasure maps" are shown in Figs.~\ref{fig:TM_150}. The maps are in Galactic coordinates and
centered on the Galactic anti-center (GA), since for EECRs heavier than protons the directions behind the
Galactic center (GC) are unreachable due to the strong deflections.  Color coding is used for the time
dispersion $\tau_{d, \text{GMF}}$. Transparency is assigned to the magnification map, where $M=0$ corresponding
to fully transparent and $M\geq 1$ to solid color. The host galaxy candidates are shown as circles, their color
corresponds to the total $\tau_{d}$, and their transparency channel is set to $\min(a_\text{GZK}, M)$ where
$a_\text{GZK}$ is the attenuation factor due to the GZK effect. (Sources farther than $d_{95\%}$ are not shown
for clarity.) Due to the deflections within the GMF, parts of the sky are inaccessible to observations since
there are no valid trajectories that can connect the Earth to a source. One needs to consider the location and
zenith angle coverage in case of a ground based detector. In particular, this is important for heavier
compositions where the accessible regions of the sky don't coincide with the exposure function at Earth.

These treasure sky maps  demonstrate that to estimate the visibility of specific host galaxies within the field
of view of an observatory requires anisotropic, four-dimensional modeling. The GMF plays a crucial role even at
these extreme energies. Thus, neglecting the GMF's impact or approximating deflections by isotropic smearing
kernels is inappropriate for source searches under the assumption of light or heavy nuclear composition. We
have compiled the figures for all  the tested models in a public data release \cite{EECR_treasure_maps_repo}.

\flashtalkDone{Marco Muzio}{Source Constraints from $\nu$\&p with GCOS}
How ultrahigh energy cosmic ray (UHECR) sources are distributed in the universe remains an open question. This \textit{source evolution} represents both a challenge and opportunity to learn about the sources of UHECRs. The UHECR spectrum is sensitive to the evolution at low-redshifts, $z < 1$, while neutrinos produced by UHECR interactions can be observed from all redshifts. Together these messengers will be able to constrain the UHECR source evolution.

The Pierre Auger Observatory has shown that the UHECR spectrum is of mixed composition. However, there is still evidence that a proton component may exist at the highest energies: Auger analysis of the highest-energy \Xmax distribution is compatible with a $\sim 1-10\%$ proton fraction~\cite{PierreAuger:2023xfc}, the Auger combined fit results in a non-negligible proton fraction above $10^{19.7}$~eV~\cite{PierreAuger:2022atd}, and phenomenologically such a component significantly improves the quality of fit~\cite{Muzio:2019leu}.

Using the elaborated Unger-Farrar-Anchordoqui (UFA) source model~\cite{Unger:2015laa,Muzio:2019leu,Muzio:2021zud}, we consider two distinct UHECR source populations: a baseline population responsible for the bulk of the UHECR data, and a population producing protons above $10$~EeV. For a given source evolution, we fix baseline population's parameters to give the best-fit to UHECR data~\cite{Verzi:2020opp,Yushkov:2020nhr}. We then uniformly sample the pure-proton population's parameter space and its normalization relative to the baseline population, so as to explore the full range of model possibilities. We discard models which are not compatible with UHECR, neutrino, and gamma-ray data \cite{Verzi:2020opp,Yushkov:2020nhr,IceCube:2018fhm,IceCube:2021rpz,Fermi-LAT:2014ryh,Fermi-LAT:2015otn,Rautenberg:2021vvt,PierreAuger:2021mjh,PierreAuger:2022uwd}.

\begin{figure}[t!]
    \centering
    \includegraphics[width=0.49\textwidth]{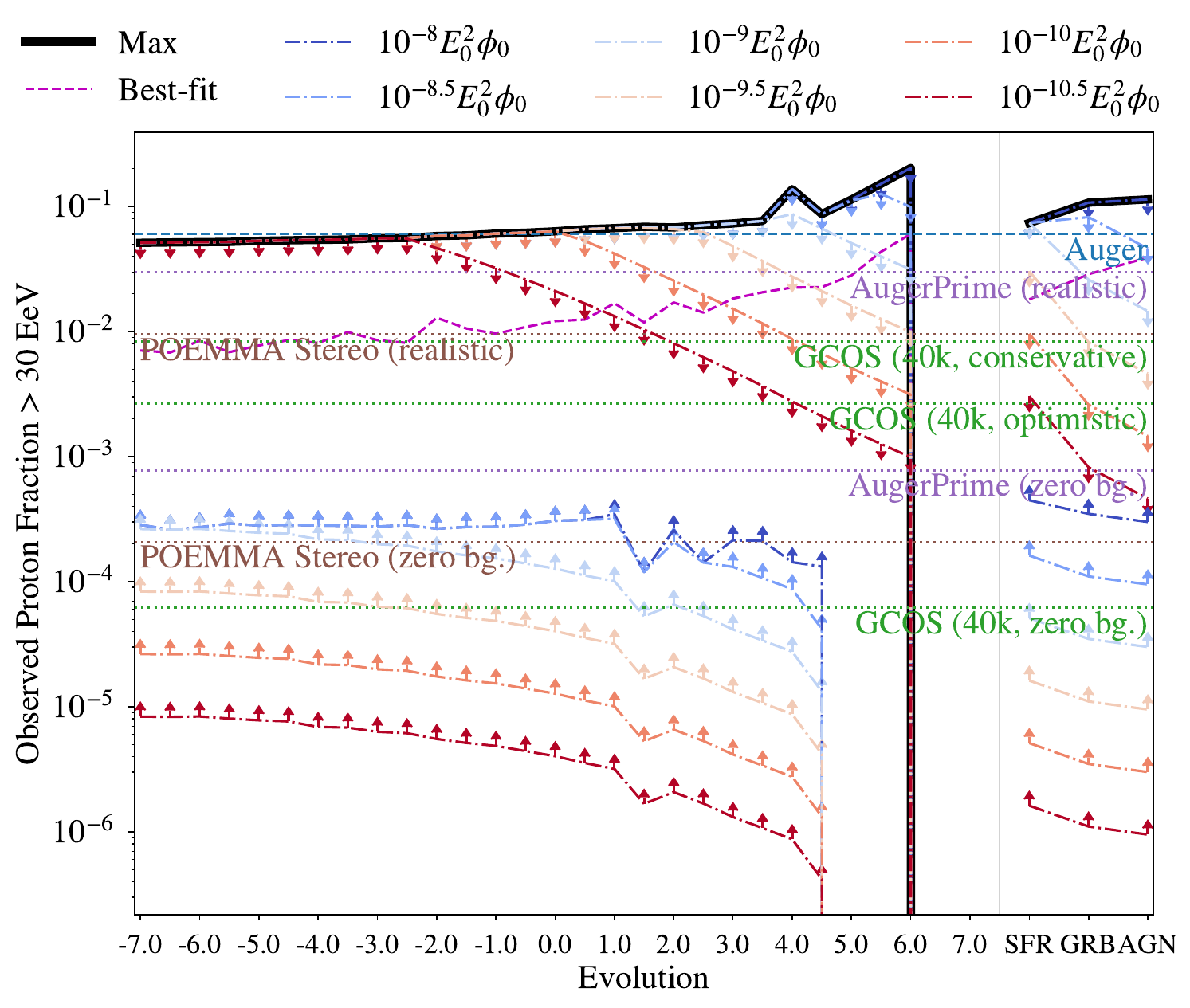}
    \hfil
    \includegraphics[width=0.49\textwidth]{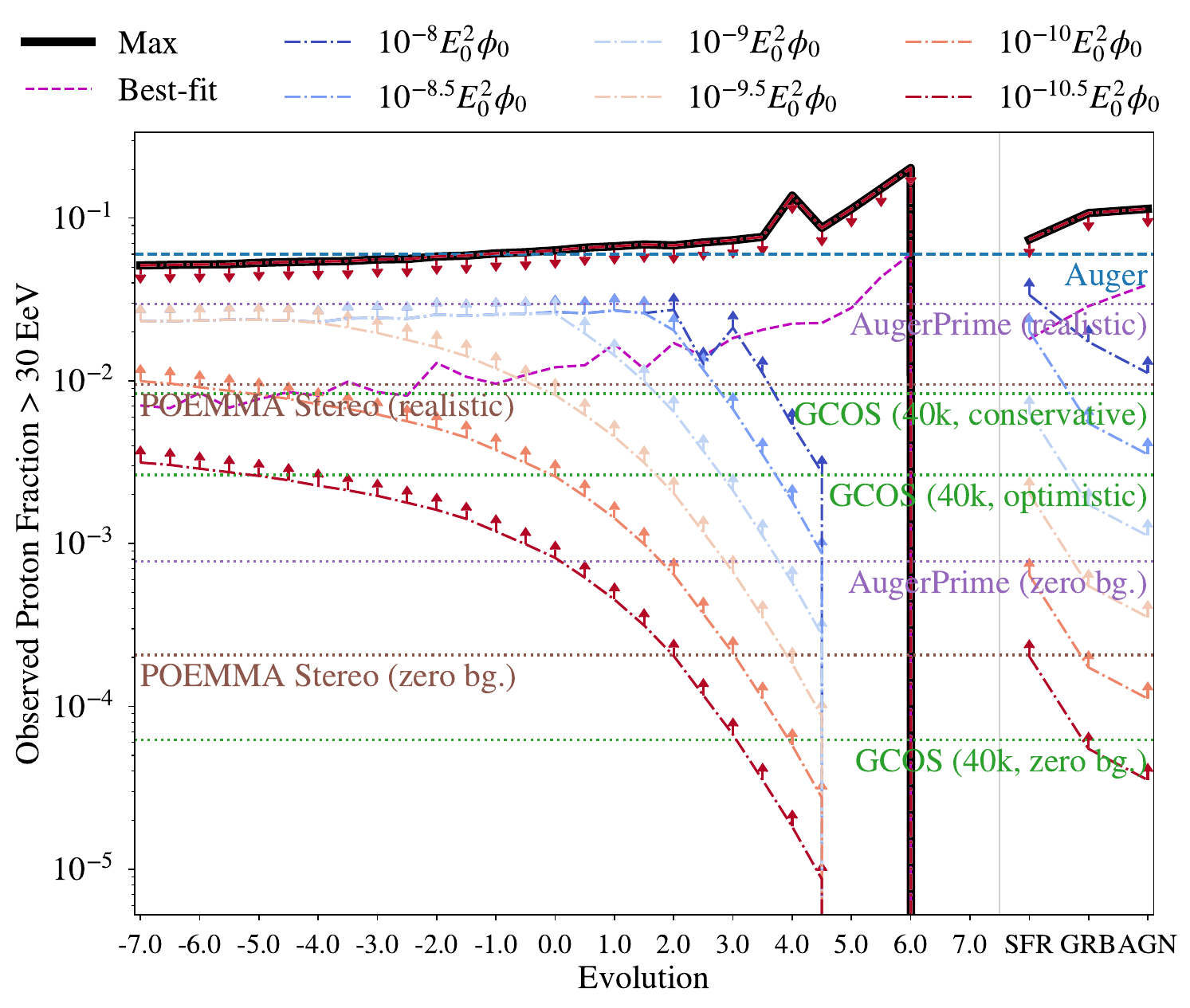}
    \caption{The range of observed proton fractions above $30$~EeV compatible with a various levels of neutrino flux at $1$~EeV (left) and $10$~EeV (right) measured in $E_0^2\phi_0=$~GeV$/$cm$^2/$s$/$sr. }
    \label{fig:obsp_generalCase}
\end{figure}

We find that for a purely cosmogenic neutrino spectrum, the observed proton fraction and $1$~EeV neutrino flux are strongly correlated, with very little dispersion~\cite{Muzio:2023skc} -- in agreement with the results of \cite{vanVliet:2019nse}. In this case, GCOS will be be able to place stringent constraints on the evolution of such a population when combined with data from near-future neutrino observatories. In the general case, where we allow for a significant astrophysical neutrino flux, the observed proton fraction and the $1$~EeV neutrino flux still correlate well but have a much larger dispersion. This large dispersion means that only an upper-bound on the source evolution can be set by combining the observed proton fraction with $1$~EeV neutrino flux measurements. However, we find that if the $10$~EeV neutrino flux is also measured these three observables will allow GCOS to place upper- and lower-limits on the evolution of such UHE proton sources (see Fig.~\ref{fig:obsp_generalCase}).

\flashtalkDone{Markus Risse}{Photons with GCOS}
UHE photons are unique messengers and nicely complement  UHE neutrinos and nuclear cosmic rays.
They are the only gauge bosons we can access as EAS primaries.
Photons are neutral and massless, so source pointing is possible.
Particularly, the search for transient sources can be performed by looking for time-directional correlations.
This makes photons a {\bf cornerstone of multimessenger astronomy}.
Photons are GZK messengers, being produced in photo-pion interactions of UHE protons.
With an attentuation length of $\sim$10~Mpc, UHE photons test the local universe, complementary to neutrinos.
Photons are also possible indicators of BSM physics:
large photon fractions are predicted in top-down scenarios.
An unexpected correlation with a distant transient can be a smoking gun for new physics such as Lorentz invariance violation or axions.
Taken together, {\bf UHE photons are possible game changers}.

The search for UHE photons via EAS~\cite{Risse:2007sd} profits from their interactions being dominated by electromagnetic (EM) processes.
Firstly, this reduces hadronic uncertainties. Secondly, EAS detectors used for normal CR are fine.
To trigger on photon showers, an EM detector is needed. The EM detector can further provide the shower energy and,
via $X_{\rm max}$, separation power to nuclear CR. The lack of muons strongly separates photons from nuclear primaries.
In short, {\bf what is good for composition, is very good for photons} (assuming an EM trigger exists).

\begin{figure}[!hbt]
\centering
\includegraphics[width=0.8\columnwidth]{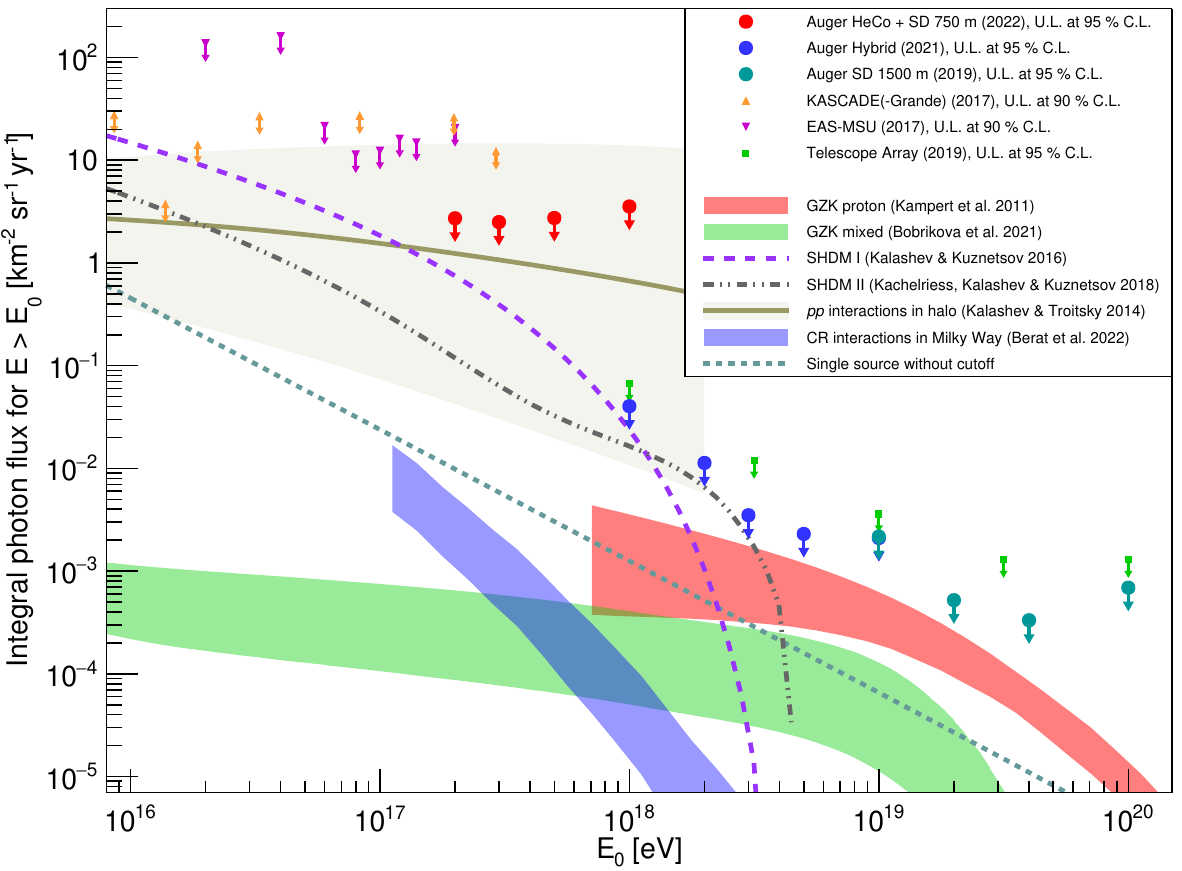}
\caption{Photon limits on diffuse fluxes and predictions (from Ref.~\cite{PierreAuger:2022uwd}).}
\label{fig_photonlimits22}
\end{figure}

Current limitations relate to the exposure (at UHE) and to separation power (towards lower energy or, more general, when larger statistics leads to background photon candidates).
The present bound at 20~EeV (Fig.~\ref{fig_photonlimits22}) scratches predictions from cosmogenic proton scenarios. With a factor 10 improvement, the sensitivity reaches mixed composition scenarios:
{\bf a detection of cosmogenic photons might be in reach with GCOS}.

In conclusion, {\bf the detection of UHE photons can be considered a primary goal of GCOS}.
Should there be earlier discoveries, the case of UHE photons only becomes stronger.

\begin{figure}[htbp]
\begin{center}
\includegraphics[clip, trim=0cm 2.8cm 0cm 2.65cm, width=0.9\linewidth]{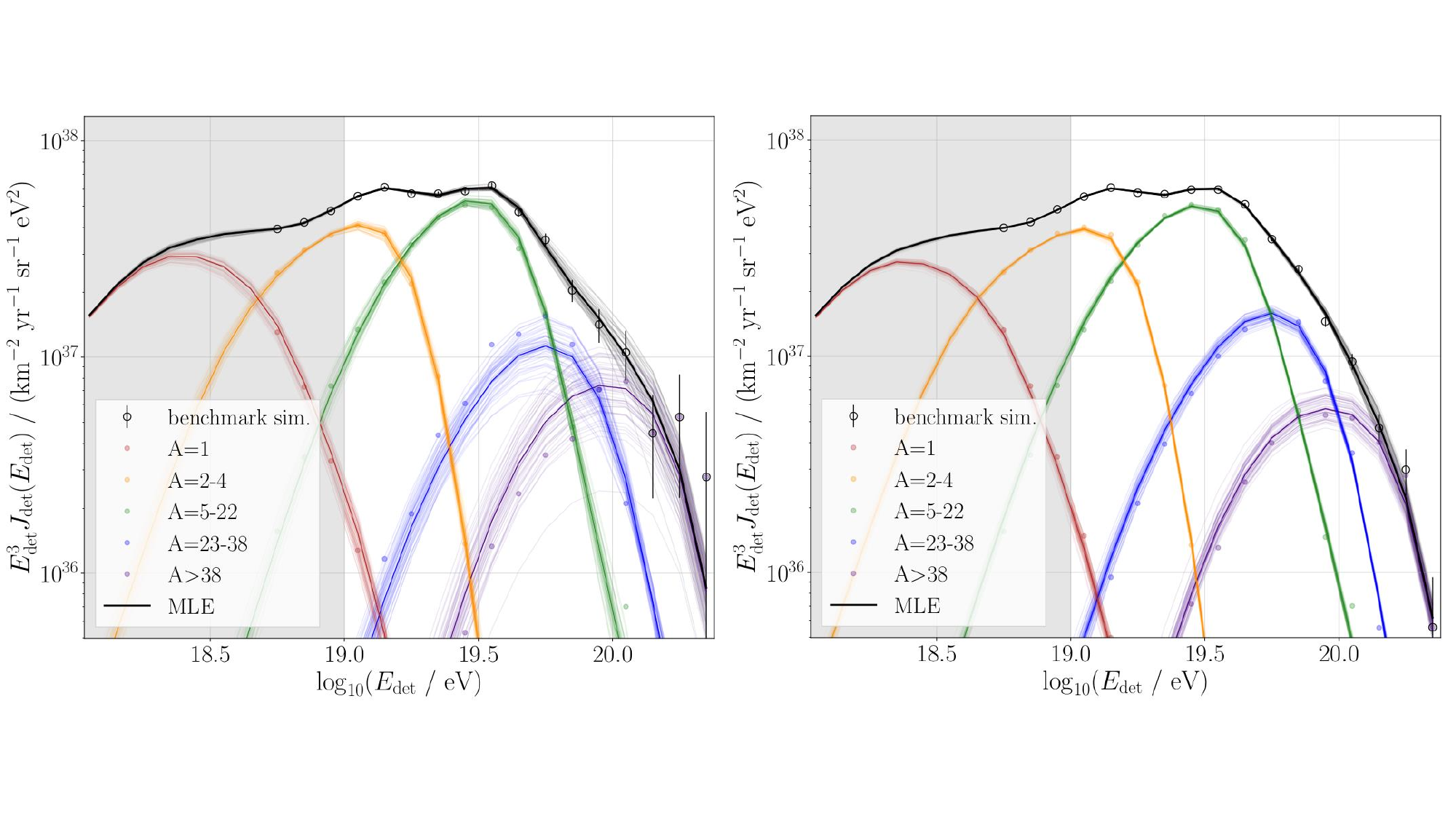}
\caption{Energy spectrum (markers) of simulation with current statistics of the Pierre Auger Observatory (\textit{left}) and 10 times increased statistics (\textit{right}). The thick lines represent the best-fit, the thin lines the uncertainty estimated by the MCMC sampler. With increased statistics, the statistical fluctuations decrease significantly, allowing for a more efficient rejection of alternative models and a reduced uncertainty on the fitted source parameters.}
\label{fig:sspec}
\end{center}
\end{figure}

\flashtalkDone{Teresa Bister}{``Combined Fit'' with GCOS statistics}
The combined fit of energy spectrum and shower depth distributions of ultra-high-energy cosmic rays allows conclusions about the properties of the cosmic-ray source distribution~\cite{PierreAuger:2021mmt, PierreAuger:2016use}. For example, if the source injection is assumed to follow a Peter's cycle, the combined fit can be used to constrain parameters of the source injection, namely the maximum rigidity, the spectral index of the injection spectrum, and the contribution of different elements. The inference of these parameters can be performed with MCMC methods, which additionally allow for the determination of parameter uncertainties via posterior distributions.\\
With an increase of statistics, as expected with GCOS, the uncertainties on the fit parameters can be significantly reduced. In an example simulation which mimics the shape of the spectrum and shower depth measurements of the Pierre Auger Observatory, the uncertainty on the spectral index could be reduced by a factor of 6 when increasing the statistics by a factor of 10. The fitted energy spectrum is depicted in Fig.~\ref{fig:sspec}.
With this increased precision in the reconstruction of possible source parameters, theoretical models (for example acceleration mechanisms) can be further constrained. Additionally, models for the source distribution or evolution can be discarded more effectively if they do not describe the more precisely measured spectrum or shower depth distributions for GCOS statistics.\\
\\
When the arrival directions are included as an additional observable, the discrimination power of the combined fit regarding different source catalogs is greatly enhanced~\cite{Bister:2021H0}.
This effect is tested on another simulation based on~\cite{Bister:2021H0} in which the anisotropies measured in the cosmic-ray arrival directions~\cite{PierreAuger:2018qvk} originate from a contribution by starburst galaxies.
With current statistics, the combined fit can identify the correct starburst galaxy model with a significance of 5.4$\sigma$ compared to an alternative model containing active galactic nuclei as sources. When again 10 times more events are simulated, the discrimination power can become as large as $>20\sigma$ (not considering detector effects).\\
In the future, the combined fit including arrival directions can further benefit from the whole-sky exposure that is foreseen for GCOS. Also, a mass-indicating observable on an event-level could allow for new possibilities on reconstructing the Galactic magnetic field parameters simultaneously with the source parameters~\cite{Wirtz:2021ifo}.

\flashtalkDone{Michael Unger}{Source Discoveries with GCOS}
\label{sec:unger2}
The Telescope Array and the Pierre Auger Observatory have reported
tantalizing evidence of intermediate-scale anisotropies in the arrival
directions of UHECRs. For instance, Auger detected a hotspot of
$N_\text{obs}=153$ events in the the Centaurus region, while an
isotropic distribution would predict only $N_\text{obs}=97.7$. If the
signal continues to grow with the current rate, its significance
is expected to reach 5~$\sigma$ at a total accumulated exposure of
(165\,000$\pm$15\,000)~km$^2$~yr~sr~\cite{PierreAuger:2022axr}, i.e.\
when the exposure reaches approximately 1.5 of the Auger Phase I
exposure. This milestone would mark the first-ever imaging of a UHECR
source. However, to truly identify the origins of UHECRs, a
large-exposure observatory like GCOS is needed to discover fainter
sources and to distinguish between different source classes.

\begin{figure}[h]
  \centering
  \def\figw{0.33}
 \includegraphics[width=\figw\linewidth]{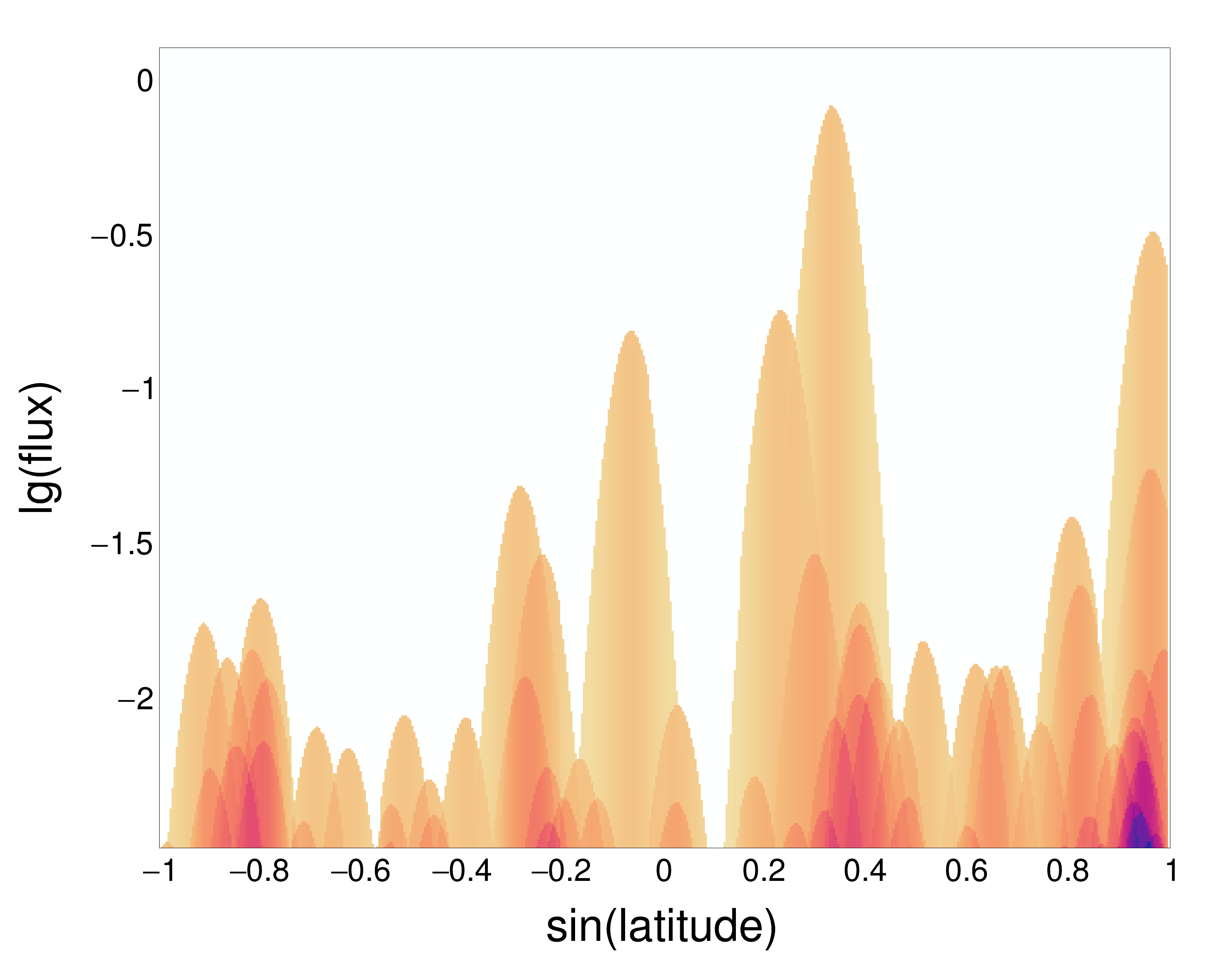}%
 \includegraphics[width=\figw\linewidth]{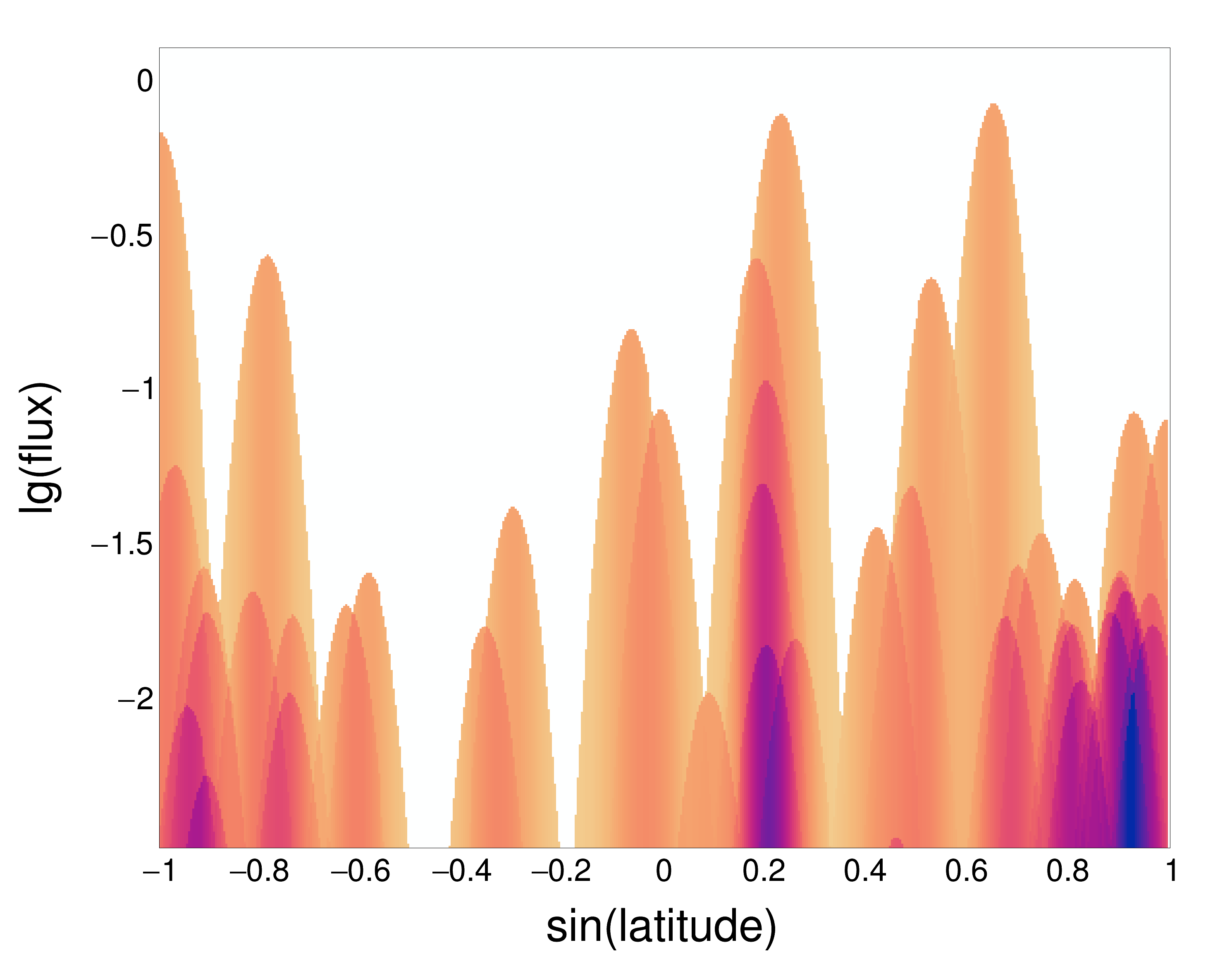}%
 \includegraphics[width=\figw\linewidth]{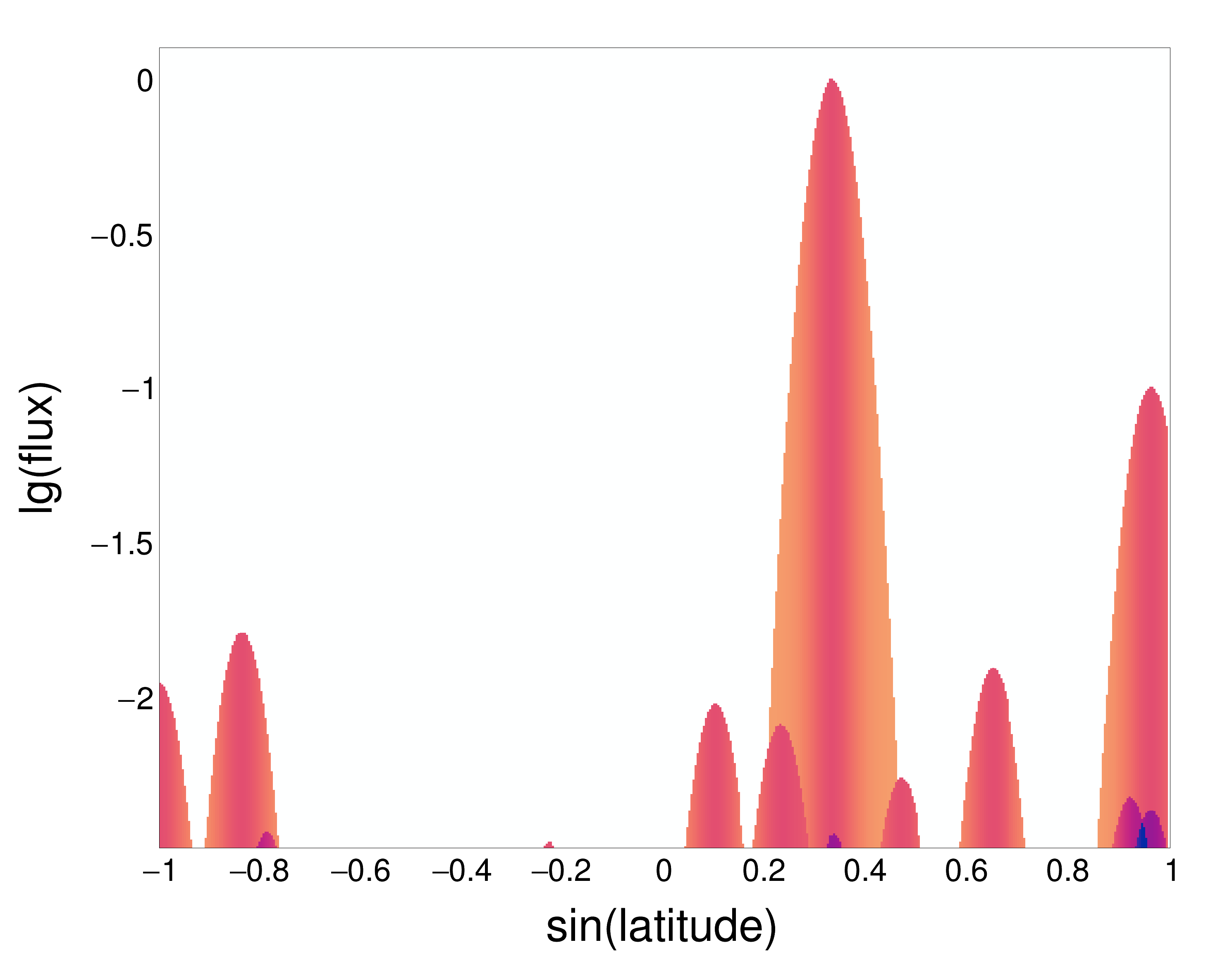}%
 \caption{Illustration of the UHECR flux from source candidates as a function of latitude. Left: active galactic nuclei (AGN),
   middle: starburst galaxies (SBG), right: powerful radio galaxies (RG). \label{fig:catalogues}}
\end{figure}

How many sources can we hope for to detect for a given exposure? To address this question, we assume that the Centaurus excess of Auger is
caused by the brightest objects in its direction, as listed in common catalogs of UHECR source candidates. We consider three source classes: starburst
galaxies~\cite{2019JCAP...10..073L} (Centaurus objects: NGC4945 and
M83, flux proxy: 1.4 GHz radio flux), SwiftBAT
AGNs~\cite{2018ApJS..235....4O} (Centaurus objects: CenA and NGC4945,
flux proxy: 14-195~keV X-ray flux) and powerful radio
galaxies~\cite{vanVelzen:2012fn,Matthews:2018laz} (Centaurus object:
CenA, flux proxy: 1.4 GHz radio flux). In each case we account for the attenuation of the flux due to UHECR energy loss following
Ref.~\cite{PierreAuger:2018qvk} using their ``composition scenario A'' at $E\geq
40$~EeV. A potential (de-)magnification of the fluxes due to the
Galactic magnetic field is not included in this calculation (see e.g.\ Ref.\ \cite{Bister:2024ocm} for magnification maps for the ensemble of magnetic field models of Ref.~\cite{Unger:2023lob}). The attenuated
fluxes are visualized in Fig.~\ref{fig:catalogues} as
flux-``mountains'' of arbitrary width as function of Galactic
latitude.

 Current observatories will soon be able to detect the peak of the
 tallest mountain,
\begin{wrapfigure}{r}{10cm}
\centering
\vspace*{-0.2cm}
\includegraphics[width=\linewidth]{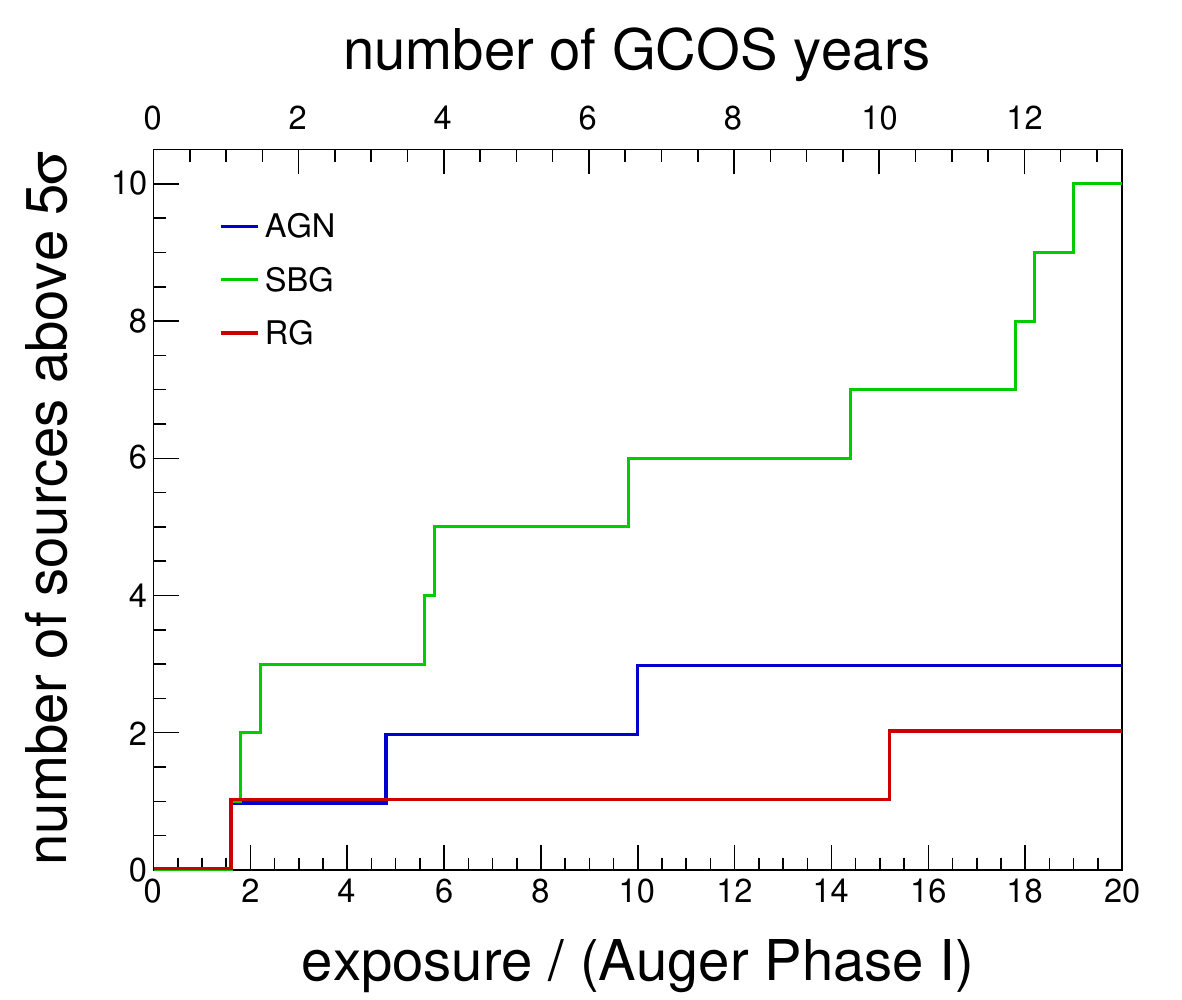}
\caption{Expected number of source images as a function of exposure for the different source classes shown in Fig.~\ref{fig:sources}. The exposure is given in units of ``Auger phase I'', i.e.\ $1.2\times 10^5$ km$^2$\,sr\,yr~\cite{PierreAuger:2022axr}. The number of years to collect the exposure with GCOS is shown at the top of the plot.\label{fig:sources}}
\end{wrapfigure}
\noindent
  while GCOS will probe the next set of lower mountain peaks.  The source
 discovery potential as a function of exposure is shown in
 Fig.~\ref{fig:sources}. For each source catalogue, GCOS is expected
 to observe a different number of sources. The distribution of source
 strength together with the positions on the sky will help to
 identify the sources of UHECR.  Note that the more source images are
 discovered, the easier it will be to match their positions to
 catalogues allowing for deflections in magnetic fields.

Within one year of operation, GCOS can confirm the hot spots of TA and
Auger at 5 $\sigma$ significance.  After several years, it will bring
us within reach of answering the long-standing question of the origin
of UHECR sources.

\subsubsection{Further Contributions}
\label{sec:futherScience}
{Glennys Farrar -- GMF Caustics at UHE}\\
{\footnotesize \url{https://agenda.astro.ru.nl/event/21/contributions/260/attachments/76/86/GCOS_071322.pdf}}
\\

\newpage
\subsection{Particle detector}

\begin{figure}
    \centering
    \includegraphics[page=1, width=0.45\textwidth]{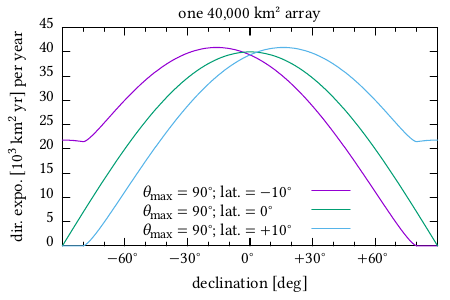}\hfil
    \includegraphics[page=2, width=0.45\textwidth]{plotArmando.pdf}
    \caption{Directional exposures of hypothetical SD arrays with~$\theta_{\max} = 90^\circ$ (left) and~$85^\circ$ (right).  Note how curves other than the left green one vanish in a region near $\delta = -90^\circ$, $+90^\circ$, or both.}
    \label{fig:expo1}
\end{figure}

\flashtalkDone{Armando di Matteo}{location (exposure)}
\label{sec:dimatteo2}
The directional exposure of a surface detector (SD) array to ultra-high-energy cosmic rays (UHECRs) depends on its geographic latitude~$\lambda$ and maximum zenith angle~$\theta_{\max}$, as described in Section~2~of Ref.~\cite{Sommers:2000us}.  Full-sky exposure with one array is only possible if both~$\theta_{\max} = 90^\circ$ and~$\lambda = 0^\circ$; any array with~$\theta_{\max} < 90^\circ$ or~$\left|\lambda\right| > 0^\circ$ will have a blind spot around one or both celestial poles, as shown in Fig.~\ref{fig:expo1}.
Using an array in the southern hemisphere and one in the northern hemisphere, it is possible to achieve full-sky coverage even with~$\theta_{\max} < 90^\circ$.  The total coverage is reasonably homogeneous when $\lambda \sim \pm35^\circ$.

\begin{figure}
    \centering \includegraphics[page=12,
    width=0.45\textwidth]{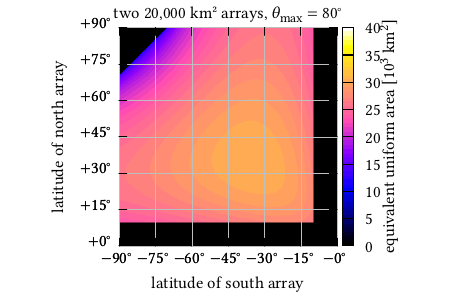}\hfil \includegraphics[page=16,
    width=0.45\textwidth]{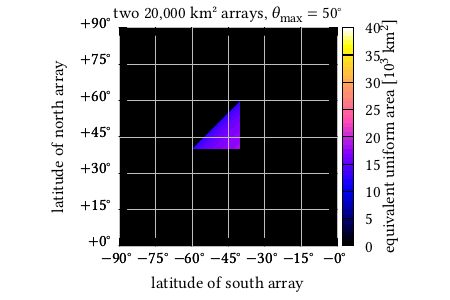} \caption{The equivalent
    uniform area (inversely proportional to the square of expected
    statistical uncertainties, see text) of a pair of SD arrays
    with~$\theta_{\max} = 80^\circ$ (left) and~$50^\circ$ (right) as a
    function of their latitudes} \label{fig:expo3}
\end{figure}

With large~$\theta_{\max}$, there is a wide range of latitudes at
which the arrays can be located while retaining near-optimal coverage,
whereas with small~$\theta_{\max}$ the latitudes need to be more
fine-tuned, as shown in Fig.~\ref{fig:expo3}, where the equivalent
uniform area for two values of the maximum zenith angle is shown.
This is the area a detector with uniform exposure would need to have
to achieve the same statistical uncertainties. It is inversely
proportional to the integral over the sky of the inverse directional
exposure.
Hence, it is recommended for GCOS to use detectors sensitive to near-horizontal showers, such as water Cherenkov or radio ones, rather than scintillators alone.

Other possibilities which would result in acceptable full-sky coverage include placing $\sim 90\%$ of the SDs at~$\lambda \sim +15^\circ$ and the rest at~$\lambda \sim -80^\circ$, or placing $\sim 40\%$ of the SDs on the equator and the rest in two equal-size arrays at $\lambda \sim \pm 50^\circ$ (see \url{https://agenda.astro.ru.nl/event/18/contributions/145/} for details and more examples).

\flashtalkDone{Ioana Maris}{How many surface particle detectors do we need and what spacing?}
\label{sec:maris1}

\begin{figure}[tb]
    \hspace*{-3ex}\includegraphics[width=0.34\textwidth]{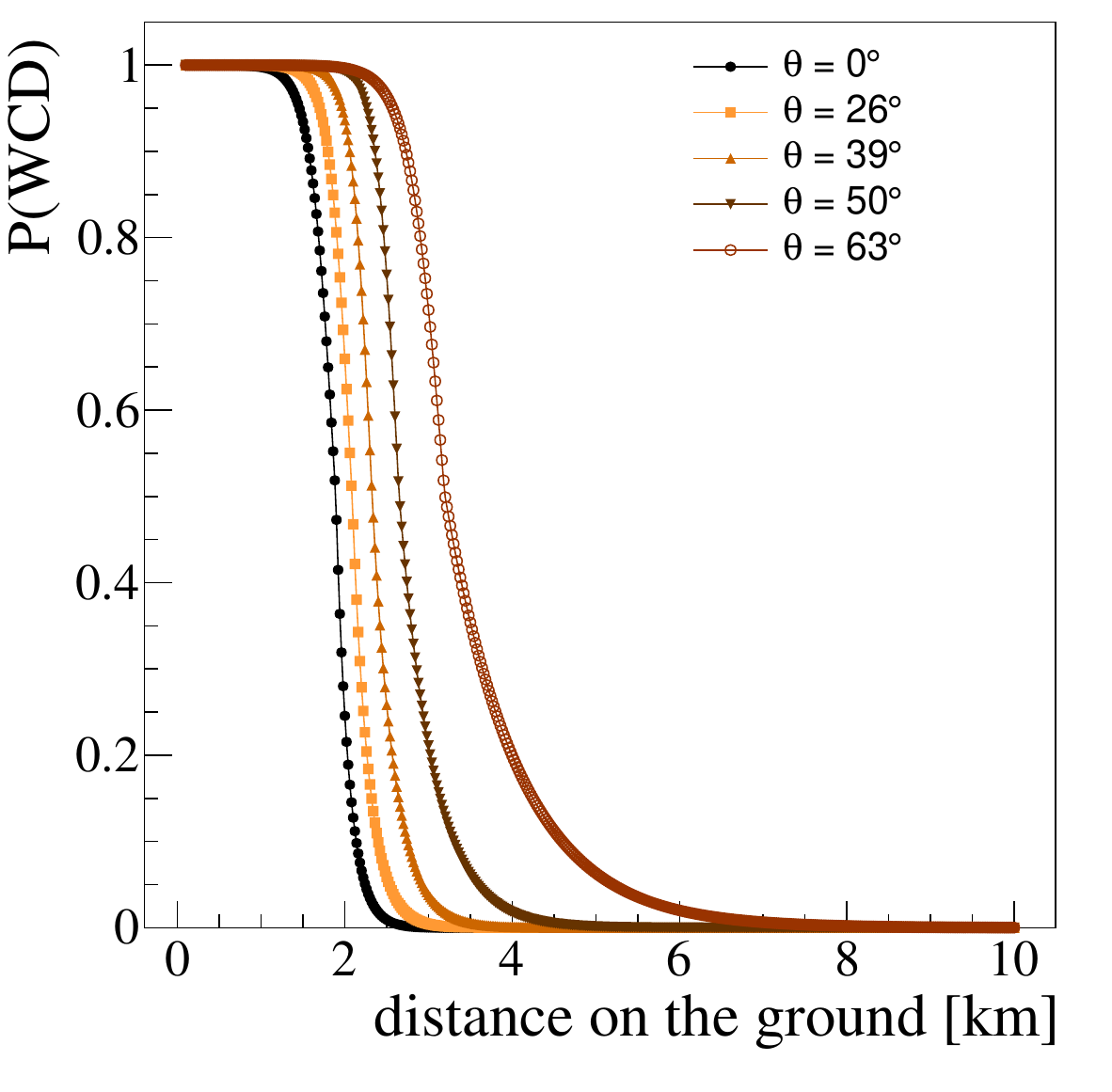}
    \includegraphics[width=0.34\textwidth]{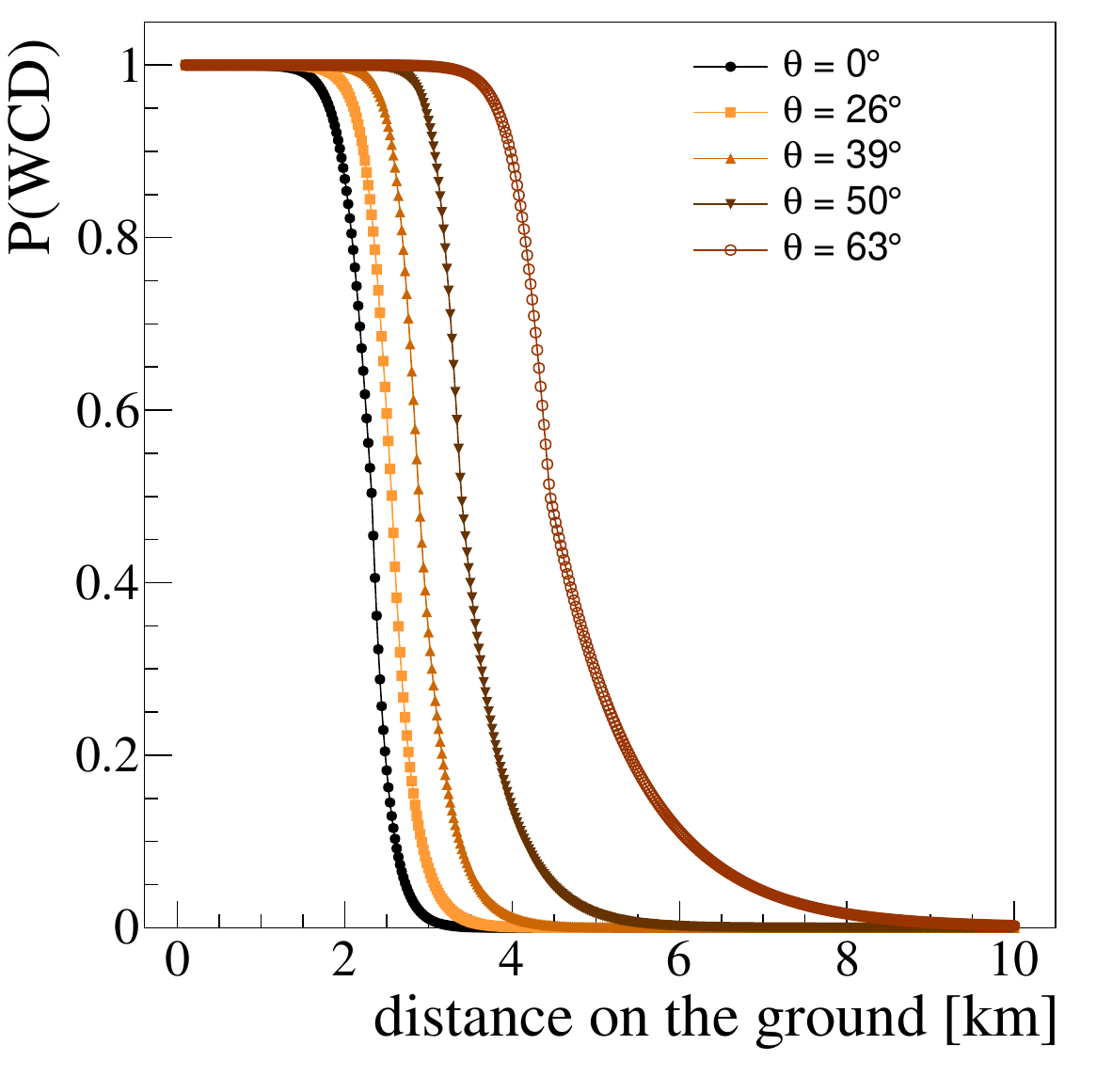}
     \includegraphics[width=0.34\textwidth]{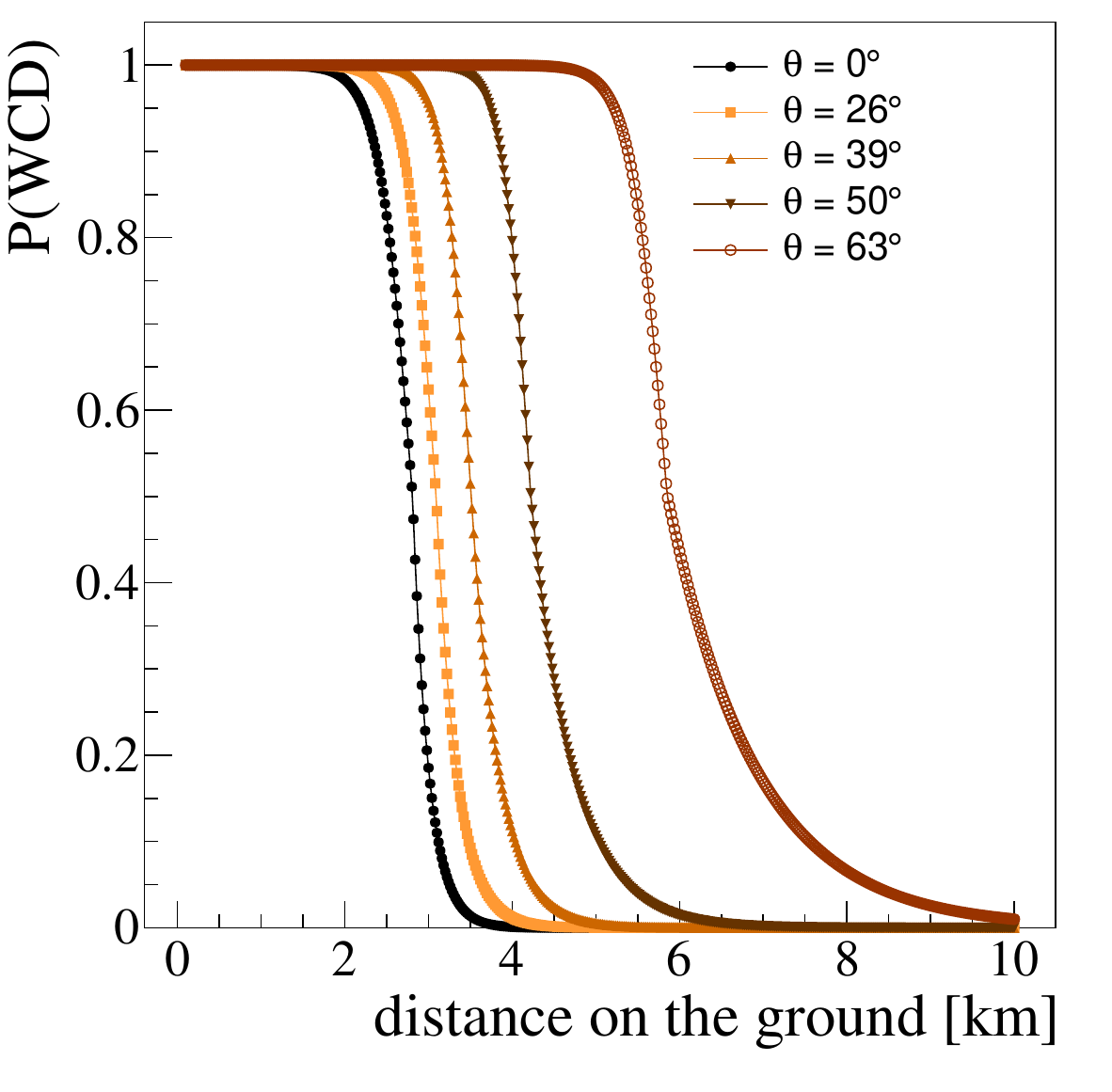}
    \caption{The lateral trigger probabilities for proton initiated air-showers at lg(E/eV)=[19,19.5,20] (left, center, right).}
     \label{fig:ltps}

\end{figure}

\begin{figure}[tb]
    \hspace*{-3ex}\includegraphics[width=0.34\textwidth]{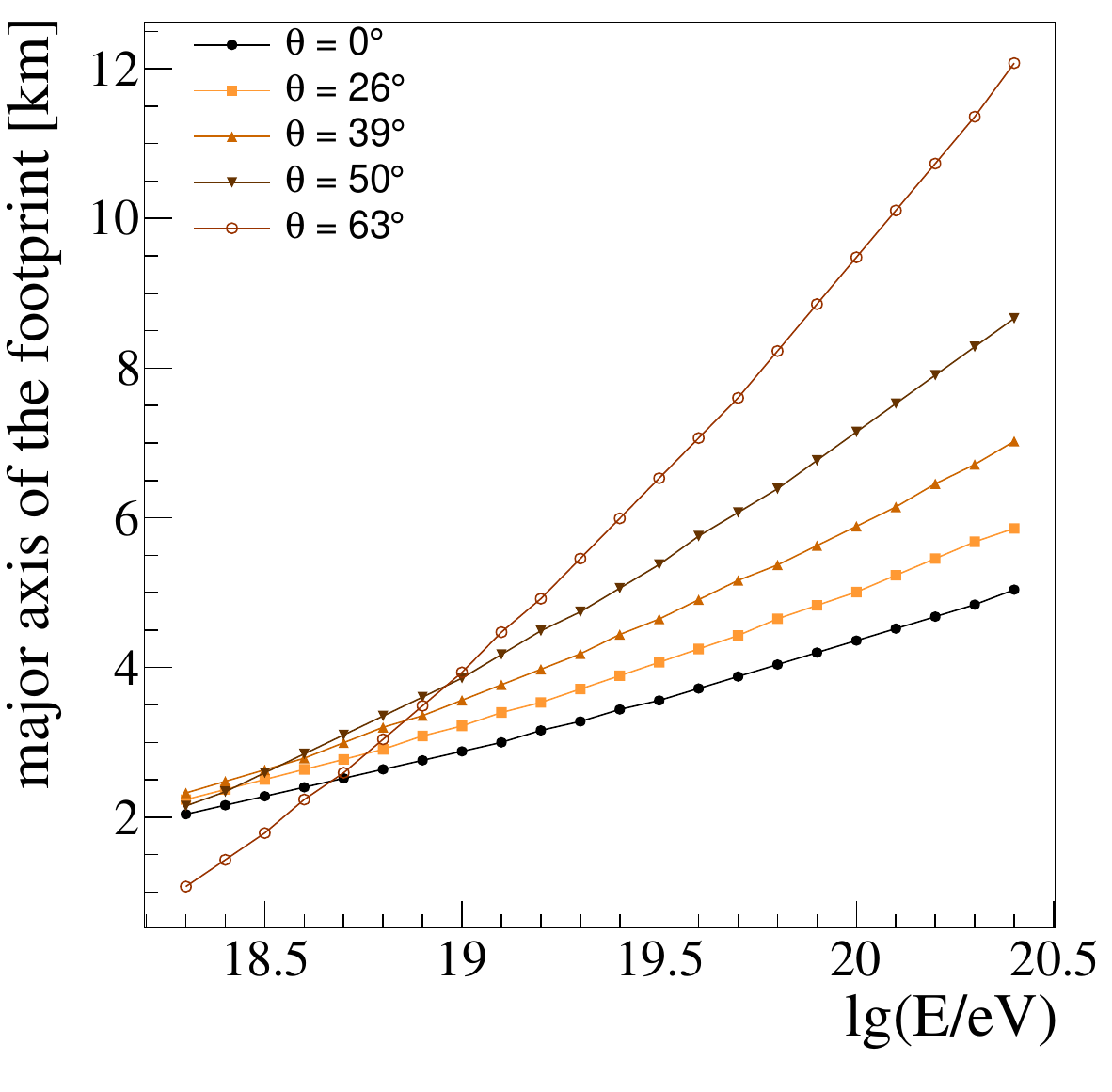}
    \includegraphics[width=0.34\textwidth]{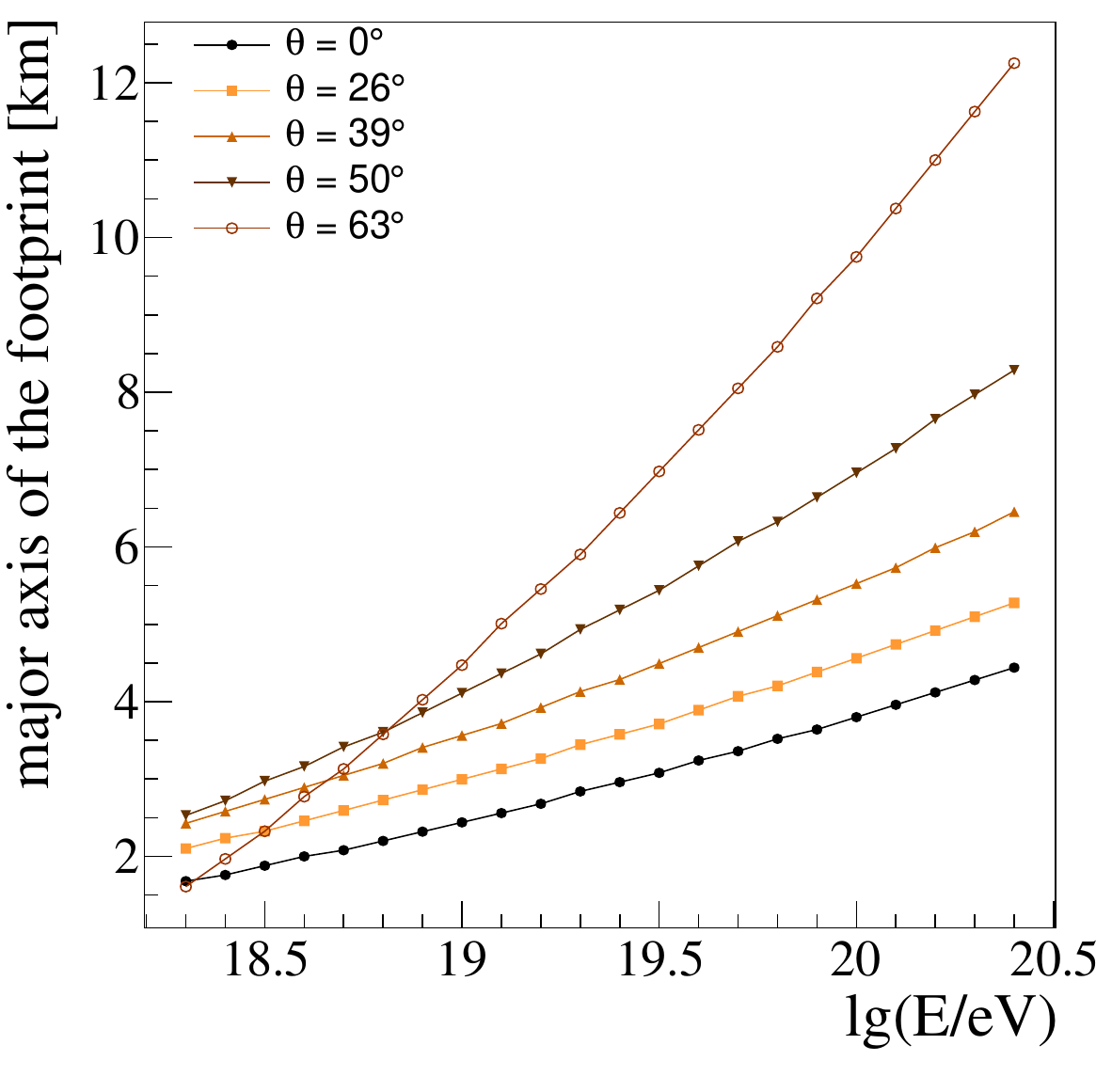}
     \includegraphics[width=0.34\textwidth]{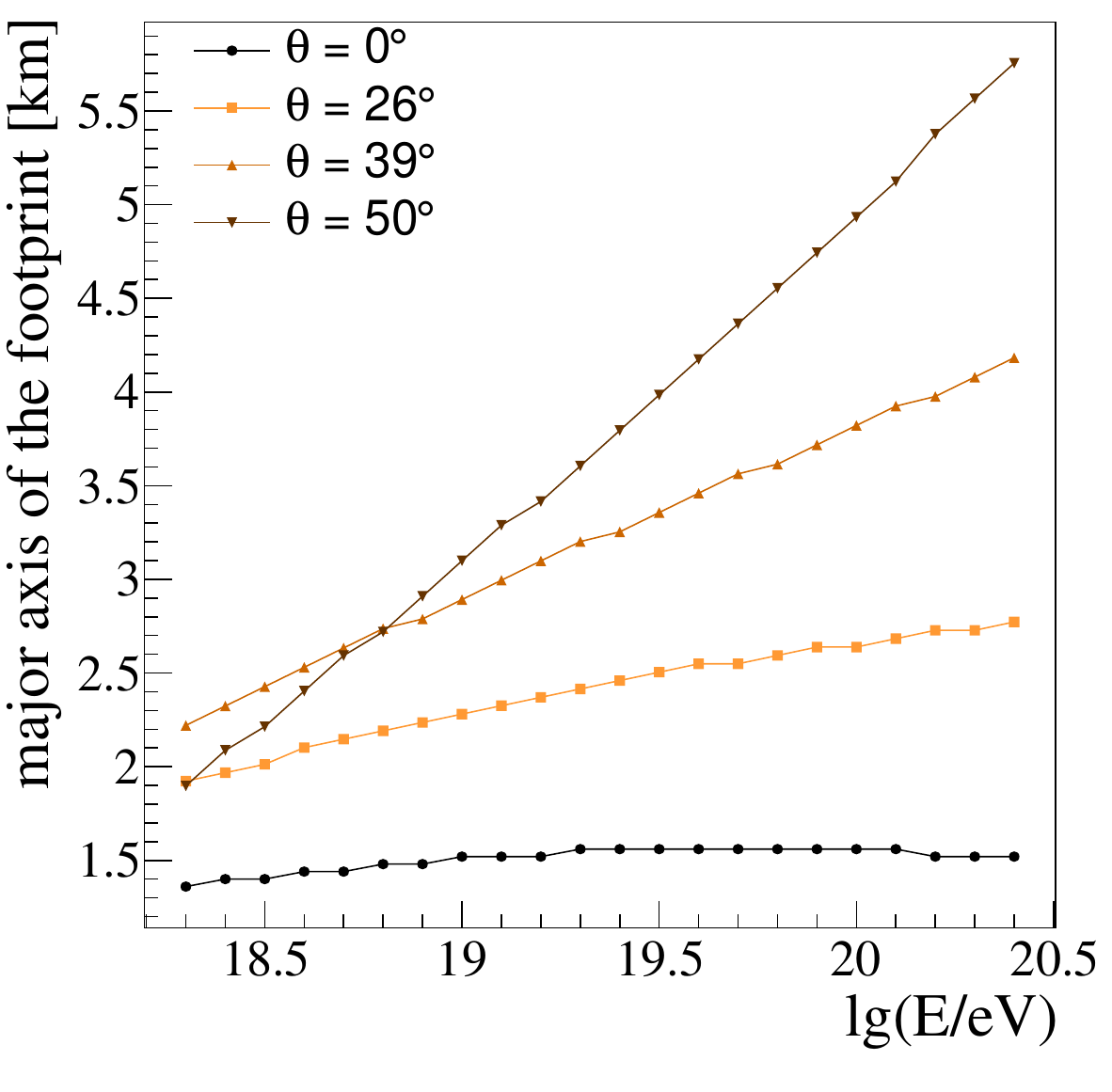}
    \caption{The size of the footprint on the ground (major axis, defined as more than 98\% trigger efficiency) for iron (left), proton (center) and photon (right).}
    \label{fig:majoraxis}
\end{figure}

To estimate the number of required detectors a water-Cherenkov detector is assumed in this study, similar to the surface detectors of the Pierre Auger Observatory.  The requirements for GCOS are full trigger efficiency at energies above 10~EeV.

\begin{wrapfigure}{r}{0.33\textwidth}
    \includegraphics[width=0.33\textwidth]{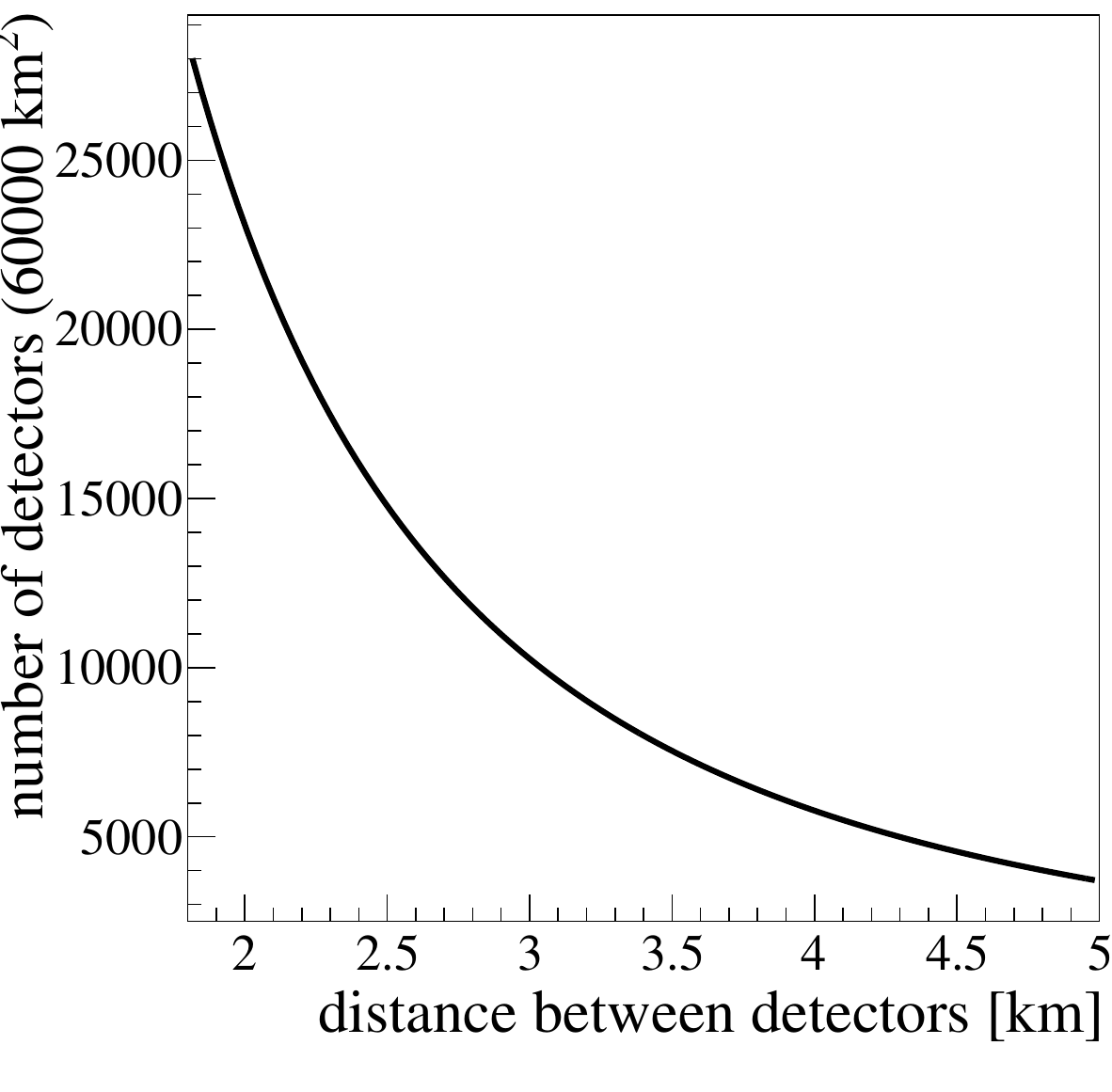}
  \caption{Number of detectors needed to cover a surface of 60,000 km$^2$ as a function of the distance between detectors for a triangular grid array.}
  \label{fig:det}
\end{wrapfigure}

The probabilities of having a triggered station as a function of the distance to the air-shower axis are shown in Fig.~\ref{fig:ltps} for proton-initiated air showers in the case of a simple time-over-threshold trigger at the station level. As can be seen, the probability of triggering a station decreases very fast at distances larger than about 2~km for vertical showers at 10 EeV, while for inclined showers it starts declining at around 3~km. To better understand the shower footprint the largest distances on the ground that air showers can reach were deduced as a function of energy and arrival directions.
This is illustrated in Fig.~\ref{fig:majoraxis} for iron, proton, and photon primaries. The maximum distance we will be able to allow between detectors to obtain full efficiency for the lighter components at 10~EeV cannot exceed 2.2~km. One might increase the spacing and trade trigger efficiency for costs up to 3~km, but that will drastically reduce the trigger efficiency for photons and affect the science case for the searches of ultra-high-energy photons.
The number of detectors to cover an area of 60,000~km$^2$ using a triangular grid as a function of the distance between the detectors is shown in Fig.~\ref{fig:det}. An impressive number of detectors are needed, with more than 15,000 detectors required if the final separation is smaller than 2.5~km. From this, we can conclude that it will not only be tough to build them but also that it will require industrial production and failure-free (maintenance-free) detectors.

\flashtalkDone{Pierre Billoir}{Exploiting the geomagnetic distortion of very inclined showers }
\par At large zenith angles ($\theta > 70^\circ$) the electromagnetic
component of atmospheric showers is extinguished. Moreover, the
surviving muons have a long path down to the ground, so they are
sensitive to the magnetic field of the Earth. The shape of their
footprint at ground level reflects the history of their production, so
it is sensitive to the hadronic interactions and it provides some
constraints on the models used to simulate the hadronic cascade.
\par In this study, we have extracted the muons from CORSIKA showers
and then followed them in an independent module, neglecting the low
energy photons and electrons, which have no chance to generate later muons
reaching the ground. From this sample, we have extracted the position
$X^\mu_{\rm max}$ of the maximum. Using an auxiliary sample of
``complete'' showers, we have checked that this quantity is tightly
correlated to the usual \Xmax, measured by fluorescence or radio detector. \\

\par The distortion is shown in the transverse plane in Fig.5 of Ref.~\cite{Billoir:2015cu} for
different values of $\theta$  and $B_T$, the transverse component of
the magnetic field. Fig.7 of Ref.~\cite{Billoir:2015cu} shows that, for a given position along the
shower axis, the lateral density of muons is a nearly exponential function of $\sqrt{r}$ modulated by a shape in
$\cos(2(\psi-\psi_B))$, where $\psi$ is the azimuthal angle in the
transverse plane and $\psi_B$ the direction of the transverse field in
this plane. This suggests the following empirical parameterization:
$$ f(r,\psi) =
\exp\left(\lambda(\rho)+\alpha(\rho)\cos(2(\psi-\psi_B))+\beta\cos(\psi)\right)
~~~~{\rm with}~~~~\rho=\sqrt{r/r_{\rm ref}-1} $$
The term in $\beta$ is introduced to account for the forward/backward
asymmetry on the footprint on earth; in practice it does not play an
important role.
The functions $\lambda$ and $\alpha$ may by expanded at degree 2:
$$ \lambda(\rho) = \lambda_0+\lambda_1\rho+\lambda_2\rho^2~~~~
\alpha(\rho) = \alpha_0+\alpha_1\rho+\alpha_2\rho^2 $$
When fitting this function to data obtained with a given hadronic
model, we find a clear quasi-linear correlation between the fitted
values of $\alpha_0$,$\alpha_1$ and $\lambda_0$ and $X^\mu_{\rm max}$
with a universal shape (independent of the nature and the energy of the primary)
An example is shown in Fig.10 of Ref.~\cite{Billoir:2015cu}.
\par This means that these parameters provide an indirect measurement
of \Xmax if the hadonic model is supposed to be known. Moreover, Fig.12 of Ref.~\cite{Billoir:2015cu}
shows that this ``universal'' line is different for different hadronic models, that is, the dependence of
$\alpha_0$, $\alpha_1$ and $\lambda_0$ on $X^\mu_{\rm max}$
may provide a test of validity of the model, independently of the
composition of the flux of UHECR.\\
\par This opportunity could be exploited within a big UHECR detector
providing a measurement of \Xmax (e.g. fluorescence or radio) of
inclined showers before they reach the ground, by adding a ``small'' muon
counters (e.g.\ an array of vertical scintillators of a few 100
km$^2$). This is illustrated in Fig.14 of Ref.~\cite{Billoir:2015cu}.

\flashtalkDone{Eric Mayotte}{Mass sensitivity from muon separation with WCDs and SSDs}
Water Cherenkov Detectors (WCD) are, essentially, light-tight bodies of water monitored by photo-sensitive devices. WCDs record the flashes of Cherenkov light produced when charged particles with high Lorentz factors pass through them. The amount of light created by each particle in this process is proportional to its path length through the detector. Electrons and positrons, with their high rate of energy loss in water, often do not fully transit the detector and leave shorter, lower-intensity, signals. Muons, however, pass through the detector volume regularly, leaving larger signals. The upshot of this is that muonic and electromagnetic components of extensive air showers (EAS) elicit different responses in a WCD.

Surface Scintillation Detectors (SSD) are, generally, plates of scintillators placed in a light-tight enclosure connected to a PMT through optical fibers. The internal geometries of the detectors, the compositions of the scintillators and optical fibers, and the response of the monitoring PMTs are chosen to maximize signal efficiency while preserving the arrival time structure of the signals generated when particles strike the detector~\cite{PierreAuger:2016qzd}. Generally, SSDs are too thin to significantly slow particles striking them, which in turn means they have a roughly equal response to all components of an EAS~\cite{Taboada:2020spx}.

\begin{figure}
  \begin{center}
   \vspace{-8mm}
    \includegraphics[width=0.5\textwidth]{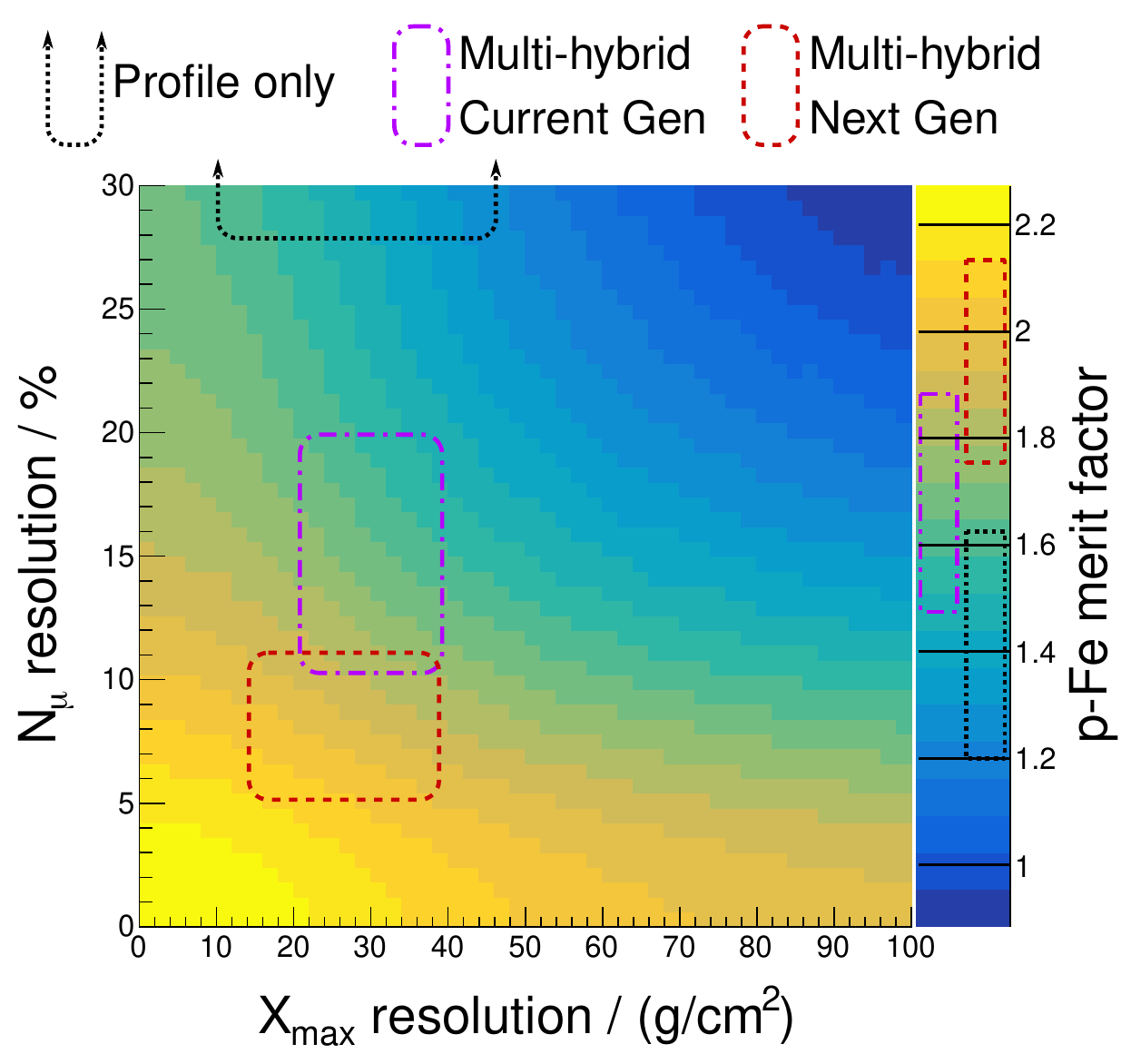}
  \end{center}
  \vspace{-6mm}
  \caption{\footnotesize Proton-iron separability, as described by the \textit{merit factor}, as a function of detector resolution on \xmax and $\lg(N_\mu)$ using the Sibyll2.3d hadronic interaction model. The merit factor is calculated as $mf = (\mu_p - \mu_{Fe}) / \sqrt{\sigma_p^2 + \sigma_{Fe}^2}$ with $\mu$ and $\sigma$ being the mean value and widths of an optimized linear combination of \xmax and $\lg(N_\mu)$. Predicted performances for generic detector configurations indicated. An SSD + WCD approach is labeled 'Multi-hybrid Current Gen'. Plot prepared by M.~Unger and annotated by E.~Mayotte for the Snowmass UHECR white paper~\cite{Coleman:2022abf}.}
  \label{fig:Merit}
  \vspace{-4mm}
\end{figure}

The differing response of a WCD to each shower component, when combined with the uniform response of an SSD, provides an opportunity to extract the muonic component of a given shower. As an illustration only, the following gives a simplified, intuitive example of how this could be achieved:
\begin{enumerate}
    \item Use the mean signal for a Minimum Ionizing Particle (MIP) transiting an SSD, $\langle S_{\rm mip} \rangle$, the total SSD signal, $S_{\rm ssd}^{\rm total}$, and the projected area of the SSD scintillator, $A_{\rm scint}$, to estimate the density of charged particles at the SSD:
    \vspace{-2mm}
    $$\rho_{\rm charged} = \frac{N_{\rm charged}}{A_{\rm scint}} \approx \frac{S_{\rm ssd}^{\rm total}}{S_{\rm mip}} \frac{1}{A_{\rm scint}}$$.
    \vspace{-4mm}

    \item Obtain the mean signal for an electron/positron depositing its energy in a WCD using the expected electron/photon spectrum of the observed EAS $\langle S_e \rangle$ (i.e., by leveraging universality~\cite{Lafebre:2009en}), then combine it with the projected WCD area, $A_{\rm wcd}$, and $\rho_{\rm charged} $ to extract the expected WCD signal if the shower were only electromagnetic:
    \vspace{-2mm}
    $$S_{\rm wcd}^{\rm em} = \langle S_e \rangle \times \rho_{\rm charged} \times A_{\rm wcd}$$.
    \vspace{-7mm}

    \item Subtract $S_{\rm wcd}^{\rm em}$ from the total WCD signal $S_{\rm wcd}^{\rm total}$ to get the signal residual due to some of the charged particles being muons:
    \vspace{-2mm}
    $$S_{\rm wcd}^{\mu\,{\rm res}} = S_{\rm wcd}^{\rm  total} - S_{\rm wcd}^{\rm em}$$

    \item Obtain the mean signal for a muon transiting a WCD for the observed shower, $\langle S_\mu \rangle$ (e.g., from Geant4 simulation) and subtract off $\langle S_e \rangle$ to get the average residual energy for one muon passing through the water tank:
    \vspace{-2mm}
    $$\langle S_\mu^{\rm res} \rangle = \langle S_\mu \rangle - \langle S_e \rangle$$.
     \vspace{-7mm}

    \item Estimate the number muons impinging on the WCD:
    \vspace{-2mm}
    $$N_\mu^{\rm wcd} \approx \frac{S_{\rm wcd}^{\mu\,{\rm res}}}{\langle S_\mu^{\rm res} \rangle}$$
    \vspace{-5mm}
\end{enumerate}
\noindent This can then be used, in conjunction with the $N_\mu^{\rm wcd}$ from many tanks to reconstruct the total muon content of the shower. As can be seen in Fig.~\ref{fig:Merit}, the muon content of a shower can then, in theory, be used as a highly sensitive estimator of the mass of a cosmic ray primary, motivating the use of SSDs for the AugerPrime upgrade of the Pierre Auger Observatory~\cite{PierreAuger:2016qzd, Castellina:2019irv}. It is also apparent that sensitivity can be further enhanced when combined with a simultaneous measurement of \xmax{}, but only if it is fully independent from the $N_\mu$ estimation~\cite{Coleman:2022abf}.

\paragraph{Status and outlook:}In practice, extracting mass sensitive parameters by combining SSD and WCD measurements has proven to be complicated process often heavily leveraging universality~\cite{schmidt2018sensitivity, Stadelmaier:2022tbt} or the use of DNNs~\cite{Mayotte:2021zev}. Unfortunately, it has also so far proven to be more difficult than originally hoped, with only marginal increases in mass sensitivity observed in simulations~\cite{Schmidt:2019duw}, as compared to WCDs alone~\cite{PierreAuger:2021fkf, PierreAuger:2021nsq}. It is likely that this situation will improve as hadronic interaction models gain in accuracy and expertise is brought to bear on the problem of mass reconstruction with an SSD + WCD hybrid detector through the AugerPrime upgrade of the Pierre Auger Observatory~\cite{Coleman:2022abf}. However, this will require time before the degree of sensitivity is known. It would therefore be advisable to investigate alternative detector designs with better muonic / electromagnetic component separation, for example as so-called 'double liner' WCDs, in order to hedge against SSD+WCD systems providing insufficient mass sensitivity.

\flashtalkDone{Benjamin Flaggs}{Mass sensitivity}
\label{sec:flaggs}
\begin{figure}[!th]
\centering
\includegraphics[clip, width=0.69\textwidth]{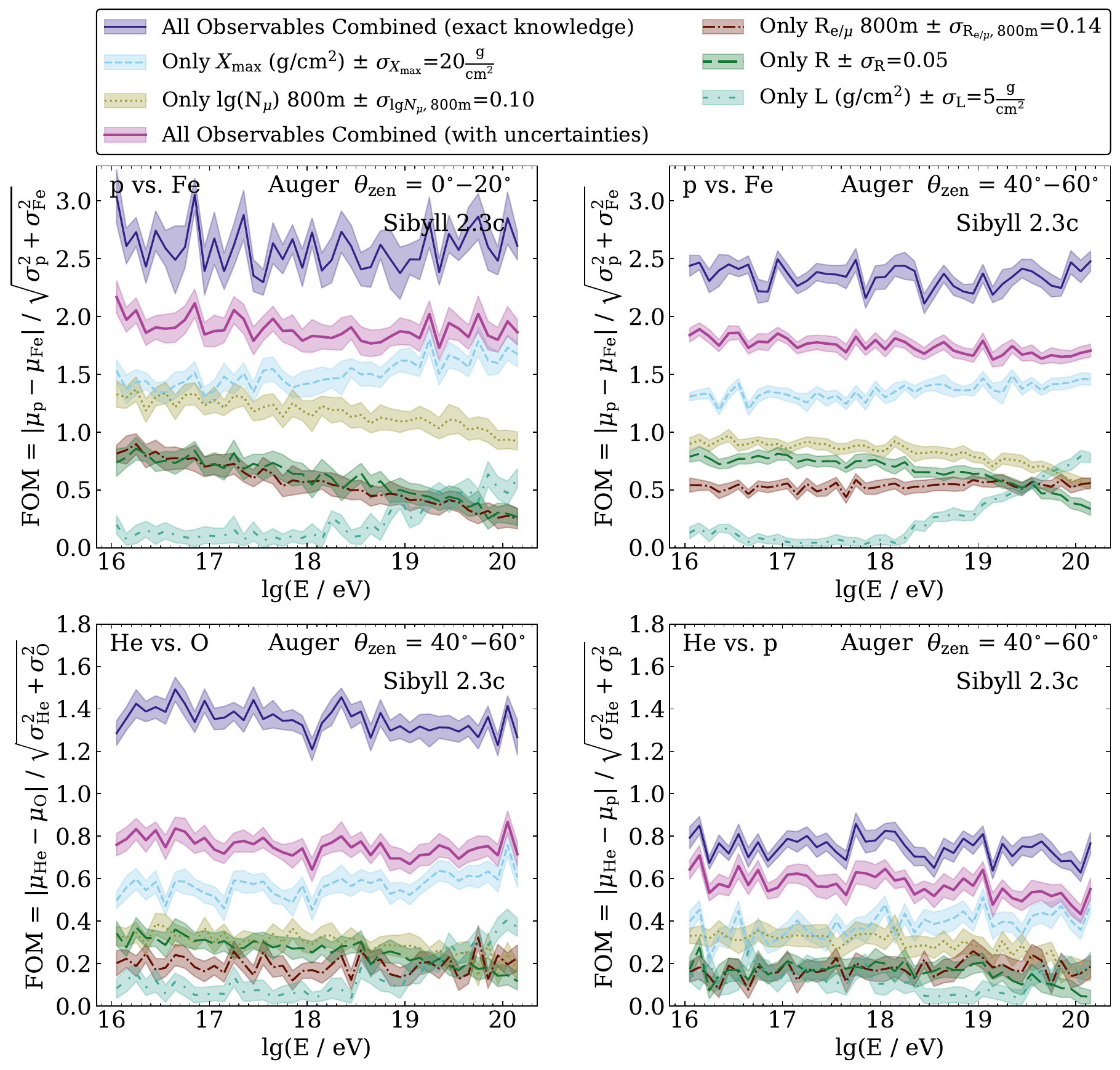}
\caption{Figure of merit (FOM) vs.\ air-shower energy for a variety of shower observables, zenith ranges, and primary particle combinations. See text for further description.}
\label{fig:mass_sensitivity}
\end{figure}

CORSIKA simulations at the location of the Pierre Auger Observatory were used to study the mass sensitivity of air-shower observables with a merit factor (FOM) calculated from a Fisher linear discriminant analysis.
Observables studied include parameters of the electromagnetic shower profile ($X_{\rm max}$, $R$, $L$) determined from a parameterized Gaisser-Hillas fit~\cite{Andringa:2011zz} to the CORSIKA longitudinal profile, along with the muon number and electron-muon ratio, both at observer level, within an 800$-$850\,m annulus from the shower axis.
Fig.~\ref{fig:mass_sensitivity} shows the FOM as a function of air-shower energy for all observables combined and individually, assuming the reconstruction uncertainties stated in the legend.
All observables were scaled by the electron number at $X_{\rm max}$ ($N_{{\rm e, max}}$) which serves as an energy reference.
A 10\% uncertainty is assumed for this energy reference.
The event-by-event mass separation was studied for different zenith ranges and primary particle combinations, as stated on the subplots in Fig.~\ref{fig:mass_sensitivity}.
The results of this analysis determine the shower observables of importance for mass composition of the ultra-high-energy cosmic rays that GCOS plans to measure, while the modularity of this analysis allows the extension to additional locations as long as a CORSIKA simulation library is available.
Therefore, these results can be used for determining the importance of both detector types and locations for GCOS. See~\cite{Flaggs:2023exc} for more details.

\flashtalkDone{Hazal Goksu}{Synergies with SWGO}
Southern Wide Field Gamma-ray Observatory (SWGO)~\cite{Hinton:2021rvp} is a proposed next-generation gamma-ray ground particle array to be built in the Southern Hemisphere. As a wide-field gamma-ray observatory, it will utilize a large area, high altitude, and a southern location for a galactic-focused science program, complementing arrays such as HAWC and LHAASO in the north. SWGO will be an array of water Cherenkov detectors with a central dense array of fill factor around 80\% and a sparser outer array. This gamma-ray observatory will be located in Pampa La Bola, at the Atacama Astronomical Park in Chile, at an altitude of 4,770 meters, near the vicinity of a number of astronomical observatories. SWGO has synergies with GCOS in some scientific goals, but mainly in aspects related to the detector design (i.e. Water Cherenkov Detector designs, light collection elements, electronics, DAQ, etc).

Cosmic ray, neutrino, and gamma-ray observatories that cover different energy ranges can be combined for multi-messenger astrophysics to study the physics of extreme environments.  While cosmic rays cannot be traced back to their sources due to their charge and the presence of magnetic fields, gamma rays, being chargeless, can be used to probe these sources of particle acceleration. SWGO will be a wide field detector array observing TeV to PeV gamma rays. It will access regions such as the Galactic Centre, Fermi Bubbles, the inner Galactic plane, and various extragalactic sources. Thanks to its wide coverage of the sky and nearly 100\% duty cycle, SWGO is ideal for studying very extended emissions including diffuse galactic emission and extended halos around PWNs and transient phenomena such as prompt phase emission from GRBs and mergers that emit gravitational wave~\cite{Hinton:2021rvp,Albert:2019afb}. SWGO will also enable studies for cosmic ray anisotropies around 0.03 to 3 PeV ~\cite{Albert:2019afb,Taylor:2021gex}. Furthermore, SWGO provides an opportunity for searching for dark matter candidates, with a sensitivity to DM in the range from ~100 GeV to a few PeV~\cite{Viana:2021smp}. These scientific prospects align with those of GCOS.

Although SWGO is optimized for extensive air showers initiated by gamma-rays, where cosmic ray showers are treated as background, both SWGO and GCOS are expected to have surface detector arrays. The unit surface detectors to be built for the two could have different optimization parameters, however, the need to mass-produce an array of tanks and bladders remains the same. Similarly, both detectors need to use light sensors, such as PMTs, DAQ systems, etc.

SWGO collaboration has been carrying out unit detector prototyping studies that can inform and be informed by similar studies from GCOS collaboration~\cite{Werner:2021hiq}. In SWGO, the main detector unit options being considered are steel tanks, rotomolded tanks, and bladders deployed in a natural or artificial lake (pond). Steel tanks, similar to the detector units of HAWC, can be outsourced from a company that is able to build tanks in any dimension needed and can be detached for easy transportation and built on-site. Rotomolded tanks are being considered for a single chamber option with an innovative "Mercedes" design. There are also companies in South America making these rotomolded tanks, such as the tanks made for the Auger observatory [5]. The option of bladders deployed in a lake does not require tanks, however, it does place extra requirements on liners, which are also needed for tanks. The lake option is no longer considered for the main array but is rather in consideration for an ultra-high extension to SWGO. The main array will be an array of tanks that house light-tight bladders and photo sensors.

The reference unit detector design of SWGO is double-layered WCDs, which were studied previously by members of the Auger and GCOS collaboration. Optically separating a WCD provides better particle identification, enabling muon identification and better background separation for SWGO. For the double-layered WCDs, two photo-multiplier tubes that are connected with a custom photo-multiplier tube support structure would be used. In addition, designs utilizing multiple smaller photo-multiplier tubes and smaller WCD units with a single chamber that use machine learning techniques (the Mercedes design) are investigated.

The bladders that would be filled with purified water and house photo-sensors need to be made up of liners that are absolutely light-tight and watertight. Moreover, the liner materials should not degrade the quality of water inside them. For the lake/pond approach, the outer layer should withstand UV radiation and should have sufficient flexibility. Studies of various liners are being carried out by the SWGO collaboration, including diffuse reflectivity tests, water degradation tests, and data taking with prototype detectors. The liner material studies performed within the SWGO collaboration are informed by the experiences of previous water Cherenkov detector arrays such as Auger, LHAASO, and HAWC, and GCOS would also benefit from these studies~\cite{Goksu:2021bfq}.

The prototyping experience of SWGO and GCOS should be informed by one another to ensure maximum efficiency. Communication between the two detector groups is important during the R\&D phase and also once the detectors are in full operation.

\flashtalkDone{Ioana Maris}{Layered water-Cherenkov detector}
\label{sec:maris2}
Water-Cherenkov detectors have been used successfully for measuring the particles from the air showers reaching the ground. The water is usually contained in a highly reflective material and the light produced by the particles passing through is measured with PMTs with a large collection area. The current water-Cherenkov detectors used at the Pierre Auger Observatory cannot distinguish between the signal produced by photons and electrons from the one produced by the muons. We proposed a simple solution in~\cite{Letessier-Selvon:2014sga}: an optical separation of the water volume, a layered water-Cherenkov detector. Given that the electromagnetic component of the air showers is mostly attenuated in about 40~cm, we separated the water volume in two, a top one with a height of 40 cm and the bottom one of 80 cm.\\

\begin{figure}[tb]
    \hspace*{-3ex}\includegraphics[width=0.5\textwidth]{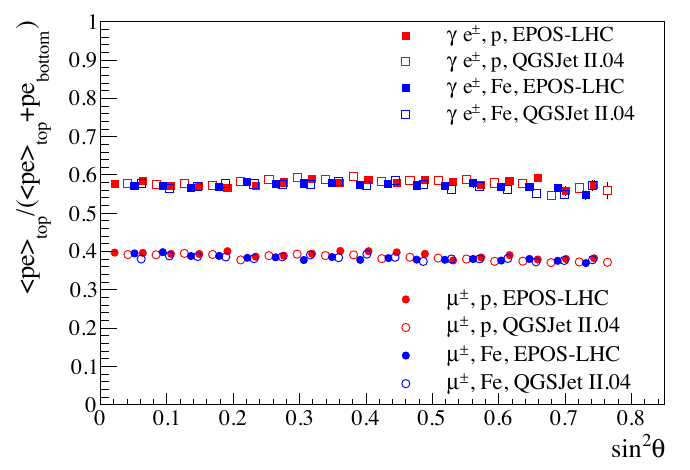}
    \includegraphics[width=0.5\textwidth]{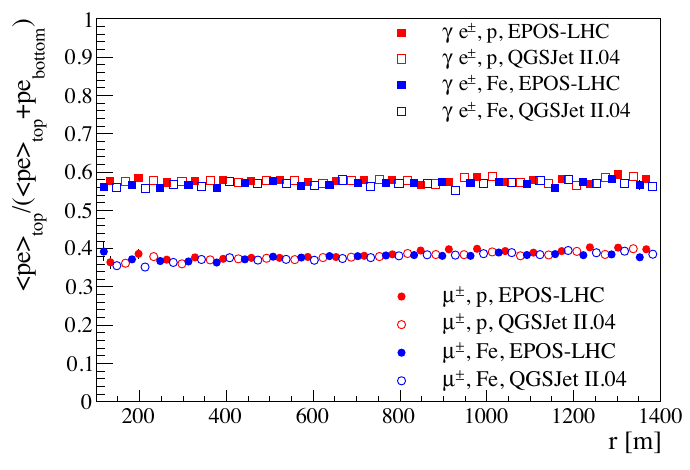}
    \caption{The fraction of photoelectrons produced in the upper layer divided by the total number of photoelectrons versus the angular direction (left) and distance to the axis of the air shower (right).}
    \label{fig:pe}
\end{figure}
\begin{figure}[tb]
    \hspace*{-3ex}\includegraphics[width=0.33\textwidth]{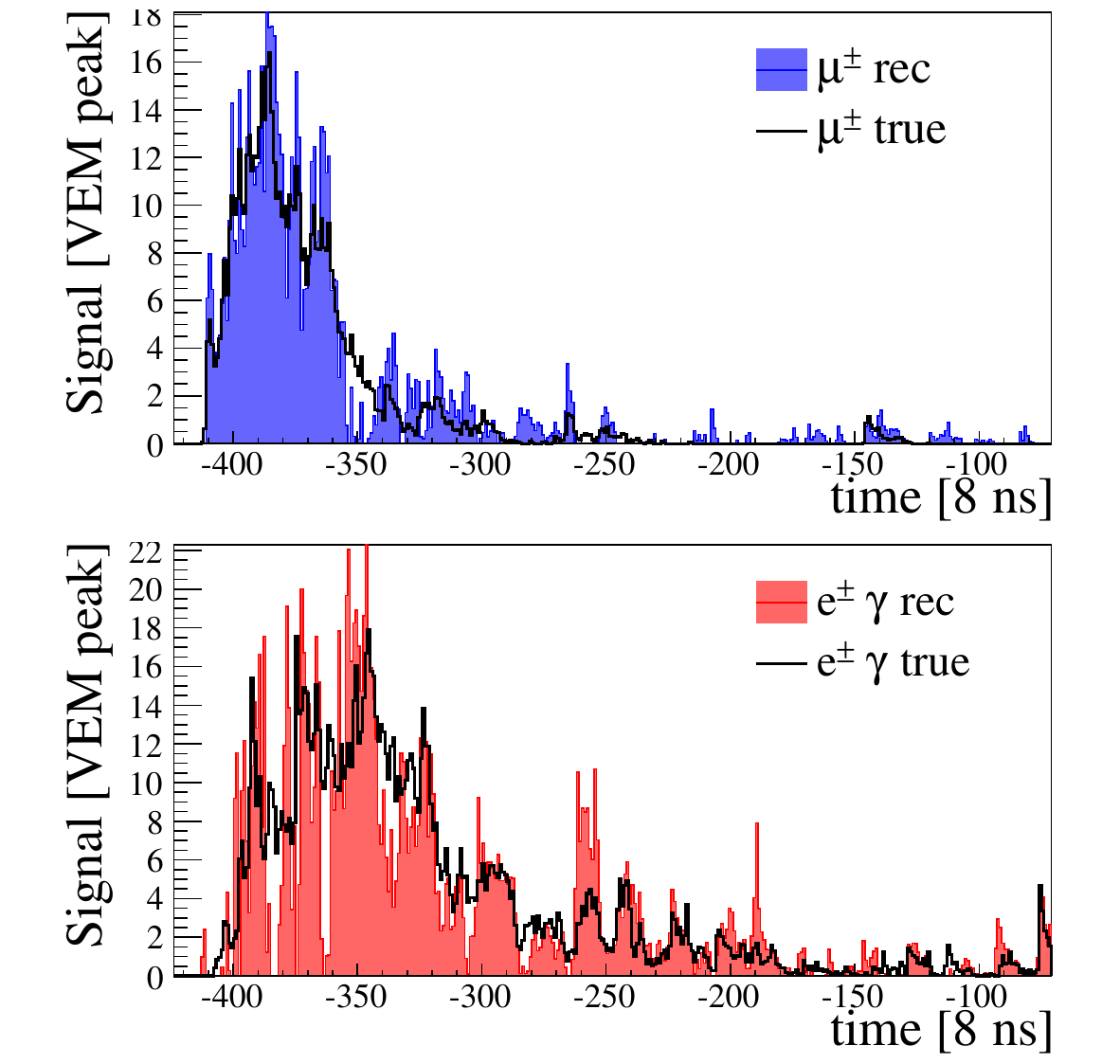}
    \includegraphics[width=0.33\textwidth]{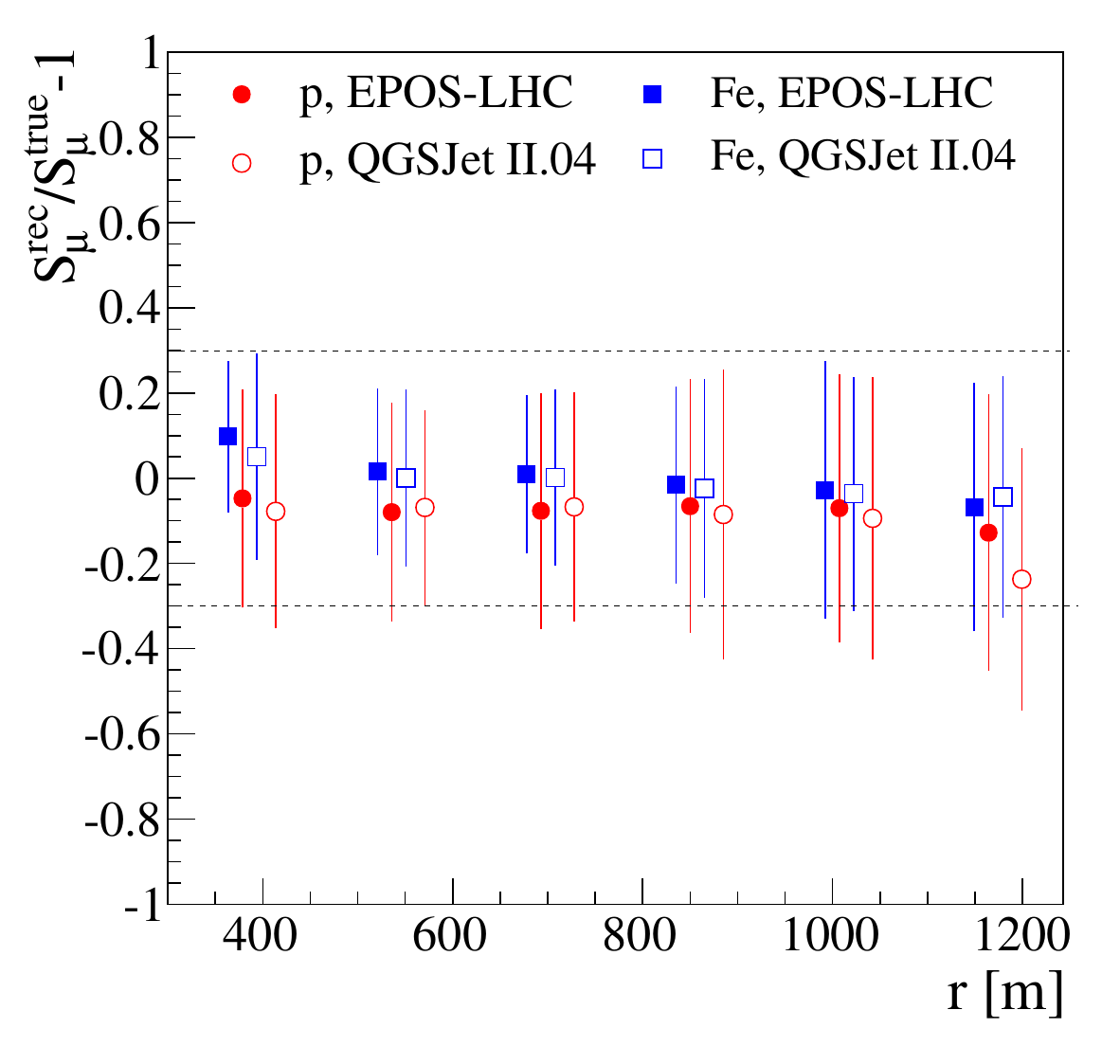}
    \includegraphics[width=0.33\textwidth]{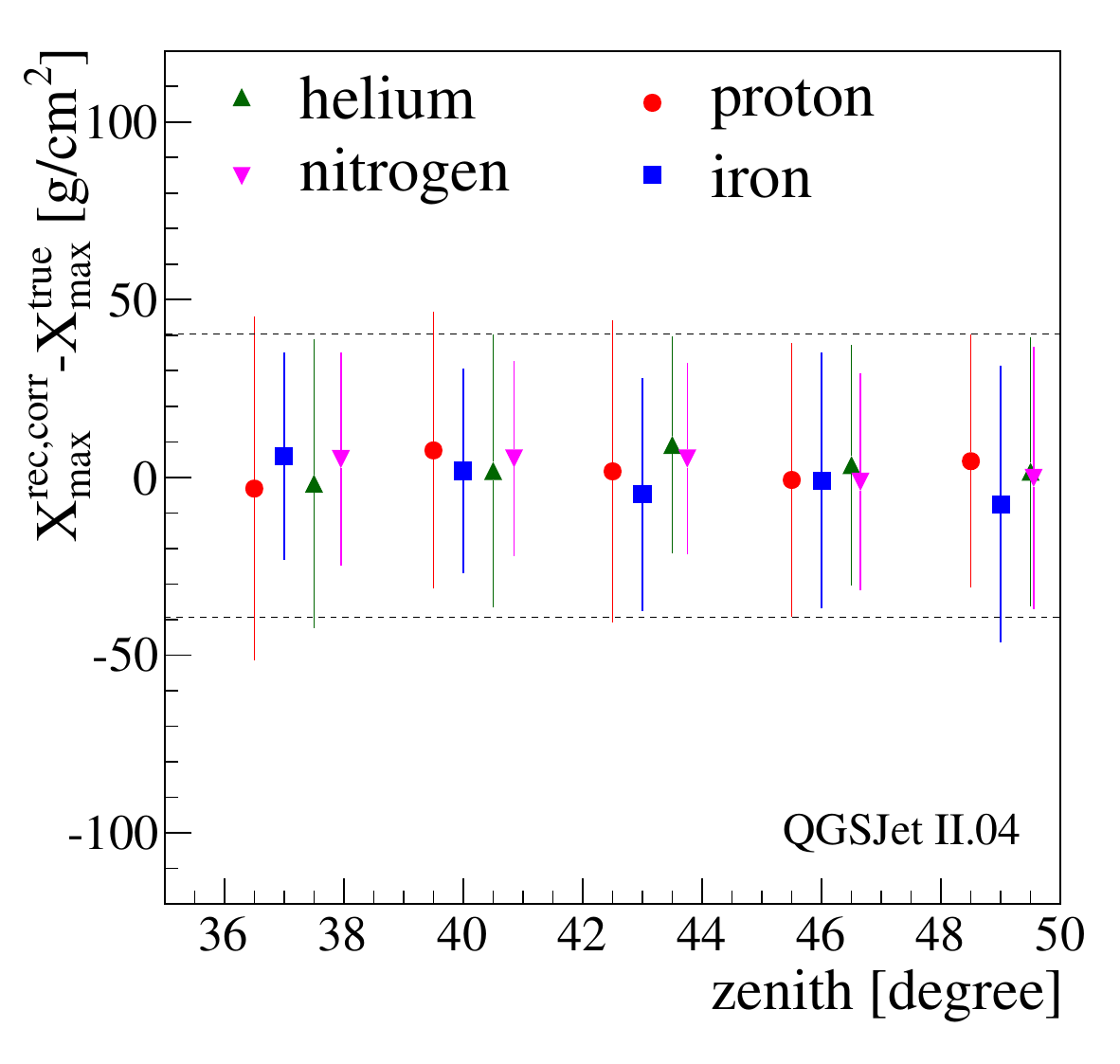}
    \caption{(left) The reconstruction of the muonic and electromagnetic components in a station. (center) Resolution of the muonic signal reconstruction. (right) Reconstruction of the $X_{\rm max}$ based on Universality. The uncertainties represent the RMS of the distributions}
    \label{fig:traces}
\end{figure}
The number of produced photoelectrons in the two layers produced by the electromagnetic and the muonic components is illustrated in Fig.~\ref{fig:pe} as a function of the sinus of the zenith angle and as a function of the distance to the axis of the air-shower. These fractions are very stable, with about 60\% of the light from the electromagnetic component measured in the top layer and about 40\% of the muonic signal.  These different responses in the two layers allow for the reconstruction of the different components in the detector from the total signal.

An example of the reconstruction is illustrated in Fig.~\ref{fig:traces}. The reconstruction of the two components is based on solving a set of two linear equations. The resolution that can be reached is about 15\% in each detector as shown in the same figure, with almost no bias. After this separation, a Universality-based algorithm can be applied to obtain the $X_{\rm max}$, or a simple lateral distribution function can be used to interpolate the muonic signal at a certain distance. The obtained resolutions can be better than 25\,g/cm$^2$ and better than 10\% on the muonic signal at 1000\,m leading to merit factors for the separation between proton and iron of better than 1.5 at 60 EeV.

\noindent
%{\footnotesize %\url{https://agenda.astro.ru.nl/event/21/contributions/276/attachments/72/81/IoanaMaris_Segmented_v2.pdf}}

\subsubsection{Further Contributions}
{Alan Watson -- Array configurations}\\
{\footnotesize \url{https://agenda.astro.ru.nl/event/21/contributions/281/attachments/69/78/Watson_GCOS_ropt.pdf}}

\noindent
{Hazal Goksu -- SWGO Unit Prototyping}\\
{\footnotesize \url{https://indico.iihe.ac.be/event/1729/contributions/3492/attachments/2119/2672/GCOS\%20Workshop\%20SWGO\%20public.pdf}}

\newpage
\subsection{Fluorescence detector}

\flashtalkDone{Bruce Dawson}{Design Considerations for Auger FD}
\label{sec:dawson}
\begin{wrapfigure}{r}{0.5\textwidth}
  \begin{center}
    \includegraphics[width=0.48\textwidth]{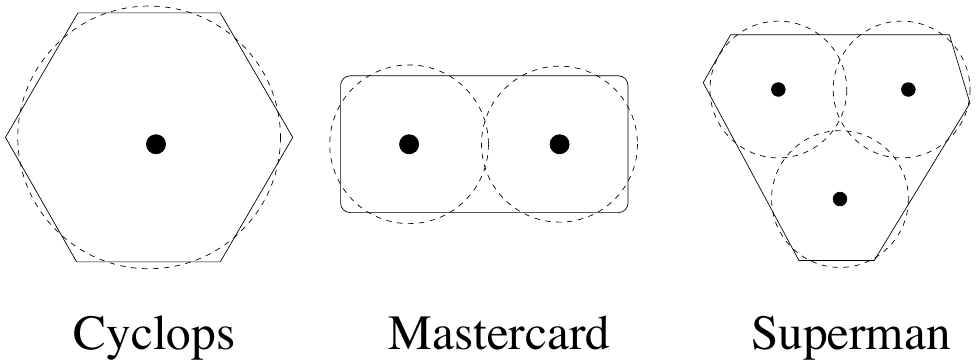}
  \end{center}
  \vspace{-5mm}
  \caption{Options from Design Report (1st ed.)}
\end{wrapfigure}
The first edition of the Pierre Auger Project design
report~\cite{PierreAuger:1996cxg} was published in October 1995, just
before the site selection in November of that year.  At that time, the
SD array area was set at 3000\,km$^2$, the need for a hybrid (SD/FD)
design was established~\cite{Sommers:1995dm,Dawson:1996ci}, and stereo
FD observations were not a high priority because of the good geometry
reconstruction available with FD hybrid.  Also, an FD elevation range
of $0-30^\circ$ was assumed, based on the HiRes experience.  The
design report contained an optimization exercise to decide on the
number of FD stations (1, 2 or 3) for the so-called reference design
(Figure~1).  Inputs included the minimum pixel signal-to-noise ratio
needed to achieve 10\% energy resolution and \Xmax resolution of
20\,\gcm, and the cost of mirrors, pixels, electronics and site
preparation.  The pixel signal-to-noise can be written
$$ S/N \propto \sqrt{\frac{A}{\Omega}}\frac{1}{R^2}\exp(-R/\xi) $$ for
mirror area $A$, pixel solid angle $\Omega$, distance to the shower
$R$ and atmospheric attenuation length $\xi$.  Assuming a good
knowledge of $\xi$, the optimization returned a single site known as
Cyclops 3000, with forty eight 4.4\,m diameter mirrors, $15^\circ
\times 15^\circ$ cameras with 1\deg diameter pixels, and a total of
10,800 channels.

Eighteen months later, the second edition of the Design
Report~\cite{PierreAuger:1997} revised the reference design to reduce
the risks of a single ``far-sighted'' eye, particularly related to
knowing sufficiently well the aerosol attenuation over a 30\,km path
length.  Two designs with smaller telescopes were proposed,
``Superman'' and ``Hexagon'', each composed of 135 telescopes with
1.5\,m diameter mirrors, $16^\circ \times 14^\circ$ cameras, 1.5\deg
pixels and a total of 16,335 pixels.  The difference was in the
arrangement of telescopes, with Superman having three full-azimuth
eyes within a superman logo-shaped array, versus a hexagonal array
with a central full eye and six 120\deg-azimuth eyes on the vertices
of the hexagon.  The two designs were assessed to have very similar
performance.  The overall cost of each was somewhat more than the Cyclops
design, but without the perceived risk associated with aerosols.  The
larger pixels (1.5\deg vs. 1\deg) were sufficient for the
determination of the shower-detector plane, while having the advantage
of fewer boundary crossings.

The design converged in the period 1998-2000, with the major
innovation being the adoption of Schmidt optics (including correction
lens)~\cite{Cordero:1996,Matthiae:1998} for the telescopes, reducing
coma aberration and allowing $30^\circ \times 30^\circ$ cameras and a
reduction of $4\times$ in the number of telescopes.  While not a
cost-saver (mirrors need to be $\sim 4\times$ larger), the improved
optics was a boon for data quality.  Faced with the actual
Malarg\"{u}e site, a challenge was finding elevated sites for the eyes
to minimize problems with the aerosol boundary layer and
fog~\cite{Dawson:2001nd}.  There was no central elevated site, leading
to the current arrangement of four 180\deg-azimuth sites on the array
boundary~\cite{PierreAuger:2009esk}.  Unfortunately this sacrificed
some stereo aperture, but the cost-savings satisfied the perceived
imperative that the FD costs should be $<50$\% of the Observatory cost.
This very brief summary has omitted many of the design ideas that were
discarded, including ``Plan B''~\cite{Sommers:1996a} (a similar idea
to FAST and CRAFFT), and Dual Mirrors~\cite{Sommers:1996b} (bigger
pixels, two mirrors per telescope, offset FoV).

\flashtalkDone{Shoichi Ogio}{Design Considerations for TA FD}
The fluorescence detector (FD) telescopes currently in operation at the Telescope Array (TA) experiment were either transferred from the HiRes experiment or were newly built, inheriting the design philosophy, specifications, and features of HiRes. This section describes the telescopes newly built for TA, focusing on their features. For a detailed description, please see the reference by H. Tokuno et al. in 2012\cite{HT2012}.

The TA experiment was originally planned as an array of ten stations with 40 telescopes per station.
The reflectors were stacked two floors high, designed to reduce land occupancy, i.e., to reduce the impact on the natural environment.
In the current TA experiment, twelve telescopes per station with this structure are installed at two stations.

Atmospheric fluorescence signals detected by the cameras are processed, triggered, and recorded by three types of VME modules.
Each of the three modules has a distinct function and was designed to operate in the following simple sequence with respect to a common clock.
The Signal Digitizer / Finder (SDF) digitizes and buffers the analog signal from the preamplifier output. At the same time, it performs hit detection (Level 1 trigger)  on a pixel-by-pixel basis.
For each telescope, the Level 1 trigger information is collected by the Track Finder (TF), which determines the air shower track image on the camera
and sends the Level 2 trigger to the Central Trigger Distributor (CTD)\cite{YT}.  The CTD collects Level 2 triggers from all telescopes in the station,
makes and sends the final trigger decision for all telescopes and the SD array.
In addition, the CTD has a GPS module and a reference clock that synchronizes all digital electronics in the station.

The PMT camera of the TA experiment FD has several features as
\begin{itemize}
\item  The PMTs are aligned so that the photocathode surface of the camera is flat, in accordance with a simple optical design that avoids the use of optical correction plates.

\item Since each PMT is designed to have a negative high voltage applied to it, the PMT anode and preamplifier are DC coupled, and the DC current of the PMT output is measured. In addition, Zener diodes are used in the latter stage of the voltage divider circuit instead of resistors to suppress fluctuations in the voltage applied to the dynodes in the latter stage.

\item We initially considered installing a so-called "Mercedes" reflector to eliminate dead zones between PMTs, but finally decided against using it for several reasons, such as applying a negative high voltage to the PMTs.

\item The three PMTs per camera are absolutely calibrated in the laboratory prior to their installation with an absolute light source utilizing a 337.1 nm nitrogen laser\cite{SK}. In addition, a small (4 mm diameter) pulsed light source consisting of YAP (${\rm YAlO_{3}}$ scintillator) and 50 Bq ${\rm ^{241}Am}$ is attached to the photocathode of each absolutely calibrated PMT to monitor its performance during observations\cite{HT2009}\cite{BK}.
\end{itemize}

Finally, the following comment is based on the author's personal impressions. The double-decker structure of the telescopes is not a good structure because of the complexity of the installation and subsequent works, the wider opening of the station building, the larger area of the opening shutters and windows, the higher strength required for them, and the operability of the telescope.

\flashtalkDone{Pierre Sokolsky }{The Snake array}
The Snake Array was a design proposal that arose out of the controversy between AGASA spectrum results showing no GZK cutoff and the HiRes results with a clear evidence of a spectral break. A larger-aperture air-fluorescence detector with the ability to determine both
\begin{wrapfigure}{r}{5.5cm}
    \includegraphics[width=\linewidth]{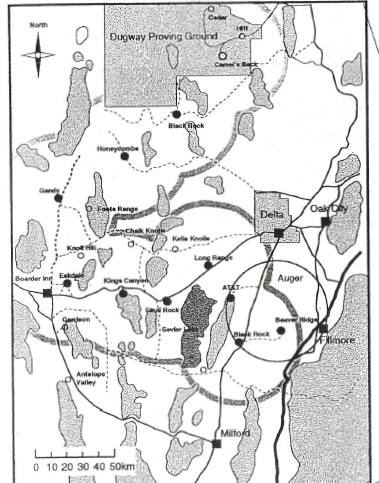}
\caption{Possible sites for Snake Array detector stations from surveys by Lawrence Wiencke and Shigeru Yoshida. The Long Ridge site coincides with one the current TA FD stations.}
\label{fig:snake1}
\vspace*{-0.2cm}
\end{wrapfigure}
spectrum and composition with similar resolution to HiRes but able to collect ten times the HiRes or
AGASA statistics was thought to be the next step.
The design was the brainchild of discussions between Gene Loh, Masahiro Teshima, myself, and others. Under consideration was a HiRes-like detector extended  in a long chain of stations. Such a linear array could provide good stereo reconstruction of events while minimizing redundant aperture. Was such a huge detector practicable? Significant work was done by Lawrence Wiencke and Shigeru Yoshida in finding possible sites on hilltops in NW Utah in the remote Snake Valley, bordering the state of Nevada (see Fig.~\ref{fig:snake1}). The detector stations, each with 14 mirrors, would be spaced by 30 to 40 km apart,  guaranteeing stereo aperture and energy and Xmax resolution  similar to HiRes. The significant advantage to this design was its re-use of tried and true HiRes technology: same mirrors, phototubes, and electronics. This made the cost estimate straightforward and reliable. The known HiRes site preparation and housing costs could also be factored in.
Monte Carlo simulation of the aperture of such a detector (using tested and verified HiRes programs) yielded a total ten-site aperture of $2.2\times 10^4$~km$^2$~sr (see Fig.~\ref{fig:snake2}). Assuming a 15\% on time, the actual detector aperture becomes $3.3\times 10^3$~km$^2$~sr, similar to that of the Auger SD.  The detector aperture becomes fully efficient and quite flat above 10 EeV.
Cost estimates were based on extrapolating known HiRes expenditures. A pair of 14 mirror stations would require 840 k\$ for mirrors and frames, 1680k\$ for electronics, 280k\$ for housing, for a total of 3.3M\$. Note that the electronics cost estimate quoted here is based on a proposed redo of the TA system using 2014 technology. Assuming a contingency of 1M\$, the pair of sites could be build for between 4 and 4.5 M\$. A ten station Snake Array would thus cost 45-50 M\$ and would yield ~ 1000 well-reconstructed events above 30 EeV in a ten year run ( based on HiRes quality-cut experience).
The Snake Array idea was not abandoned because of any perceived flaws but because it did not address the issue of understanding the systematic difference between the HiRes and AGASA results. This required co-siting the two detector approaches and led to the current Telescope Array design.
\begin{figure}[!h]
  \centering
  \includegraphics[width=0.8\linewidth]{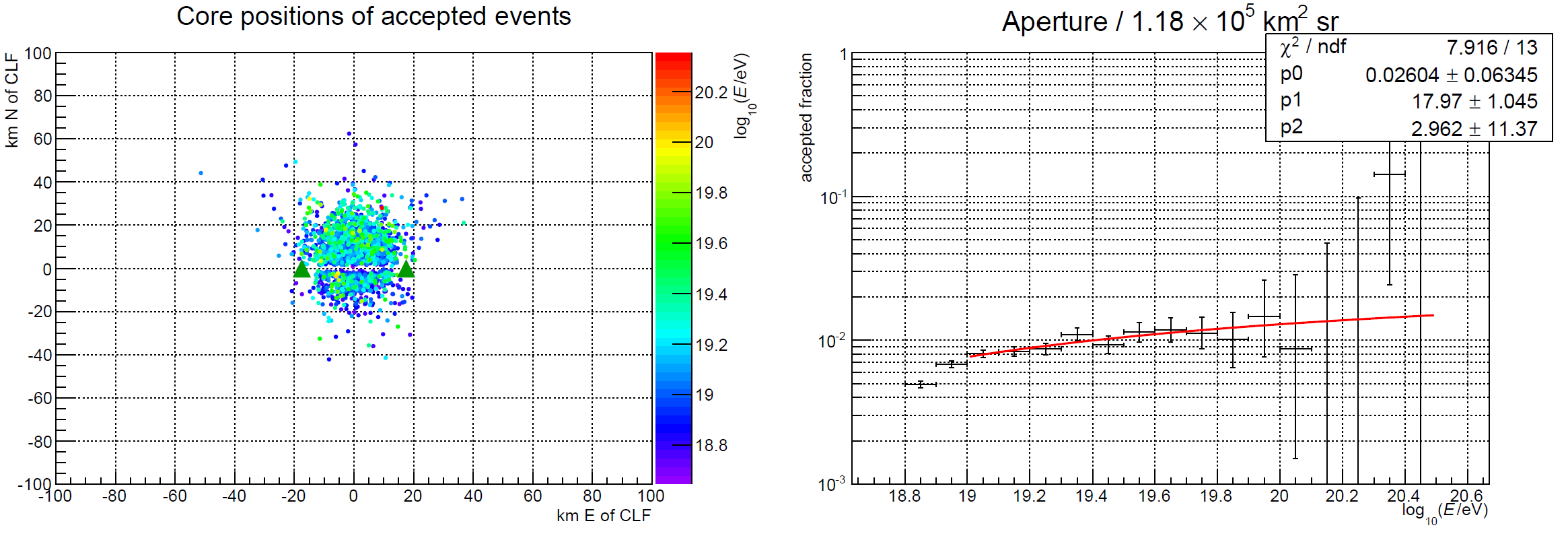}
  \caption{Core position distribution and aperture of a pair of Snake Array stations spaced 35 km apart. The two 14 mirror arrays are pointed 40 degrees up from the line joining the two detectors. This maximizes the high energy stereo reconstruction.\label{fig:snake2}}
\end{figure}

\flashtalkDone{Toshihiro Fujii}{Low-Cost Fluorescence Detector Array for GCOS}
\label{sec:fujii}
A low-cost fluorescence detector array is a promising solution to achieve an unprecedented exposure at the highest energies with a detection of $X_{\max}$, which is the most reliable parameter for the mass composition measurement.
\begin{figure}[h]
\centering
\includegraphics[width=1.0\linewidth]{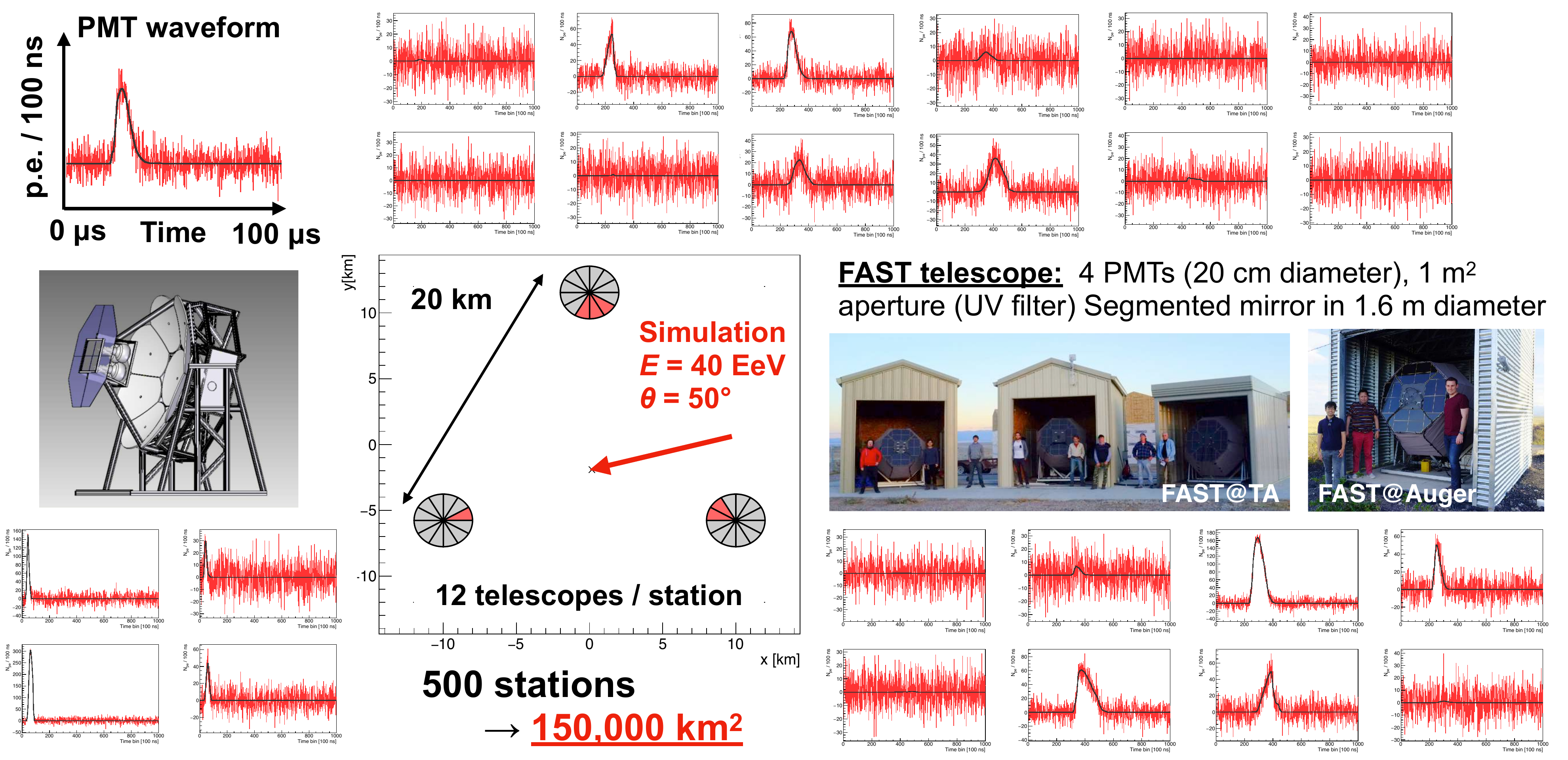}
\caption{The Fluorescence detector Array of Single-pixel Telescopes: a possible solution for a future giant ground array~\cite{bib:fast}. The traces show simulated signals emitted from a UHECR with an energy of 40\,EeV and a zenith of 50$^{\circ}$. The number of
stations and corresponding area quoted at the bottom are for a FAST-only array without a surface detector.}
\label{fig:fast}
\end{figure}
The Fluorescence detector Array of Single-pixel Telescopes (FAST)\footnote{https://www.fast-project.org} features compact FD telescopes with a smaller light-collecting area and far fewer pixels than current-generation FD designs, leading to a significant reduction in cost that allows for the production of more FD units~\cite{FAST:2021bkj,Malacari:2019uqw,bib:fast}.
In the FAST design, a 30$^{\circ}$ $\times$ 30$^{\circ}$ field-of-view is covered by four 20\,cm photomultiplier-tubes (PMTs) at the focal plane of a compact segmented mirror of 1.6\,m diameter~\cite{bib:fast_optics}.
Its smaller light-collecting optics, smaller telescope housing, and fewer number of PMTs significantly reduces its cost.
Each FAST station would consist of 12 such telescopes, covering 360$^{\circ}$ in azimuth and 30$^{\circ}$ in elevation.
These stations would be deployed in a triangular grid with a 20\,km spacing.
A relatively short distance of the array spacing will reduce an uncertainty of the atmospheric transparency,
or possibly evaluate it from a event-by-event light-balance analysis.

Figure~\ref{fig:fast} shows the simulated waveforms from a UHECR shower detected in a three-fold coincidence by such an array.
To achieve an order of magnitude larger exposure than current observatories, 500 stations covering 150,000\,km$^2$ are required for a FAST-only array without a surface detector, after accounting for the standard FD duty-cycle and additional moon-night operation.
A fully autonomous and stand-alone operation will be capable of enlarging the duty cycle to be 20\%.
\begin{figure}
  \centering
  \includegraphics[width=.45\linewidth]{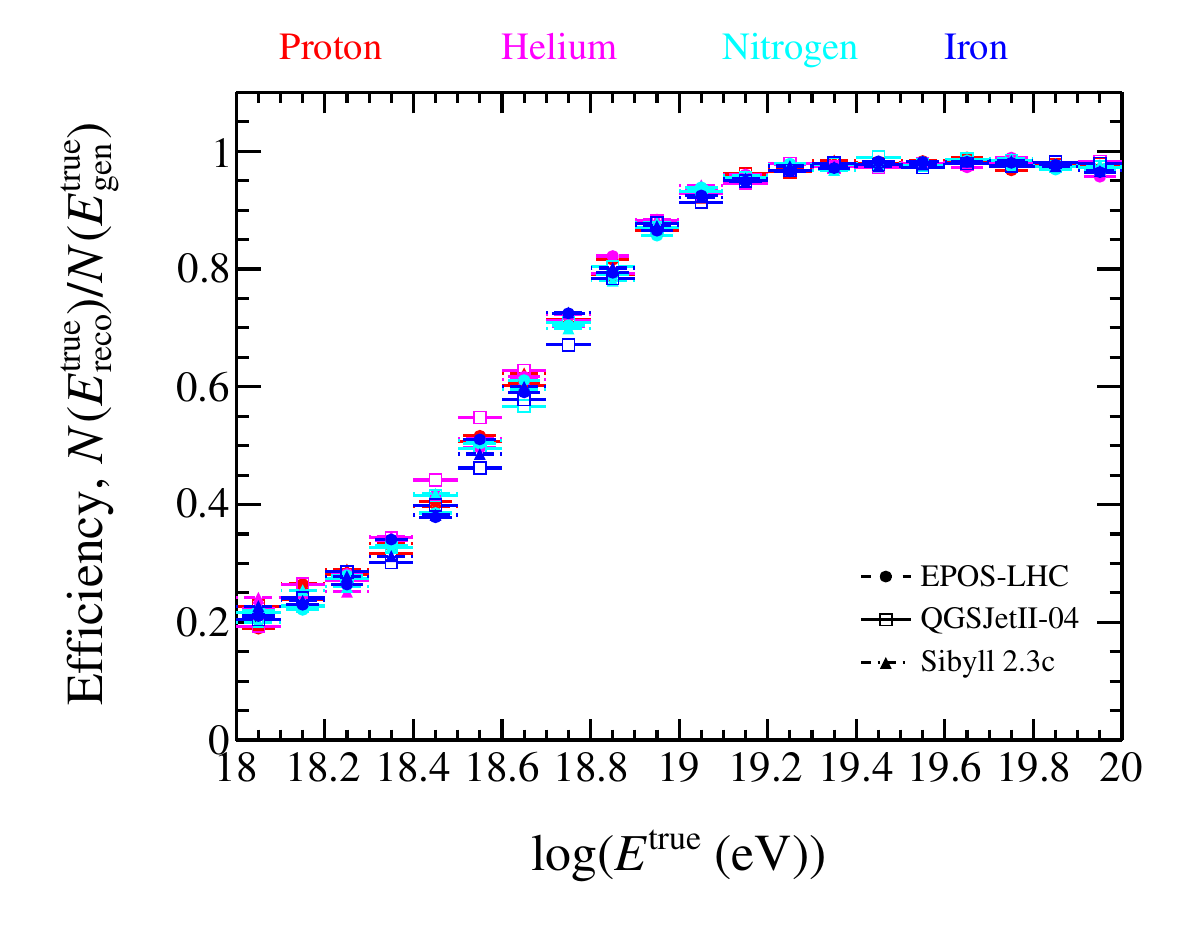}\hfill\includegraphics[clip,rviewport=0 -0.07 1 1.1,width=.43\linewidth]{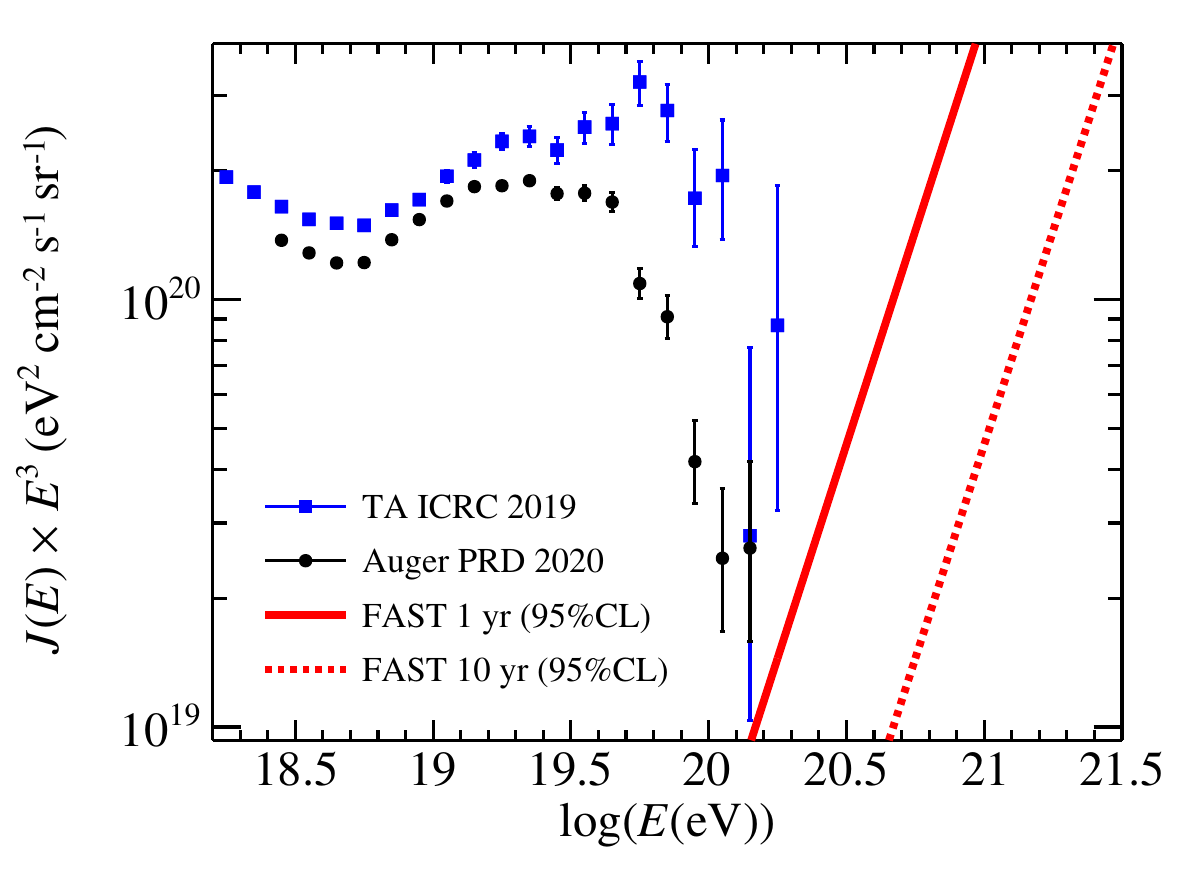}
  \caption{(a) Trigger efficiency for 3-fold coincidence with a hypothetical FAST. (b) Expected 95\% confidence-level detectable sensitivities of the energy spectrum with the full-sized FAST of 500 stations compared to the spectra reported from TA~\cite{Ivanov:2020rqn} and Auger~\cite{Aab:2020rhr}.}
  \label{fig:spec}
\end{figure}

Figure~\ref{fig:spec} shows preliminary performances of the FAST evaluated by a detector simulation~\cite{Malacari:2019uqw}.
The trigger efficiency for 3-fold coincidence is shown in Figure~\ref{fig:spec}(a), indicating a 100\% efficiency above 20\,EeV.
Figure~\ref{fig:spec}(b) is an expected sensitivity on energy spectrum with a full-sized FAST, assuming an effective exposure of 90,000\,km$^2$\,sr per year to estimate our detectable flux at 95\% confidence level. A full-sized FAST will be capable of extending UHECR measurements with a $X_{\max}$ detection beyond 300\,EeV.
The FAST will provide an unprecedented exposure exceeding current experiments by an order of magnitude, including mass composition analysis using $X_{\max}$.
Alternatively the FAST is a cost-effective method to provide a calorimetric energy determination and a mass composition sensitivity for a surface detector array.

\flashtalkDone{Jose Bellido}{FAST Reconstruction and its Performance}
A FAST telescope has been designed to detect high energy cosmic rays at a much lower cost. For that, the mirror is relative small, and it uses only four large PMTs. Therefore, a standard event reconstruction can not be performed with the FAST telescopes. A standard reconstruction identifies a shower detector plane and then it uses the time information of many pixels to fit the shower evolution over the pixels to identify the shower axis geometry.

For the FAST event reconstruction, it will be required that a shower is seen by at least three FAST telescopes located on a triangular grid. The shape and time information of the pulses (at the three FAST telescopes) will be used to reconstruct simultaneously the geometry, \Xmax and energy of the shower. The FAST reconstruction is performed in two steps~\cite{Thomas-Albury:2020fse}. The first step is a rapid reconstruction using a learning machine algorithm. A second step performs a Top Down (or inverse MC) reconstruction. The Top Down reconstruction uses as a first guess the result from the learning machine algorithm. The reason for that is because the Top Down reconstruction requires heavy computational work, but a first guess close to the correct values, reduces significantly  the searching over the entire phase space.

This learning machine algorithm uses the centroid time (signal-weighted time average), the total signal (pulse integral) and the pulse height. The centroid time provides information about the relative time of arrival of signals at each PMT. The total signal provides information about total signal measured from the shower as well as the relative signal between PMTs. The pulse height provides additional information about the shape of the signal pulse including asymmetry.

The learning machine algorithm was developed using 500,000 events simulated within a small core region of 1 km of radius, where 80\% of the events were used for training and 20\% for validation~\cite{Thomas-Albury:2020fse}. Table ~\ref{table:table1} shows preliminary results for the performance of the two steps. The performance for the learning machine algorithm (first step) is already acceptable, but a second step using the Top Down reconstruction, improves significantly the performance. The resolution for the Top Down reconstruction corresponds to the reconstruction uncertainty estimated for a particular event.

\begin{table}[hb]
\begin{center}
\footnotesize
\begin{tabular}{ p{2.5cm} p{3.cm} p{3.5cm} p{3.cm} }
\hline
parameter & resolution \newline learning machine  \newline (1st step) & Reconstruction bias \newline learning machine \newline  (1st step) & resolution \newline top down \newline (2nd step) \\
\hline
Shower axis & ~2.76\deg & $<$ 0.75\deg & 0.3 \deg\\
Core &  ~256 m   & $<$ 50 m & ~45 m\\
\Xmax & ~60 \gcm & $<$ 30 \gcm &   ~10 \gcm\\
Energy    & 25\% & $<$ 15\% &  2.5\%\\
\hline
\end{tabular}
\end{center}
\caption{Performance for the learning machine and Top Down reconstruction algorithms.}
\label{table:table1}
\end{table}

\begin{figure}[h]
\centering
\includegraphics[width=\linewidth]{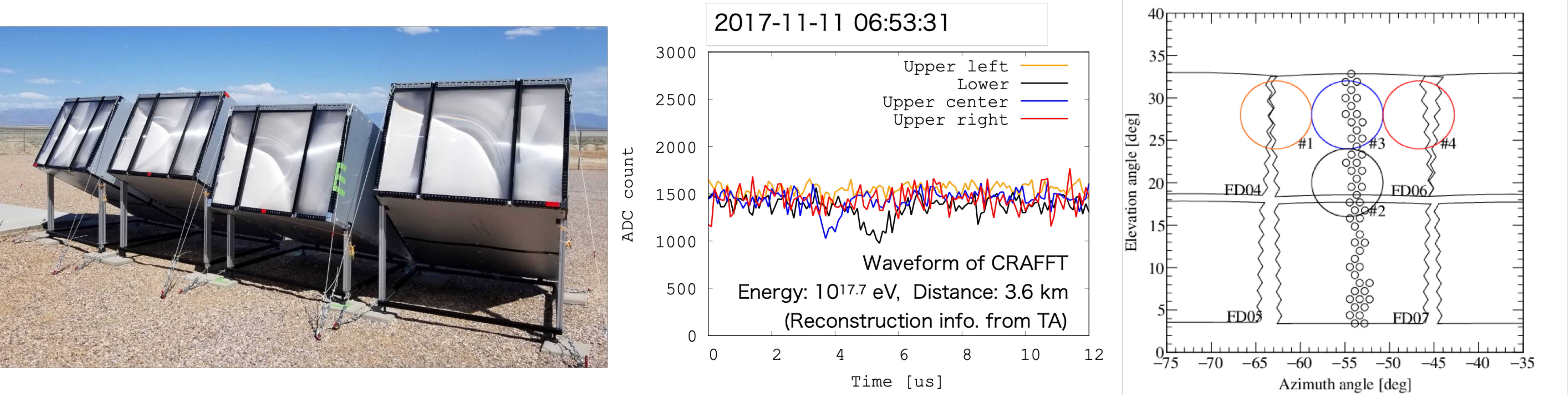}
\caption{Left: CRAFFT detectors deployed at TA FD BRM site. Right: Waveform of air shower event observed CRAFFT and F.O.V. of CRAFFT compared with TA FD event display.}
\label{fig:crafft}
\end{figure}

\flashtalkDone{Yuichiro Tameda }{CRAFFT Concept and Design}
\label{sec:tameda}
The CRAFFT (Cosmic Ray Fluorescence Fresnel Lens Telescope) project is developing an fluorescence detector (FD) for future ultra-high energy cosmic ray observations~\cite{YT2019}.
%Large-scale experiments will be realized in the future, such as the GCOS concept.
FD is one of the most promising detectors because it can make calorimetric energy measurements and $X_\mathrm{max}$ measurements.
However, conventional FDs are relatively expensive compared to surface detectors.
Therefore, CRAFFT is developing a low-cost FD.
The concept and features of CRAFFT are described below.
Manufacturing cost can be reduced by it's simple structure.
This structure covered by the Fresnel lens and galvalume steel plate doesn't need a building and all the equipment can be installed inside as shown in Fig. \ref{fig:crafft}.
Since there are no obstacles between the lens and the focal point, the light collection efficiency is high.
By realizing an automatic observation system equipped with an environmental monitor, we can reduce not only manufacturing costs but also operating costs.
For this simple FDs, observation from multiple locations is required,
because the field of view per pixel is large and geometrical reconstruction in monocular is relatively difficult.
Ensuring S/N is also an issue.
Thus, although there are some points to be overcome,
development of low-cost FDs is a good strategy for future large-scale

CRAFFT has successfully detected cosmic ray air showers as shown in Fig. \ref{fig:crafft}.
The event rate of 8 events / 10 nights in the test observation with four prototypes is consistent with the assumption of more than $10^{17}\,\mathrm{eV}$.
The observed waveforms are well reproduced by simulations.
The detector capability of the CRAFFT was as expected.
Reconstruction methods based on waveform fitting have also been developed
from simulation studies.
It was found that $\chi^2$ of the waveform fitting converges around the true geometry even for monocular mode.
Here, the waveform height is normalized and $X_\mathrm{max}$ is fixed.
It is shown that reconstruction by waveform fitting is possible in principle.
%%The six fitting parameters are zenith, azimuth, core x, core y, energy, and $X_\mathrm{max}$.
We are studying the efficient algorithms for fitting.
We also expect that multipoint observations can efficiently constrain the shower geometry.

The prototype had one pixel per telescope,
though a new detector configuration is being considered to increase the field of view and improve reconstruction accuracy while keeping costs down.
Using sixteen PMTs with 5 inch diameter, the field of view per unit would be about 24 degrees in azimuth and elevation.
The accuracy of geometrical reconstruction is estimated to be about 3 degrees in the direction of arrival, and the core position could be determined at less than 200 meters.
Using the newly configured telescope and a fully automated system (a one-year long test has already been conducted),
test observations are planned to be made at the TA experimental site.
The six new telescopes will provide the field of view of the current TA FD one station.
Observations at the TA experimental BRM site will allow cross-checking with TA FD.
In addition, the telescope will be installed at the TA CLF site to perform stereo observations.
In the future, we would like to install a 360-degree FD station with 20 km spacing to achieve an aperture 10 times larger than that of the TAx4 experiment.

\flashtalkDone{Michael Unger}{Low-Elevation Fluorescence Telescopes for GCOS}
\label{sec:unger}
One of the major advantages of the detection of extensive air showers
with the fluorescence technique is the capability of high-quality
observations of the longitudinal shower development over large
distances. Therefore, it is possible to cover large areas with a single
optical telescope. For instance, the typical viewing distance for the
Schmidt-optics of the fluorescence telescopes of the Pierre Auger
Observatory is 30~km at $10^{19}$~eV and 45~km at
$10^{20}$~eV~\cite{PierreAuger:2010swb}. A cost-efficient and
maintenance-friendly design for a fluorescence detector (FD) in a
hybrid array can therefore be achieved by a design in which air
showers are detected across the array by telescopes installed at only
one or a few sites.

\begin{figure}[h]
\centering
\includegraphics[width=0.54\linewidth]{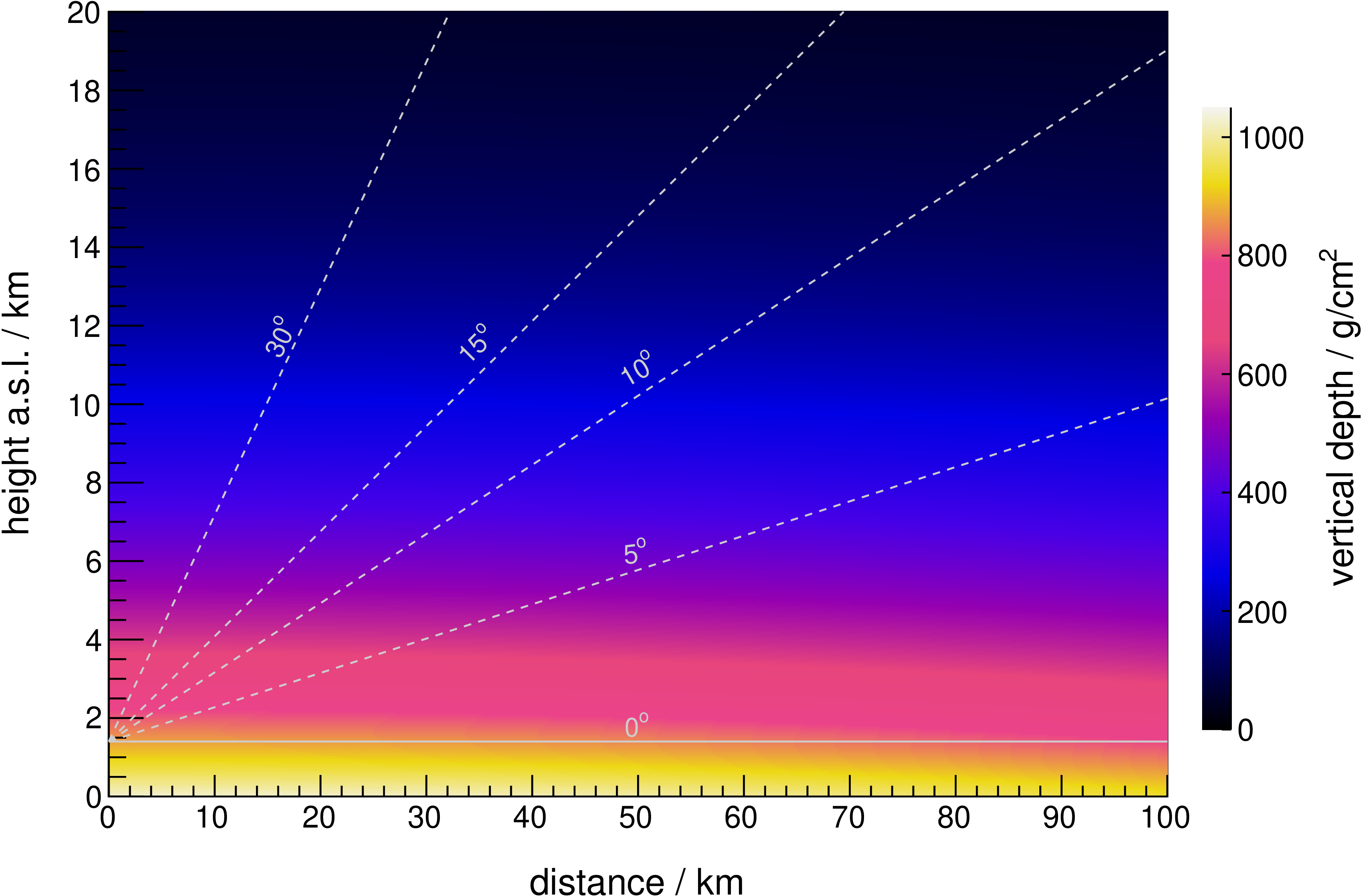}\hfill\includegraphics[width=0.45\linewidth]{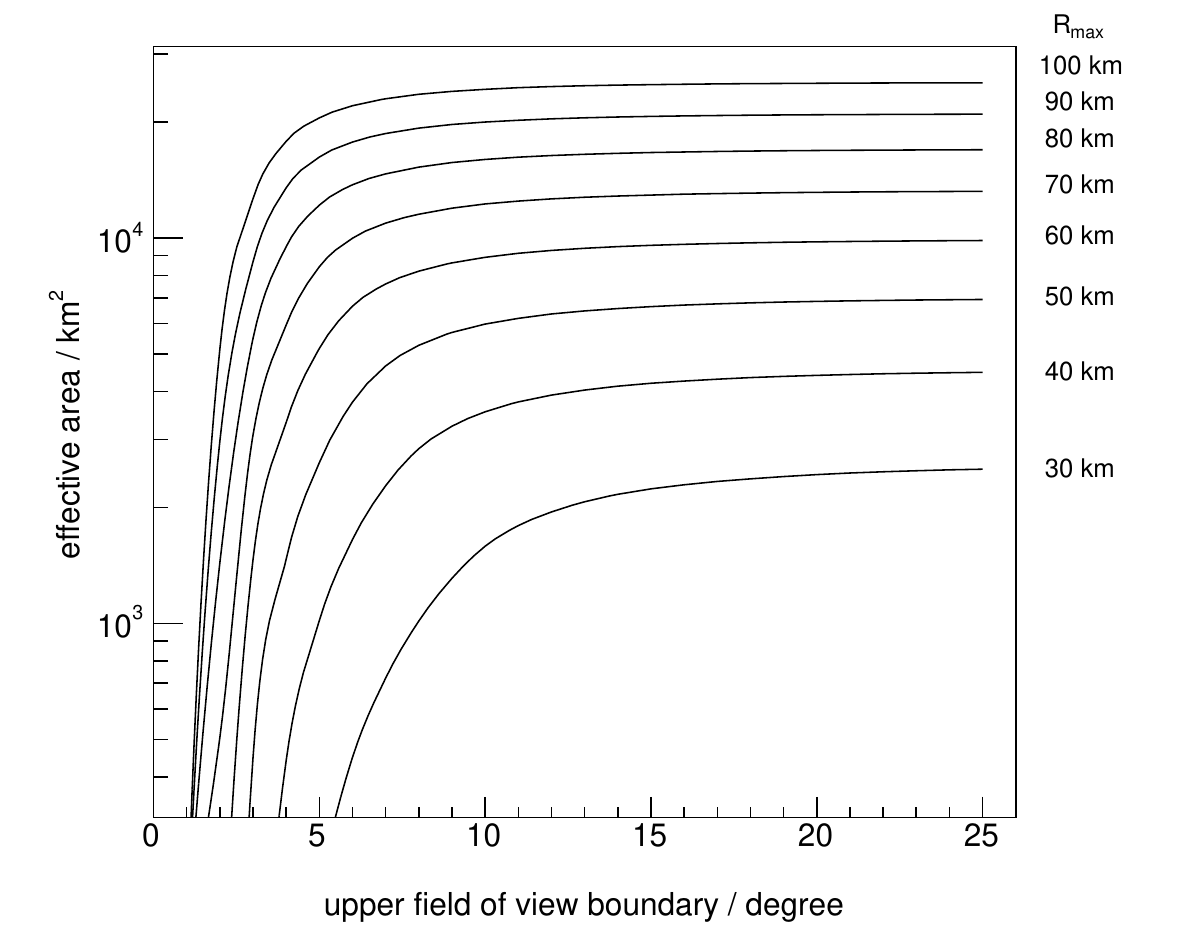}
\caption{Left: Illustration of the viewable vertical depth range. Right: Area observed by one telescope depending on maximum observable radius and elevation of the upper field of view.}
\label{fig:fov}
\end{figure}

An important design parameter for the telescopes is their field of
view range.  A large range is needed to detect close-by low-energy air
showers (e.g.\ in HEAT or TALE), but far-away high-energy showers can
be viewed within a very narrow elevation range close to the
horizon. This is illustrated in the left panel of Fig.~\ref{fig:fov},
where the viewable vertical depth range is illustrated.  The typical
slant-depth range needed to observe UHE showers is $X_\text{low}\leq
700$~\gcm $X_\text{up}\geq 900$~\gcm\cite{PierreAuger:2014sui}. The available
area for showers with zenith angles up to $60^\circ$ that fulfill this requirement is shown in the right panel of Fig.~\ref{fig:fov}. As can be seen, at UHE, where the maximum viewable distance $R_\text{max}$ is large, an upper elevation
boundary of $\leq 10^\circ$ is sufficient. This is much smaller than than the
maximum elevation of $\leq 30^\circ$ of current telescopes and therefore, a telescope design that is optimized for ultrahigh energies can be more cost-effective.

For this purpose, we studied telescope optics inspired by the MACHETE design~\cite{Cortina:2015xra} with a large azimuthal coverage and a limited field of
view in elevation. As detailed in Ref.~\cite{bscScherneKIT}, a possible
design of such a Low-Elevation Fluorescence Telescope (LEFT) could consist of six telescopes with a $(60^\circ\times 10^\circ)$ field of view in azimuth and elevation and an effective aperture of $A=15~\text{m}^2$. The camera consists of 1500 SiPM pixels with 5~cm in diameter ($0.6^\circ$).

This FD design is compared to the the ones of FAST and Auger in the table below. Due to the large aperture-to-pixel ratio ratio of $\sqrt{A/\Omega_\text{pix}}=13$ and
the correspondingly low signal-to-noise ratio, see Sec.~\ref{sec:dawson}, a maximum viewable distance of $R_\text{max}=60$~km could be achievable. The cost for one 360$^\circ$ LEFT site is calculated using the estimates for Trinity given in Ref.~\cite{otte2019}.
\begin{table}[h]
  \footnotesize
\centering
  \begin{tabular}{lcccccc}
    & $\sqrt{A/\Omega_\text{pix}}$ & $R_\text{max}$/km  & stations & cost/station & total cost \\
    & (m/deg$)^{1/2}$ & \footnotesize at 10$^{20}$ eV &  per  40k km$^2$ & (M\$) & (M\$) \\\hline
    LEFT&  13 & 60  & 3.5 &  2  & 7\\
    Auger  &  1.2 & 45  & 6.3 & 2.4 & 15 \\
    FAST   &  0.07 & 25  & 20.4 & 0.4 & 8
  \end{tabular}
\caption{Comparison of telescope designs.}
\end{table}

\subsubsection{Further Contributions}
{Masaki Fukushima -- Fluorescence Yield by sFLASH}\\
\mbox{\footnotesize \url{https://agenda.astro.ru.nl/event/21/contributions/250/attachments/77/87/sFLASH_220714_fukushima.pdf}}
\\
{Francesco Salamida -- Ideas for FD Optimization}\\
{\footnotesize \url{https://agenda.astro.ru.nl/event/21/contributions/247/attachments/61/70/salamida.pdf}}

\newpage
\subsection{Radio detector}

\flashtalkDone{J\"org H\"orandel}{General considerations}

The main benefits of Radio Detection for GCOS include:
\begin{compactitem}
\item radio detection provides a clean measurement of the electromagnetic air shower component;
\item radio detection is a good tool for horizontal air showers;
\item from the radio measurements an independent energy scale can be derived, based on the calorimetric measurement of the electromagnetic shower component;
\item the measurement for the electromagnetic component can be combined with muon measurements from a particle detector/water-Cherenkov detector to determine the mass of the incoming particle;
\item photons initiate air showers which contain a large electromagnetic shower component and almost no muons -- radio detectors can serve as triggers for photon-induced showers.
\end{compactitem}
In the following, we use the Radio Detector (RD) of the Pierre Auger Observatory as an example to illustrate the benefits of radio measurements.

{\bf Highly-inclined air showers} \cite{Aab:2018ytv} with zenith angles
$\Theta>60$\deg traverse a big amount of atmosphere until they are detected.
The thickness of the atmosphere in horizontal direction amounts to about 35
times the column density of the vertical atmosphere. Thus, the e/m shower
component is mostly absorbed and only muons are detected with the WCDs of the
SD. The atmosphere is transparent for radio emission in our band ($30-80$~MHz)
and radio measurements are an ideal tool for a calorimetric measurement of the
e/m component in HAS.

%\begin{figure}[ht]
\begin{wrapfigure}{r}{0.8\textwidth}
 \includegraphics[width=0.8\textwidth]{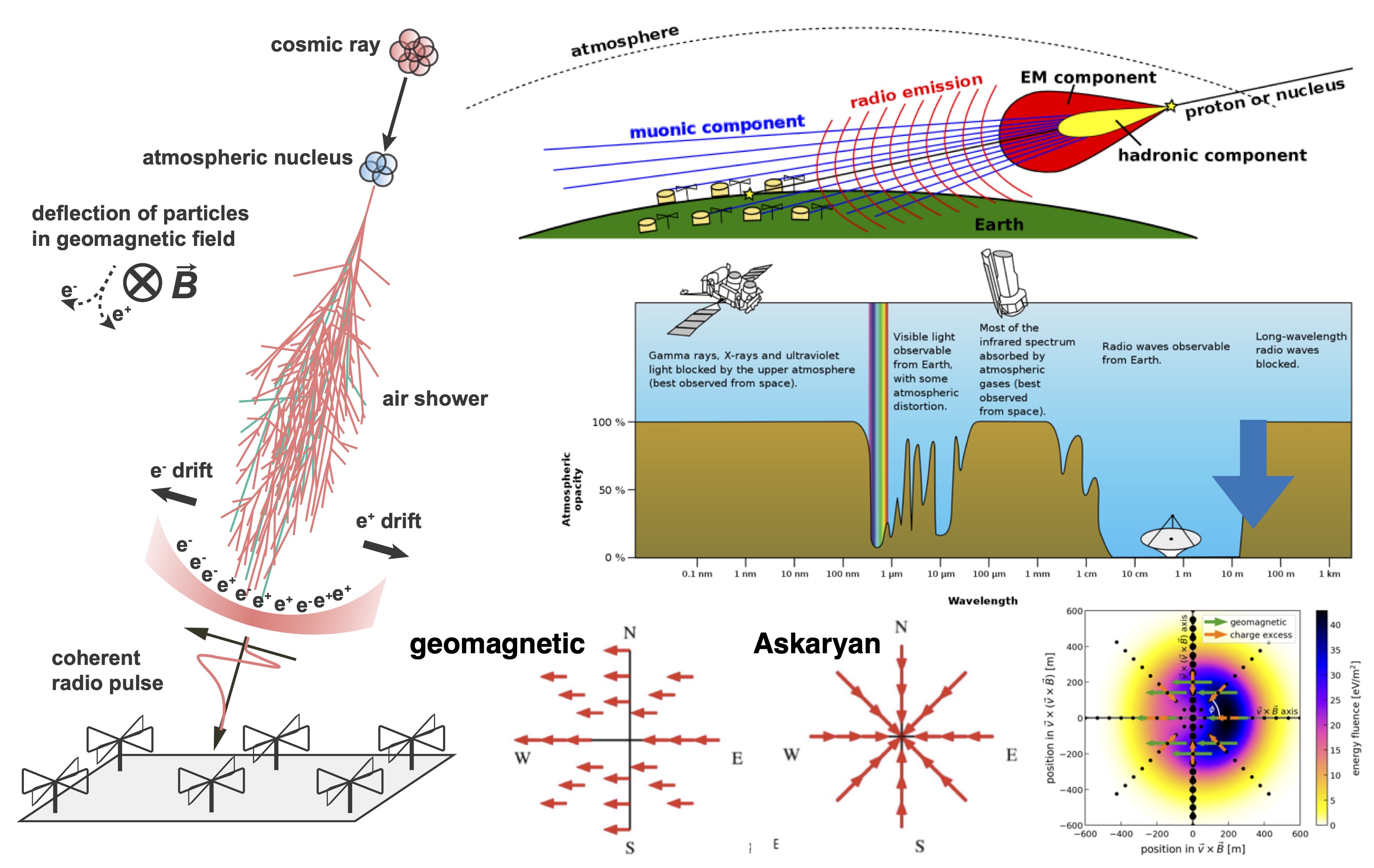}
 {\sl The basic principles of the radio detection of horizontal air showers, see text.}
 \label{radio-detection}
\end{wrapfigure}
The {\bf basic principles} of the radio detection of horizontal air
showers (HAS) with the RD are illustrated in the figure.
The radio emission in EASs originates from different processes. The dominant
mechanism is of geomagnetic origin \cite{kahnlerche,allanrev,Ardouin:2009zp}:
electrons and positrons in the shower are deflected in opposite directions by
the Lorentz force exerted by the magnetic field of the Earth. The generated
radio emission is linearly polarised in the direction of the Lorentz force
($\vec{v} \times \vec{B}$), where $\vec{v}$ is the propagation velocity vector
of the shower (parallel to the shower axis) and $\vec{B}$ represents the
direction and strength of the Earth magnetic field. A secondary contribution to
the radio emission results from the excess of electrons at the front of the
shower (Askaryan effect) \cite{askaryanexcess}. This excess is built up from
electrons that are knocked out of atmospheric molecules by interactions with
shower particles and by a net depletion of positrons due to annihilation. This
charge excess contribution is radially polarised, pointing towards the shower
axis. The resulting emission measured at the ground is the sum of both
components.
Interference between these components may be constructive or destructive,
depending on the position of the observer/antenna relative to the shower. The
emission is strongly beamed in the forward direction due to the relativistic
velocities of the particles. Additionally, the emission propagates through
the atmosphere, which has a non-unity index of refraction that changes with
height. This gives rise to relativistic time-compression effects, most
prominently resulting in a ring of amplified emission around the Cherenkov
angle \cite{Nelles:2014dja}. By precisely measuring the polarisation direction
of the electric field at various positions within the air shower footprint the
relative contribution of the main emission processes has been measured
\cite{Schellart:2014oaa,Aab:2014esa}.

The {\bf End-to-end calibration of the signal chain} of a radio detector is illustrated, using the Radio Detector of the Pierre Auger Observatory as an example. The active
components of the RD electronics, the low-noise amplifiers (LNAs)
and the digitisers have been thermally cycled to simulate ageing and take care
of eventual changes of the performance as a function of time.
All units have been end-to-end calibrated in the laboratory, recording
parameters like gain and phase shifts as a function of frequency.
The antenna pattern of the Short Aperiodic Loaded Loop Antenna (SALLA)
\cite{kroemer2009} has been simulated with the NEC software.
The antenna pattern has been verified through in-situ measurements (see e.g.\
\cite{PierreAuger:2017xgp}) in the field with a drone, carrying a reference
antenna in a defined pattern above a RD station.
\begin{wrapfigure}{r}{0.8\textwidth}
 \includegraphics[width=0.8\textwidth]{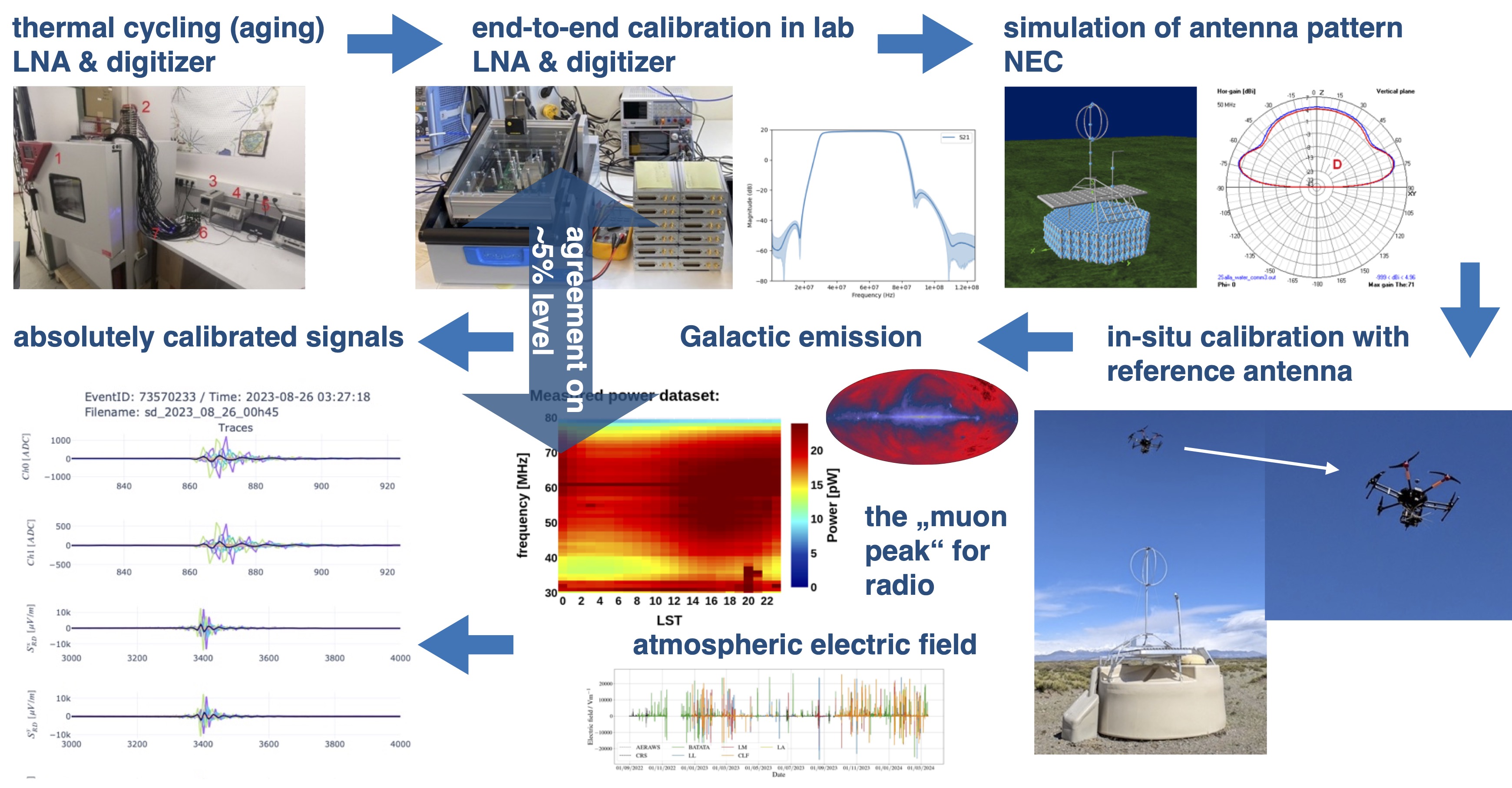}
 {\sl Calibration concept of the RD, see text.}
 \label{calibration}
\end{wrapfigure}
The diffuse Galactic radio emission is well measured \cite{Busken:2022mub},
and is used as a standard reference signal to calibrate the RD in-situ
\cite{PierreAuger:2021bwp}.
The galactic radio emission is recorded periodically on each station locally and
is used to correct for potential time-dependent changes in the parameters
(such as e.g.\ potential gain changes of the electronics as a function of
temperature).
It should be noted that the absolute calibration of the full electronics chain
(performed in the laboratory) agrees within 5\% uncertainty with the parameters
obtained from the Galactic emission. This demonstrates our excellent
understanding of the complete signal chain.
The atmospheric electric fields are continuously monitored at 5 positions in
the SD array. This allows to generate a veto against strong atmospheric
electric field during thunderstorms \cite{Schellart:2015kga}, which would
distort the energy measurements of a shower.
This yields absolutely calibrated time traces for each antenna and polarisation
direction.
A study with AERA demonstrates the long-term stability of the radio detection technique
\cite{PierreAuger:2023hun}. Only marginal deviations have been found over a
period of 10 years.

\flashtalkDone{Tim Huege}{Merits of radio detection within GCOS}

Radio detection would be a very valuable addition to GCOS. The best way the radio technique can contribute will, however, depend on the layout of a GCOS surface particle detector array, with which the radio detector will need to share common infrstructure for power and communications.

In the Pierre Auger Observatory, we are currently equipping each of the surface detector stations on the 1.5\,km grid with a SALLA antenna sensitive in the 30-80\,MHz range. This enables direct and accurate (to within 6\%) measurements of the energy in the electromagnetic component of air showers with zenith angles in the range from $\sim 65^{\circ}$ to $\sim 85^{\circ}$, contributing an aperture of $\sim 1300$\,km$^2$\,sr \cite{Huege:2023pfb} for contained events with a radio signal detectable in at least three radio antennas. Combined with a measurement of the muonic component by the surface detector, this will provide a high-quality data set for mass-composition studies \cite{Huege:2023pfb}, a goal that is also central to GCOS.

\begin{figure}[ht]
    \includegraphics[width=0.49\textwidth]{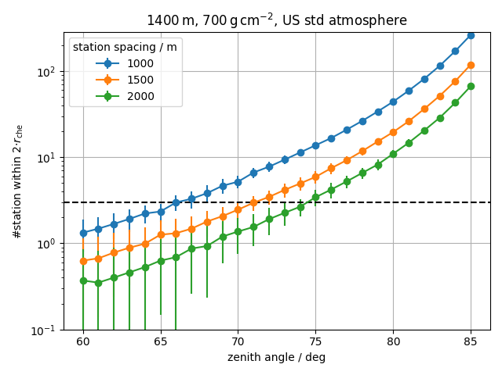}
    \includegraphics[width=0.49\textwidth]{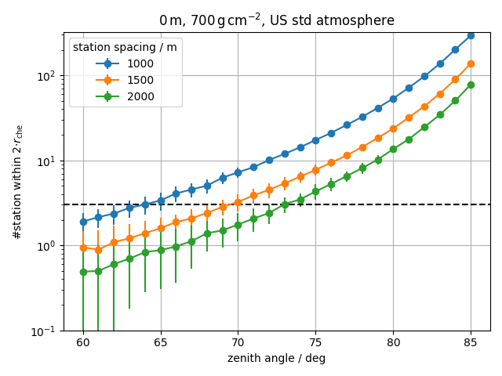}
    \caption{Multiplicity of expected radio detections in the presence of Galactic background in the 30-80\,MHz band for various detector spacings and for arrays at 1400\,m altitude (left) and at sea level (right), respectively. Figures made by Felix Schlüter (private communication).} \label{fig:multiplicity}
\end{figure}

If for GCOS a particle detector grid larger than 1.5\,km will be used, however, this will reduce the zenith angle range accessible to the radio detector to only the most inclined showers, and thus significantly reduce the achievable aperture, especially for contained events where a factor of $\cos \theta$ enters due to projection effects. The limitations are demonstrated in figure \ref{fig:multiplicity}, which shows the expected multiplicity of detectable radio signals as a function of zenith angle for different array spacings for a measurement in the 30-80\,MHz band for radio detector arrays at 1400\,m altitude (left) and sea level (right), respectively. A grid well beyond 1.5\,km does not lend itself well to coincident radio detection with three or more antennas.

A way out of this could be ``local clusters'', i.e., groups of (say) three radio antennas deployed in the vicinity of every surface detector station, connected to the surface detectors by cables and thus still profiting from the power and communications infrastructure of the surface detector array. If these antennas measured broad-band, say from 30\,MHz up to 1\,GHz, spectral information in the radio signal could be used to constrain the event geometry (in particular core position) and thus enable derivation of an energy estimate from a radio measurement with the local cluster of antennas only, as previously demonstrated by ARIANNA \cite{Barwick:2016mxm}. Coincident detection with radio antennas across local clusters would then not be required. The achievable quality for the reconstruction of the electromagnetic energy as a function of antenna cluster size would need to be worked out with a dedicated simulation study.

Application of the radio-interferometric technique \cite{Schoorlemmer:2020low}, which has the potential to determine the depth of shower maximum for inclined air showers, requires not only time-synchronization of the distributed detectors on a level of 1\,ns (or better for higher frequencies than 80\,MHz) but also a relatively high number of antennas measuring radio signals \cite{Schluter:2021egm} per event, so again does not lend itself well to grid spacings of larger than 1.5\,km and can likely not be applied to measurements with only local clusters.

A very strong case for radio detection I see in terms of an infill array on a grid of 1.5\,km or denser. Such a radio infill array would allow us to set the energy scale of the surface detector on the basis of first-principle electrodynamics calculations \cite{PierreAuger:2016vya} in combination with an absolute detector calibration exploiting Galactic radio emission \cite{Busken:2022mub}. Such an infill array would also allow us to accurately track aging effects in surface and fluorescence detectors as radio antennas have been demonstrated to exhibit no relevant aging \cite{PierreAuger:2023hun}. These two goals alone would merit deployment of a radio detector within GCOS.

\flashtalkDone{Tomas Fodran}{EAS array visualizer}

The presented Python package - \textit{EAS array visualizer}\cite{Fodran_EAS_array_visualizer} displays analytically calculated energy fluence at stations on a projected station array. Currently, options for square, hexagonal, octagonal and circular array layouts with triangular and square stations layout are available. Array size is set by the number of stations on the side (or in diameter for the circular layout) and station distance. After creating the array, the total number of stations is shown above the array figure, and the array dimensions can be checked by looking at the x and y scale in kilometres. Users can also load their customarily made arrays. The custom array is encoded in a four-column CSV file where the first column is the station number, and the remaining three are the station X, Y and Z coordinates. The array library contains one such file - an array layout of Pierre Auger Observatory. The plan is to fill the library also with arrays of other observatories such as TA. The magnetic field vector is also configurable, and its direction in the X-Y plane can be displayed on the array figure.

After selecting the air shower properties - energy, azimuth, zenith, and core position, the energy fluencies are calculated for all the stations, and the results are shown using a colour scale on the array figure and also on a more quantitative histogram which shows the energy fluence distribution of the triggered stations (threshold is set to 5 eV/m$^2$). Above the histogram, the number of triggered and untriggered stations is shown.

Currently, only energy fluence from hadronic showers can be displayed. The energy fluences are calculated with the function $\textit{LDF\_geo\_ce}$ from the \textit{geoceLDF} python package\cite{geoceLDF} using average \xmax values measured by Pierre Auger Observatory\cite{PierreAuger2014, Sanchez-Lucas2017}. The \xmax values are provided to the program by a separate file which can be altered if desired. The coordinate transformations are done by the radio tool package\cite{radiotools}. A future goal is to implement the neutrino toy model by Carvalho and Khakurdikar\cite{Washington2022}, and the parameterized frequency spectrum of radio emission by Martinelli et al.\cite{Sara2022}.

The application serves several purposes. Its primary purpose is new array projections. One can quickly check how many stations will be triggered for different shower geometries and energies for different types of arrays with different station spacing and layout. Moreover, this tool can be used before air shower simulations to get a rough idea of the expected result. Lastly, it is a tool that can cosmic ray EAS beginners play with to build an intuitive feeling for EAS.
The package can be downloaded \href{https://github.com/F-Tomas/EAS_array_visualizer}{here}.

 \begin{figure}[!htb]
    \centering
    \includegraphics[width=0.78\linewidth]{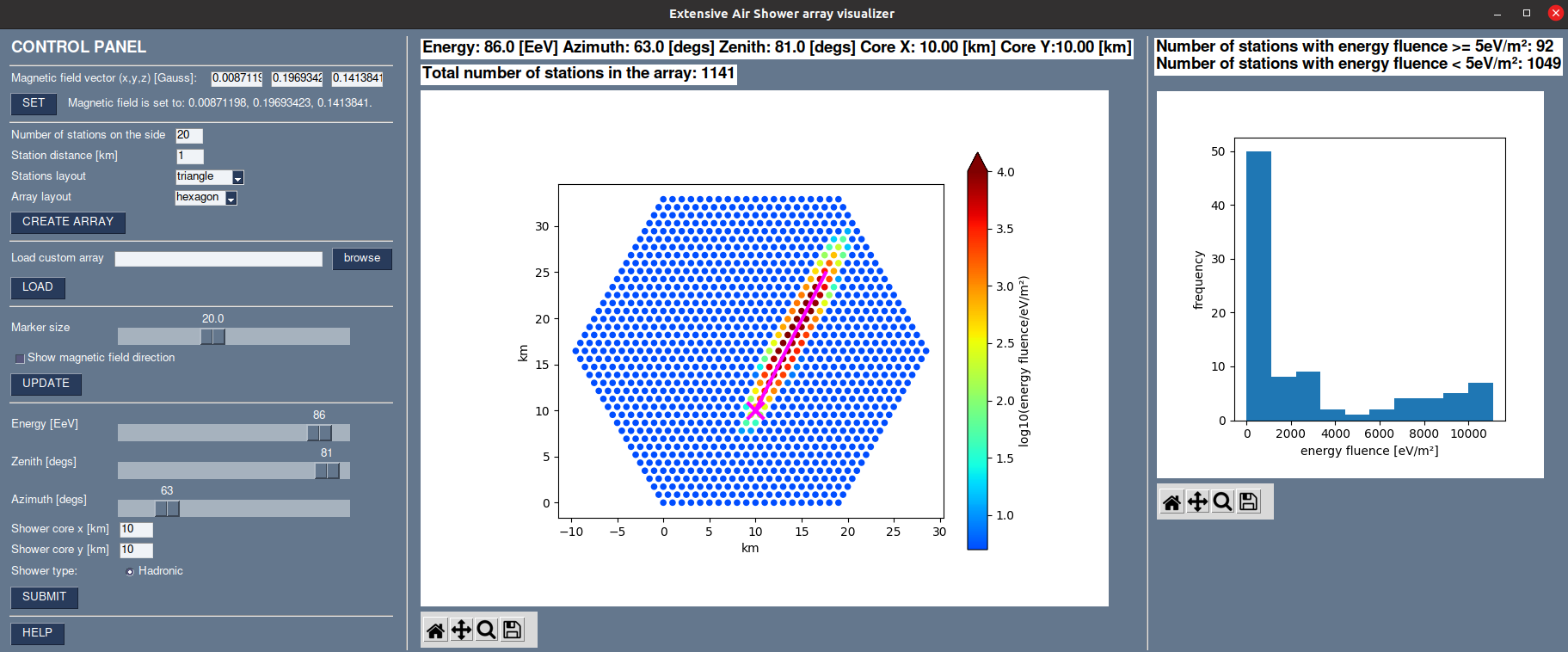}
    \caption{Screenshot of the \textit{EAS array visualizer}.}
\end{figure}

\subsubsection{Further Contributions}
{Julian Rautenberg -- Considerations about triggers and photons}\\
{\footnotesize \url{https://agenda.astro.ru.nl/event/21/contributions/283/attachments/64/73/RD_flash.pdf}}
\\
{Bjarni Pont  -- Hybrid lessons from a radio perspective} \\
{\footnotesize \url{https://indico.iihe.ac.be/event/1729/contributions/3494/attachments/2120/2642/20230610_GCOSworkshop_RadioCalibration.pdf}}
\\

\bibliographystyle{uhecr}
\bibliography{references}

\end{document}